\documentclass[sigconf, nonacm]{acmart}
\usepackage{balance} 
\usepackage{bm}
\usepackage{etoolbox}
\usepackage{marginnote}
\makeatletter
\let\oldmarginnote\marginnote
\renewcommand*{\marginnote}[1]{%
	\begingroup%
	\ifodd\value{page}
	\if@firstcolumn\normalmarginpar\fi
	\else
	\if@firstcolumn\else\normalmarginpar\fi
	\fi
		\oldmarginnote{}
	\endgroup%
}
\makeatother

\usepackage{color,soul}
\setul{0.5ex}{0.3ex}
\definecolor{Green}{rgb}{0,0,0}
\setulcolor{white}

\newcommand\vldbdoi{10.14778/3425879.3425887}
\newcommand\vldbpages{163 - 175}
\newcommand\vldbvolume{14}
\newcommand\vldbissue{2}
\newcommand\vldbyear{2021}
\newcommand\vldbauthors{\authors}
\newcommand\vldbtitle{\shorttitle} 
\newcommand\vldbavailabilityurl{https://github.com/tgbnhy/fast-kmeans}
\newcommand\vldbpagestyle{empty} 

\usepackage{booktabs}
\usepackage[linesnumbered,ruled,vlined]{algorithm2e}
\usepackage{colortbl,makecell}
\usepackage{url}
\usepackage{microtype}
\usepackage{booktabs} 

\usepackage{tikz}
\usetikzlibrary{tikzmark}
\usetikzlibrary{decorations.pathreplacing}
\usetikzlibrary{arrows}
\newcommand{\kmeans}{{\small{\textsf{$k$-means}}}\xspace}

\newcommand{\bao}[1]{\textrm{\textcolor{blue}{Bao says: #1}}}

\newcommand{\sheng}[1]{{\color{red}{#1}}}%
\newcommand{\shengnew}[1]{{{#1}}}
\newcommand{\myparagraph}[1]{\vspace{0.3\baselineskip}\noindent{\textbf{#1.}}~}

\newcommand{\var}[1]{\mbox{\emph{#1}}}

\DeclareMathOperator*{\argmin}{arg\,min}
\newcommand{\udash}[1]{%
	\tikz[baseline=(todotted.base)]{
		\node[inner sep=1pt,outer sep=0pt] (todotted) {#1};
		\draw[dashed] (todotted.south west) -- (todotted.south east);
	}%
}%
\def\D{\hphantom{1}}

\newtheorem{definition1}{Definition}

\newtheorem{example1}{Example}
\newcommand\like[1]{\begin{picture}(1,1)
	\ifnum0=#1\put(.5,.35){\circle{1}}\else
	\ifnum10=#1\put(.5,.35){\circle*{1}}\else
	\put(.5,.35){\circle{1}}\put(.5,.35){\circle*{.#1}}
	\fi\fi\end{picture}}

\newcommand{\knob}{\includegraphics[height=0.3cm]{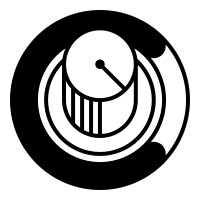}}

\newcommand{\earth}{\includegraphics[height=0.3cm]{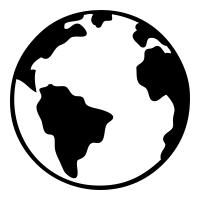}}
\newcommand{\laptop}{\includegraphics[height=0.3cm]{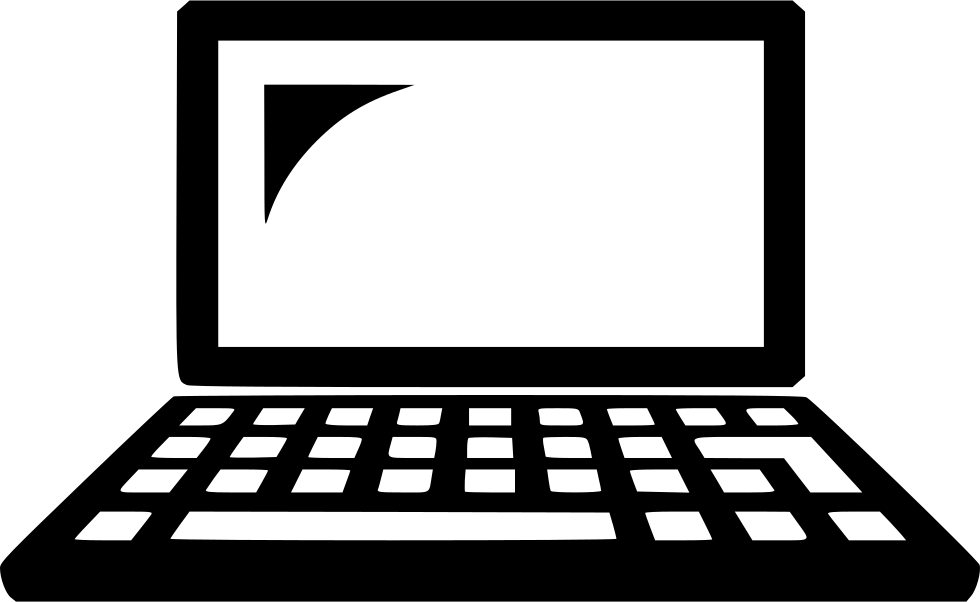}}
\newcommand{\server}{\includegraphics[height=0.3cm]{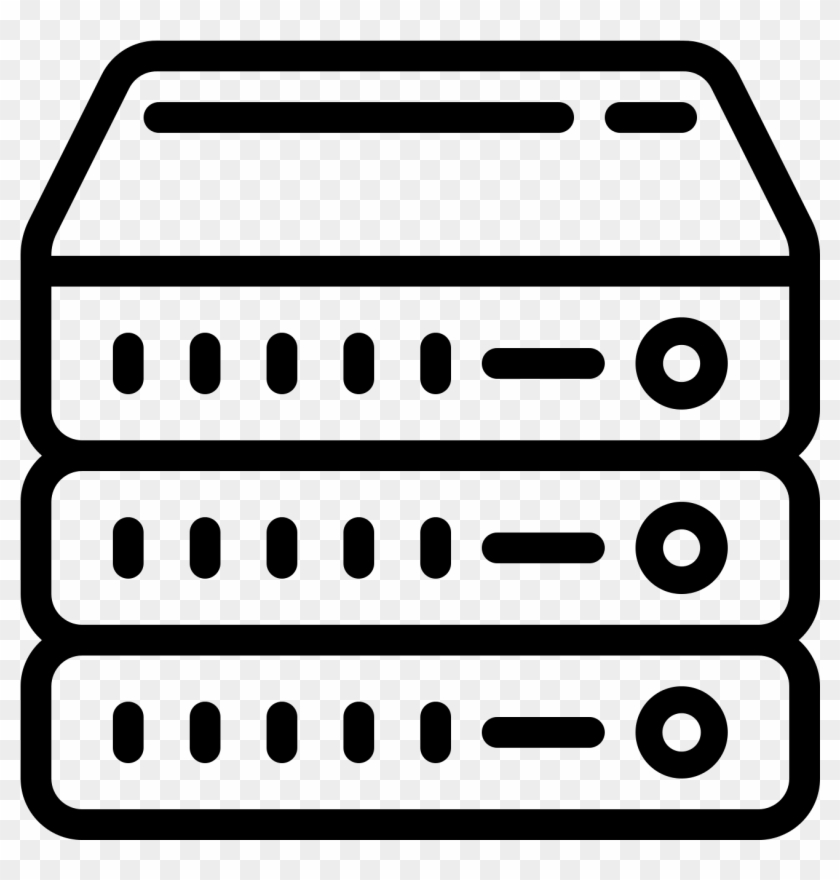}}

\newcommand{\ra}[1]{\renewcommand{\arraystretch}{#1}}

\usepackage{epstopdf}
\usepackage{mathtools}
\usepackage{cellspace}
\usepackage{makecell}
\usepackage{float}

\usepackage{multirow}

\usepackage[group-separator={,}]{siunitx}

\newcommand{\lyd}{{\small \texttt{Lloyd}}\xspace}%

\newcommand{\elkan}{{\small \texttt{Elka}}\xspace}

\newcommand{\sear}{{\small \texttt{Search}}\xspace}

\newcommand{\yiny}{{\small \texttt{Yinyang}}\xspace}

\newcommand{\ind}{{\small \texttt{INDE}}\xspace}%

\newcommand{\seq}{{\small \texttt{SEQU}}\xspace}

\newcommand{\uni}{{\small \texttt{UniK}}\xspace}%

\newcommand{\utune}{{\small \texttt{UTune}}\xspace}%

\definecolor{light-gray}{gray}{0.95}
\newcommand{\cg}{\cellcolor{lightgray}}
\newcommand{\algrule}[1][.2pt]{\par\vskip.5\baselineskip\hrule height #1\par\vskip.5\baselineskip}

\definecolor{issuecolor}{RGB}{0,166,81}

\newcounter{cN}
\usepackage{enumitem}
\setlist[itemize]{leftmargin=*}
\usepackage{subfig}
\begin{document}
\sloppy

\title{On the Efficiency of K-Means Clustering: Evaluation, Optimization, and Algorithm Selection}

\author{Sheng Wang}
\affiliation{%
	\institution{New York University}
}
\email{swang@nyu.edu}

\author{Yuan Sun}
\affiliation{%
	\institution{RMIT University}
}
\email{yuan.sun@rmit.edu.au}

\author{Zhifeng Bao}
\affiliation{%
	\institution{RMIT University}
}
\email{zhifeng.bao@rmit.edu.au}

\begin{abstract}
{This paper presents a thorough evaluation of the existing methods that accelerate Lloyd's algorithm for fast \kmeans clustering. To do so, we analyze the pruning mechanisms of existing methods, and summarize their common pipeline into a unified evaluation framework \uni. \uni embraces a class of well-known methods and enables a fine-grained performance breakdown. Within \uni, we thoroughly evaluate the pros and cons of existing methods using multiple performance metrics on a number of datasets. Furthermore, we derive an optimized algorithm over \uni, which effectively hybridizes multiple existing methods for more aggressive pruning. To take this further, we investigate whether the most efficient method for a given clustering task can be automatically selected by machine learning, to benefit practitioners and researchers.}

\end{abstract}

\maketitle

\pagestyle{\vldbpagestyle}
\begingroup\small\noindent\raggedright\textbf{PVLDB Reference Format:}\\
\vldbauthors. \vldbtitle. PVLDB, \vldbvolume(\vldbissue): \vldbpages, \vldbyear.\\
\href{https://doi.org/\vldbdoi}{doi:\vldbdoi}
\endgroup
\begingroup
\renewcommand\thefootnote{}\footnote{\noindent
	This work is licensed under the Creative Commons BY-NC-ND 4.0 International License. Visit \url{https://creativecommons.org/licenses/by-nc-nd/4.0/} to view a copy of this license. For any use beyond those covered by this license, obtain permission by emailing \href{mailto:info@vldb.org}{info@vldb.org}. Copyright is held by the owner/author(s). Publication rights licensed to the VLDB Endowment. \\
	\raggedright Proceedings of the VLDB Endowment, Vol. \vldbvolume, No. \vldbissue\ %
	ISSN 2150-8097. \\
	\href{https://doi.org/\vldbdoi}{doi:\vldbdoi} \\
}\addtocounter{footnote}{-1}\endgroup

\ifdefempty{\vldbavailabilityurl}{}{
	\vspace{.3cm}
	\begingroup\small\noindent\raggedright\textbf{PVLDB Artifact Availability:}\\
	The source code, data, and/or other artifacts have been made available at \url{\vldbavailabilityurl}.
	\endgroup
}

\section{Introduction}
\label{submission}

As one of the most widely-used clustering algorithms, \kmeans aims to partition a set of $n$ points into $k$ ($k<n$) clusters where each point is assigned to the cluster with the nearest centroid~\cite{Jain2010,Wu2008}. Answering \kmeans is NP-hard and Lloyd's algorithm \cite{Lloyd1982} is a standard approach. Essentially, it randomly initializes $k$ centroids, then assigns each point to the cluster with the nearest centroid and refines each centroid iteratively. In each iteration, it needs to compute $n \cdot k$ distances in the assignment step and access $n$ data points in the refinement step. Such intensive computations make the Lloyd's algorithm slow, especially in partitioning large datasets. 

{Accelerating the Lloyd's algorithm for \kmeans clustering has been investigated for more than 20 years since the first work was published \cite{Pelleg1999}.
Most of the existing acceleration methods focus on how to reduce intensive distance computations, which can be broadly divided into two categories: 1) the index-based methods that group and prune points in batch \cite{Pelleg1999,Moore2000,Kanungo2002,Broder2014}, and 2) the sequential methods that scan each point one by one and utilize a bound based on triangle inequality to avoid calculating certain distance~\cite{Phillips2002,Elkan2003,Hamerly2010,Drake2012,Drake2013,Ding2015,Hamerly2015a,Newling2016b,Rysavy2016a,Bottesch2016,Kwedlo2017,Curtin2017,Xia2020}.
	\begin{figure}
	\centering
	\scalebox{1}{\includegraphics[width=0.42\textwidth]{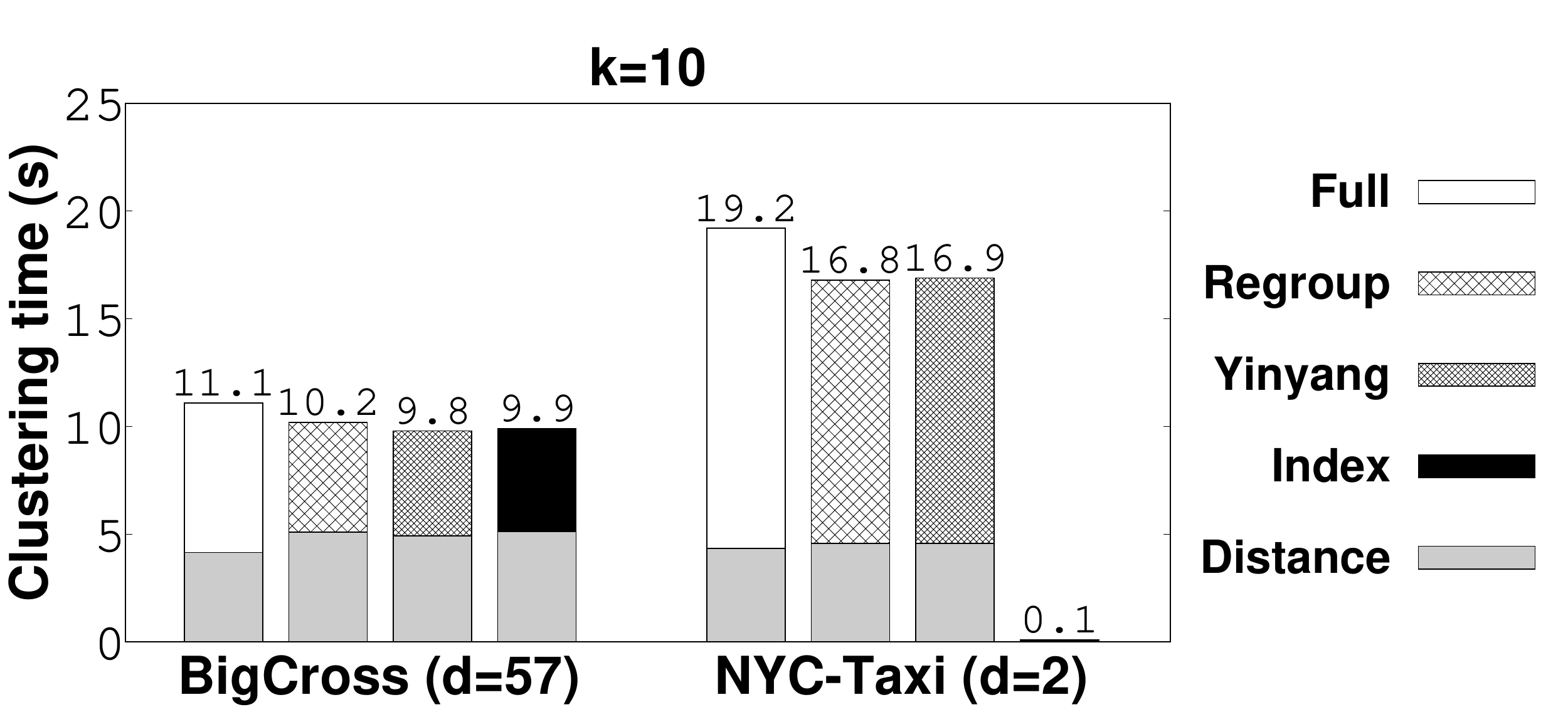}}
	\vspace{-1em}
	\caption{The performances of representative methods, {\small\texttt{Regroup}}, {\small\texttt{Yinyang}}, {\small\texttt{Index}} and {\small\texttt{Full}} (using multiple pruning mechanisms) on two clustering datasets, where $d$ is the dimensionality of the dataset. The gray bar (``{\small\texttt{Distance}}'') denotes the time taken by each method to compute distance.}
	\vspace{-1.5em}
	\label{fig:motivation}
\end{figure}
\vspace{-0.5em}
\subsection{Motivations}
\label{sec:moti}
\vspace{-0.5em}

 \myparagraph{Conducting a thorough evaluation of existing \kmeans algorithms} Whilst a large body of methods have been proposed to accelerate the Lloyd’s algorithm for \kmeans clustering, there is still a lack of thorough evaluation on the efficiency of these methods. Moreover, there seems to be some misunderstanding on the performance of certain methods in the literature. For example, the index-based method \cite{Kanungo2002} was interpreted to be slower compared to the sequential methods (e.g., {\small\texttt{Yinyang}} \cite{Ding2015}, {\small\texttt{Regroup}} \cite{Rysavy2016a}) when the dimensionality of dataset is greater than 20 \cite{Ding2015}, and hence was discarded by the machine learning (ML) community in its most recent studies \cite{Ding2015,Newling2016b,Rysavy2016a}. However, we show in Figure~\ref{fig:motivation} that the index-based method is in fact relatively fast and has the potential to significantly accelerate large-scale clustering when using a proper data structure. This motivates us to conduct a fair and more thorough efficiency evaluation on existing methods.

 In fact, most existing studies considered reducing the number of distance computations as the main goal to improve the efficiency of their methods. However, a method that computes fewer number of distances does not simply guarantee to have a shorter computational time. For example in Figure~\ref{fig:motivation}, the {\small \texttt{Full}} method, which is armed with multiple pruning techniques, has the least number of distance computation, but overall is the slowest on the BigCross dataset. \marginnote{W2@R3 \ref{sec:r3}}\ul{This is because other metrics, such as the number of data accesses and the time taken to compute a bound for pruning, also contribute to the computational cost.} \ul{To identify the key metrics, it is essential to analyse the pruning mechanisms of existing methods and extract a unified framework, such that existing methods can well fit to enable a fine-grained performance breakdown of existing methods and in turn a more comprehensive evaluation.}
	
\myparagraph{Selecting the best \kmeans algorithm for a given task} Fast \kmeans clustering for an arbitrary dataset has attracted much attention \cite{Lulli2016}. Unfortunately, there is no single algorithm that is expected to be the ``fastest'' for clustering all datasets, which is also in line with the \textit{``no free lunch" theorem} in optimization \cite{Wolpert1997}. That calls for an effective approach that is able to select the best algorithm for a given clustering task. However, existing selection \shengnew{criteria} are still based on simple rules, e.g., choosing the index-based method when the dimensionality of dataset is less than 20. Given the complex nature of clustering, they are unlikely to work well in practice. 
%
Given the large amount of data collected from our evaluations, it is natural to apply ML to learn an optimal mapping from a clustering task to the best performing algorithm. Note that the idea of using ML for algorithm selection \cite{Munoz2015} has been explored before, e.g., \textit{meta-learning} \cite{Vanschoren2018} and auto-tuning in database management \cite{Li2018g,VanAken2017}. However, as we will see shortly, it is nontrivial to apply this technique to \kmeans clustering because problem-specific features have to be carefully designed to describe datasets to be clustered. 

}
\subsection{Our Contributions}
{In this paper, we design a unified experimental framework to evaluate various existing \kmeans clustering methods, and design an ML approach to automatically select the best method for a given clustering task.} More specifically, we make the following contributions:

\begin{itemize}
	
	\item We review the index-based and sequential methods and describe their pruning mechanisms in Sections~\ref{sec:kmeans} and~\ref{sec:seq}.
	
	
	
	\item \marginnote{D2@Meta \\\ref{sec:meta}}\ul{Inspired by the common pruning pipeline of existing methods, we design a unified evaluation framework} \uni in Section~\ref{sec:frame}, that \ul{enables us to compare existing methods more fairly and comprehensively.} Some key insights obtained are: 1) the index-based method can be very fast even for high-dimensional data, when equipped with a proper data structure such as Ball-tree \cite{Uhlmann1991}; and 2) no single method can always perform the best across all cases. Detailed evaluations are in Section~\ref{sec:expunik}.

	\item \ul{The above further motivates us to design an adaptive setting for our} \uni, \marginnote{D2@Meta \\ \ref{sec:meta}}\ul{which applies the bound-based pruning from sequential methods to assign points in batch without scanning all centroids.} 
	In Section~\ref{sec:expuni}, we evaluate our adaptive \uni and show that it outperforms the existing \kmeans algorithms when tested on various real-world datasets.
	
	\item To take it further, we adopt ML to automatically select the best method for a given clustering task in Section~\ref{sec:auto}. This is achieved by learning from our evaluation records which contain the performance of all the methods on various datasets. An evaluation on multiple learning models is conducted in Section~\ref{sec:expauto}.
	
	
	
	
	
\end{itemize}
\section{Preliminaries}\label{sec:pre}
\vspace{-1em}

\hspace{-3.5em}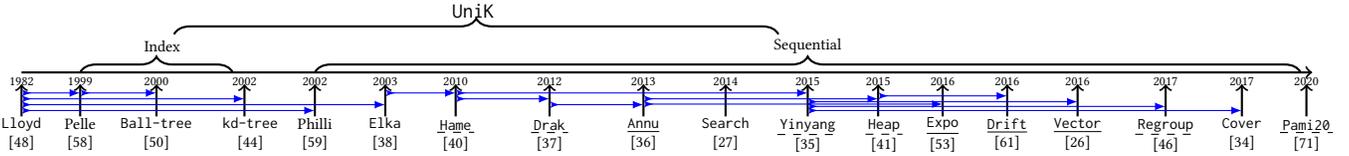
\begin{figure*}
\begin{tikzpicture}[scale=0.78]
\tikzstyle{every node}=[font=\scriptsize]
\draw [thick,->] (0,0.05) -- (22,0.05);
\node[align=center] at (0,-0.1) {{\tiny 1982}};
\node[align=center] at (0,-1) {\texttt{Lloyd} \\\cite{Lloyd1982}};
\draw [thick,->] (0,-0.7) -- (0,-0.15);

\node[align=center] at (2.4,0.5) {Index};
\draw [thick,decorate,decoration={brace,amplitude=6pt,raise=0pt}] (1,0.05) -- (3.6,0.05);

\node[align=center] at (13.4,0.5) {Sequential};
\draw [thick,decorate,decoration={brace,amplitude=6pt,raise=0pt}] (5,0.05) -- (21.8,0.05);

\node[align=center] at (1,-0.1) {{\tiny 1999}};
\node[align=center] at (1,-1) {Pelle \\ \cite{Pelleg1999}};
\draw [thick,->] (1,-0.7) -- (1,-0.15);

\node[align=center] at (2.3,-0.1) {{\tiny 2000}};
\node[align=center] at (2.3,-1) {\scriptsize \texttt{Ball-tree} \\ \cite{Moore2000}};
\draw [thick,->] (2.3,-0.7) -- (2.3,-0.15);

\node[align=center] at (3.8,-0.1) {{\tiny 2002}};
\node[align=center] at (3.9,-1) {\scriptsize \texttt{kd-tree} \\\cite{Kanungo2002}};
\draw [thick,->] (3.8,-0.7) -- (3.8,-0.15);


\node[align=center] at (5,-0.1) {{\tiny 2002}};
\node[align=center] at (5,-1) {\scriptsize Philli \\\cite{Phillips2002}};
\draw [thick,->] (5,-0.7) -- (5,-0.15);

\node[align=center] at (6.2,-0.1) {{\tiny 2003}};
\node[align=center] at (6.2,-1) {\texttt{Elka} \\\cite{Elkan2003}};
\draw [thick,->] (6.2,-0.7) -- (6.2,-0.15);

\node[align=center] at (7.4,-0.1) {{\tiny 2010}};
\node[align=center] at (7.4,-1) {\udash{\texttt{Hame}} \\\cite{Hamerly2010}};
\draw [thick,->] (7.4,-0.7) -- (7.4,-0.15);

\node[align=center] at (9,-0.1) {{\tiny 2012}};
\node[align=center] at (9,-1) {\udash{\texttt{Drak}} \\\cite{Drake2012}};
\draw [thick,->] (9,-0.7) -- (9,-0.15);

\node[align=center] at (10.6,-0.1) {{\tiny 2013}};
\node[align=center] at (10.6,-1) {\underline{\texttt{Annu}} \\ \cite{Drake2013}};
\draw [thick,->] (10.6,-0.7) -- (10.6,-0.15);

\node[align=center] at (12,-0.1) {{\tiny 2014}};
\node[align=center] at (12,-1) {\texttt{Search} \\ \cite{Broder2014}};
\draw [thick,->] (12,-0.7) -- (12,-0.15);

\node[align=center] at (13.4,-0.1) {{\tiny 2015}};
\node[align=center] at (13.4,-1) {\udash{\texttt{Yinyang}} \\ \cite{Ding2015}};
\draw [thick,->] (13.4,-0.7) -- (13.4,-0.15);

\node[align=center] at (14.6,-0.1) {{\tiny 2015}};
\node[align=center] at (14.7,-1) {\udash{\texttt{Heap}} \\ \cite{Hamerly2015a}};
\draw [thick,->] (14.6,-0.7) -- (14.6,-0.15);

\node[align=center] at (15.7,-0.1) {{\tiny 2016}};
\node[align=center] at (15.7,-1.01) {\underline{\texttt{Expo}} \\\cite{Newling2016b}};
\draw [thick,->] (15.7,-0.7) -- (15.7,-0.15);

\node[align=center] at (16.8,-0.1) {{\tiny 2016}};
\node[align=center] at (16.8,-1) {\underline{\texttt{Drift}} \\ \cite{Rysavy2016a}};
\draw [thick,->] (16.8,-0.7) -- (16.8,-0.15);

\node[align=center] at (18,-0.1) {{\tiny 2016}};
\node[align=center] at (18,-1) {\underline{\texttt{Vector}} \\ \cite{Bottesch2016}};
\draw [thick,->] (18,-0.7) -- (18,-0.15);

\node[align=center] at (19.5,-0.1) {{\tiny 2017}};
\node[align=center] at (19.5,-1) {\udash{\texttt{Regroup}}\\ \cite{Kwedlo2017}};
\draw [thick,->] (19.5,-0.7) -- (19.5,-0.15);

\node[align=center] at (20.8,-0.1) {{\tiny 2017}};
\node[align=center] at (20.8,-1) {\texttt{Cover} \\ \cite{Curtin2017}};
\draw [thick,->] (20.8,-0.7) -- (20.8,-0.15);

\node[align=center] at (21.9,-0.1) {{\tiny 2020}};
\node[align=center] at (21.9,-1) {\udash{\texttt{Pami20}} \\ \cite{Xia2020}};
\draw [thick,->] (21.9,-0.7) -- (21.9,-0.15);

\node[align=center] at (7.7,1.1) {\uni};
\draw [thick,decorate,decoration={brace,amplitude=6pt,raise=0pt}] (2.6,0.7) -- (12.9,0.7);

\draw [thin,latex' reversed-latex,blue] (0,-0.3) -- (1,-0.3);
\draw [thin,latex' reversed-latex,blue] (1,-0.3) -- (2.3,-0.3);
\draw [thin,latex' reversed-latex,blue] (0,-0.4) -- (3.8,-0.4);
\draw [thin,latex' reversed-latex,blue] (0,-0.5) -- (6.2,-0.5);
\draw [thin,latex' reversed-latex,blue] (0,-0.6) -- (5,-0.6);
\draw [thin,latex' reversed-latex,blue] (6.2,-0.3) -- (7.4,-0.3);
\draw [thin,latex' reversed-latex,blue] (7.4,-0.4) -- (9,-0.4);
\draw [thin,latex' reversed-latex,blue] (9,-0.5) -- (10.6,-0.5);
\draw [thin,latex' reversed-latex,blue] (7.4,-0.3) -- (13.4,-0.3);
\draw [thin,latex' reversed-latex,blue] (10.6,-0.4) -- (14.6,-0.4);
\draw [thin,latex' reversed-latex,blue] (10.6,-0.49) -- (15.7,-0.49);
\draw [thin,latex' reversed-latex,blue] (14.6,-0.35) -- (16.8,-0.35);
\draw [thin,latex' reversed-latex,blue] (13.4,-0.45) -- (18,-0.45);
\draw [thin,latex' reversed-latex,blue] (13.4,-0.53) -- (19.5,-0.53);
\draw [thin,latex' reversed-latex,blue] (13.4,-0.6) -- (20.8,-0.6);

\end{tikzpicture}
	\vspace{-2.5em}
	\caption{The research timeline of fast Lloyd's algorithms for \kmeans, where the blue arrows show the successive relationship, and {the underlines differentiate two types of sequential methods covered in Section~\ref{sec:less} (dash lines) and \ref{sec:tightbound}.}}
	\vspace{-1.5em}
\label{fig:timeline}
\end{figure*}

\begin{figure}
	\centering
	\scalebox{1}{\includegraphics[height=2.8cm]{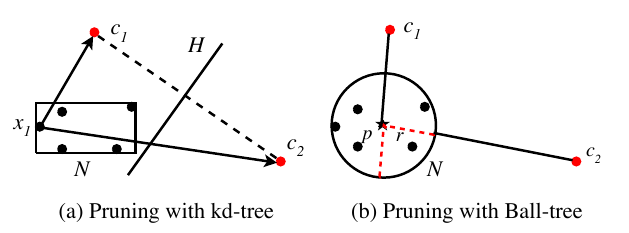}}	
	\vspace{-1em}
	\caption{Example of assignment of node $N$ in kd-tree and Ball-tree, based on a hyperplane $H$ and a ball with radius $r$.}
	\label{fig:node}
	\vspace{-1.2em}
\end{figure}

\noindent Given a dataset $D=\{x_1, x_2, \cdots, x_n\}$ of $n$ points, and each point has $d$ dimensions, \kmeans aims to partition $D$ into $k$ mutually exclusive subsets $S=\{S_1, S_2, \cdots,S_k\}$ to minimize the \textit{Sum of Squared Error},

\vspace{-1em}
\begin{equation}
\small
\argmin\limits_{S} \sum_{j=1}^{k}\sum_{x\in S_j}\|x-c_j\|^2,
\end{equation}
\vspace{-1em}

\noindent where $\small c_j=\frac{1}{|S_j|}\sum_{x\in S_j}x$, namely the \textit{centroid}, is the mean of points in $S_j$.

\subsection{Lloyd's Algorithm}
With the initialized $k$ centroids, the Lloyd's algorithm for \kmeans \cite{Lloyd1982} conducts the assignment and refinement in an iterative manner until all the centroids do not change.

\myparagraph{Initialization} It randomly chooses $k$ points in $D$ as the initial centroids. Normally, k-means++ \cite{Arthur2007} is the default initialization method which aims to make $k$ centroids far away from each other.
	
\myparagraph{Assignment} It needs to assign each of the $n$ data points to a cluster with the nearest centroid, and therefore requires $n\cdot k$ number of distance computations. 
	
\myparagraph{Refinement} It needs to read every data point in a cluster to update the centroid. Hence $n$ data accesses are conducted.

\vspace{-0.5em}
\begin{example1}
	Figure~\ref{fig:node} shows two centroids $c_1, c_2$ (red) and five data points (black) bounded by a node $N$.
	The assignment step in Lloyd's algorithm computes the distance from every point $x_i$ to every centroid $c_j$ to determine the nearest cluster.\footnote{\scriptsize By default, index $i$ and $j$ refer to data and cluster index respectively in the rest of this paper.}
\end{example1}

\vspace{-0.5em}
\subsection{Acceleration}
\label{sec:accl}
Given a dataset and $k$, there are four types of acceleration for fast \kmeans with the Lloyd's algorithm:
\begin{itemize}\setlength\itemsep{-0.3em}
	\item \myparagraph{Hardware Acceleration} Parallelization \cite{Zhao2009}, GPU \cite{Zechner2009}, and cache \cite{Bender2015} can accelerate it at physical level.
	
	\item \myparagraph{Approximate Acceleration} It aims to find approximate clustering results within a bounded error w.r.t. the exact result of the Lloyd's algorithm, by developing techniques like sampling \cite{Aggarwal2009} and mini-batch \cite{Newling2016a}.
	
	\item \myparagraph{Fast Convergence} \marginnote{D1@Meta \\\ref{sec:meta}}\ul{It uses efficient initialization techniques such as $k$-means++} \cite{Arthur2007,Bahmani2012}. As Celebi et al. \cite{Celebi2013a} \ul{have done an evaluation on this, it will not be our focus.}
	
	\item \myparagraph{Exact Lloyd's Algorithm} It focuses on reducing the number of distance computations in the Lloyd's algorithm, and {can be integrated with the above methods to reduce their running time.}
\end{itemize}

Clustering has been evaluated from different perspectives. For example, Muller et al. \cite{Muller2009} evaluated clustering in subspace projections of high-dimensional data; Hassanzadeh et al. \cite{Hassanzadeh2009a} proposed a framework to evaluate clustering for duplicate detection.
In contrast, this is the first work that evaluates all accelerating tricks to reduce distance computations in the exact Lloyd's algorithm.
Figure~\ref{fig:timeline} summarizes a timeline of fast Lloyd's algorithms. 

\section{{Index-based Algorithms}}
\label{sec:kmeans}

By assigning points to the nearest clusters in batch, index-based algorithms have been proposed to support fast \kmeans, 
such as kd-tree \cite{Kanungo2002,Pelleg1999} and Ball-tree \cite{Moore2000}.
Intuitively, if an index node that covers $m$ points is assigned to a cluster directly, then $m \cdot k$ number of distance computations and $m$ data accesses can be reduced.

\subsection{Typical Indexes}\label{sec:kd}
\myparagraph{kd-tree}
Indexing the dataset using kd-tree \cite{Bentley1975} can accelerate \kmeans with batch-pruning for low dimensional data~\cite{Kanungo2002,Pelleg1999}, where its intuition is presented in Figure~\ref{fig:node}(a): 
node $N$ locates in the hyperplane $H$ of $c_1$ completely, thus all the points in $N$ are closer to $c_1$ than $c_2$ and then $N$ cannot be assigned to $c_2$. Unfortunately, bisecting hyperplane in a high-dimensional space using kd-tree (Voronoi diagram is a common method) costs much time with a complexity of $\mathcal{O}(k \log k+k^{\lceil \frac{d}{2}\rceil})$ \cite{Chazelle1993}.

\myparagraph{Ball-tree}
Moore et al. \cite{Moore2000} proposed to use Ball-tree \cite{Uhlmann1991} (a.k.a. metric-tree) to group points using balls, which is a circle in a 2-dimensional space as shown in Figure~\ref{fig:node}(b). All the points in a ball $N$ are closer to centroid $c_1$ than $c_2$ if 
\begin{equation}
\label{equ:ball}
\|p-c_1\| + r < \|p-c_2\| - r,
\end{equation}
where $p$ and $r$ are the center and radius of the ball $N$. The left-hand side of Equation~\ref{equ:ball} is an upper bound on the distance from the points in $N$ to $c_1$, while the right-hand side is a lower bound on the distance from the points in $N$ to $c_2$. However, data points in a high-dimensional space are dispersed, resulting in a much bigger radius $r$ and degenerating its pruning power, i.e., it is unlikely to find a ball $N$ and two centroids $c_1$ and $c_2$ such that Equation~\ref{equ:ball} holds. In this case, using Ball-tree can be even slower than directly assigning each point to a cluster, due to the computational overhead of traversing the Ball-tree.






\myparagraph{Other Indexes}
Apart from kd-tree and Ball-tree, there are other data structures in metric space that can be used to prune distance computations \cite{Chavez2001,Samet2006}. This includes Hierarchical \kmeans tree \cite{Ross1975,Wangaclustering}, M-tree \cite{Ciaccia1997a} and Cover-tree \cite{Beygelzimer2006,Curtin2017}, to name a few. We will evaluate all these typical indexes in our experiment (Section~\ref{sec:exp_im}). 


\subsection{Pre-assignment Search}
Broder et al. \cite{Broder2014} proposed to search around a centroid $c_j$ the data points that lie within a distance threshold, and directly assign these points to $c_j$, in order to avoid certain distance computations. {For example in Figure~\ref{fig:node}(a), we can search for data points around $c_1$ within a distance threshold $\frac{\|c_1-c_2\|}{2}$, and the points found can be assigned to $c_1$ directly, as they are closer to $c_1$ than $c_2$. {Here, we can use an index like kd-tree or Ball-tree to conduct a fast \textit{similarity search} \cite{Chen2017c} to obtain those points within this distance threshold.}
Similar operation can be conducted on $c_2$. Then, we can use Lloyd's algorithm to assign the rest of data points to a cluster.}
For those points that cannot be assigned by this method, we then sequentially scan $k$ centroids to find a cluster to assign them. In this sense, it also belongs to the sequential algorithm to be introduced in Section~\ref{sec:seq}. We denote this method as \sear. 

\section{Sequential Algorithms}
\label{sec:seq}
%

Essentially, sequential algorithms try to exploit the triangle inequality to derive various forms of distance bounds, and use them to reduce the number of distance computations in the assignment step.
However, it is inevitable to access the stored data and update the distance bound of each point.
These extra costs actually account for a significant portion of the total running time.

\subsection{Triangle Inequality}
\label{sec:tri}

To check whether a point $x_i$ belongs to a cluster of centroid $c_j$,\footnote{{To facilitate our illustration, each cluster is simply represented by its centroid if no ambiguity is caused.}} we can first compare the lower bound on the distance between $x_i$ and $c_j$, denoted as $lb(i,j)$, against the upper bound on the distance from $x_i$ to its closest centroid $ub(i)$. If $lb(i,j) > ub(i)$, then $x_i$ does not belong to the cluster $c_j$ and we do not need to compute the distance between $x_i$ and $c_j$. Formally, we denote this inequality as 
\begin{equation}
\label{equ:notassign}
lb(i,j) > ub(i) \rightarrow a(i)\ne j,
\end{equation}
where $a(i)\ne j$ means that $x_i$ is not assigned to the cluster $c_j$. Since the pruning is conducted in a pairwise fashion (point, centroid), we call it \textit{\underline{local pruning}}.

{Elkan et al. \cite{Elkan2003} obtained the lower bound $lb(i,j)$ in the following way: (1) the distance between centroids is used to derive the first lower bound $lb(i,j) = \frac{\|c_{a(i)}-c_j\|}{2}$.\footnote{A similar idea was proposed by Phillips \cite{Phillips2002} one year earlier.} We call it as \textit{inter-bound}. (2) The centroid drift from the previous cluster $c_j^{'}$ to the current cluster $c_j$ is used to derive the second lower bound via the triangle inequality: $lb(i,j) = \|x_i-c_j^{'}\|-\|c_j^{'}-c_j\|<\|x_i-c_j\|$. We call it as \textit{drift-bound}. (3) Between these two bounds, the bigger one is chosen as the final $lb(i,j)$.
We denote the algorithm using both inter-bound and drift-bound as \elkan.}


The \textit{inter-bound} requires to compute the pairwise distance between centroids, and thus costs $\frac{k(k-1)}{2}$ number of computations. The \textit{drift-bound} uses $n \cdot k$ units of memory, as it stores a bound for each (point, centroid) pair. The main issues of \elkan are: 1) it uses much space; 2) the bounds derived might not be very tight. In what follows, we introduce the methods to address each of these issues. 


\subsection{Methods with a Focus of Less Bounds}
\label{sec:less}
\subsubsection{Hamerly Algorithm}
Instead of storing a lower bound for each centroid, storing a global bound can save much space.
Motivated by this, 
Hamerly \cite{Hamerly2010} proposed to store the minimum lower bound as the global bound.
Further, a global lower bound $lb(i)$ for each point $x_i$ is used before scanning every centroid, i.e., $\small lb(i)=
\max\big(\min_{j\ne a^{'}(i)}lb(i,j), \min_{j\ne a^{'}(i)} \frac{\|c_{a^{'}(i)}-c_j\|}{2}\big)$ where $a^{'}(i)$ points to previous cluster of $x_i$, $x_i$ can stay if: $
lb(i) > ub(i) \rightarrow a(i)=a^{'}(i)$.
We denote this algorithm as {\small\texttt{Hame}}, and it is known as the \textit{\underline{global pruning}}.

\subsubsection{Sort-Centers}
Beyond a single bound, Drake et al. \cite{Drake2012} proposed to store $b<k$ bounds for each point, which can reduce both the storage space and the distance computation by bound.
The $b$ points that are selected based on their closeness to the assigned centroid, so it needs to be updated frequently in each iteration.
Moreover, the selection of parameter $b$ is also highly dependable on datasets.
By default, we use a fixed ratio suggested in \cite{Drake2012}, i.e., $b=\lceil \frac{k}{4}\rceil$.
We denote this algorithm as {\small\texttt{Drak}}.

\subsubsection{Yinyang: Group Pruning}
\label{sec:yinyang}
\texttt{Hame} needs to conduct $k-1$ local pruning if the global pruning fails, i.e., $lb(i) \le ub(i)$,
Ding et al. \cite{Ding2015} proposed to divide $k$ centroids into $t=\lceil \frac{k}{10}\rceil$ groups when $k$ is much bigger than 10, and add a pruning on a group between local pruning and global pruning.
This is known as \textit{\underline{group pruning}}.

In the first iteration, $k$ centroids are divided into several groups based on \kmeans.
Each group will maintain a bound rather than each point.
However, the fixed group may lead to a loose bound with the increase of iterations.
Kwedlo et al. \cite{Kwedlo2017} regrouped the $k$ centroids in each iteration while \cite{Ding2015} only did it in the first iteration, and the grouping used a more efficient way than \kmeans, and the group bounds become tighter.
We denote these two algorithms as {\small\texttt{Yinyang}} and {\small\texttt{Regroup}}.
An index such as Cover-tree \cite{Beygelzimer2006} can also be used to group $k$ centroids \cite{Curtin2017}.



\newcommand{\pluseq}{\mathrel{+}=}
\subsubsection{Storing Bound Gaps in a Heap}
Hamerly et al. \cite{Hamerly2015a} further proposed a combination of the global lower bound $lb(i)$ and upper bound $ub(i)$ in \texttt{Hame}, 
and used their gap $lu(i) = lb(i)-ub(i)$. It 
can further reduce the space on storing bounds, 
but needs a heap for each cluster to hold it dynamically,
and the heap update incurs extra costs.
We denote this algorithm as {\small \texttt{Heap}}.
Then each point will be attached with such a gap $lu(i)$ and inserted into a queue, unless when $lu(i)\ge 0$, it will stay in the current cluster and the distance between all other points and every centroid has to be computed.

\shengnew{
\subsubsection{Using Centroid Distance Only}
\label{sec:pami20}
\marginnote{A new algorithm published on July 13 \\\ref{sec:newmethod}}
Xia et al. \cite{Xia2020} proposed to use only the centroid distance to generate a subset of centroids $N_{c_{a^{'}(i)}}$ as candidates for all the points inside. The main pruning idea is based on the radius $ra$ of each cluster, where $ra$ is the radius of the current assigned cluster $c_{a^{'}(i)}$, i.e., the distance from the centroid to the farthest point in the cluster.
In particular, each candidate centroid $c_j$ should have a distance of less than $2{ra}$ from $c_{a^{'}(i)}$; otherwise, the points would choose $c_{a^{'}(i)}$ rather than $c_j$ as the nearest centroid:
\vspace{-0.5em}
\begin{equation}
\label{equ:pami20}
N_{c_{a^{'}(i)}}=\{j:\frac{\|c_j-c_{a^{'}(i)}\|}{2}\le ra\},
\end{equation}
We name this method as {\small\texttt{Pami20}}, and it can save much space.}
\vspace{-1em}

\subsection{Methods with a Focus of Tighter Bounds}
\label{sec:tightbound}
All the above bounds are based on the triangle inequality over a point and multiple centroids.
To further tighten the bounds, L2-Norm is used by \cite{Bottesch2016,Drake2013,Hamerly2015a,Rysavy2016a}.
We denote the following four algorithms as {\small\texttt{Annu}}, {\small\texttt{Expo}}, {\small\texttt{Drift}}, and {\small\texttt{Vector}}.

\subsubsection{Annular Algorithm: Sorting Centers by Norm}
Drake et al. \cite{Drake2013, Hamerly2015a} proposed an annular algorithm to filter the centroids directly.
{Through using the \textit{Norm} of centroids, an off-line sorting can estimate a bound to determine the two closest centroids and tighten the upper and lower bounds}.
The basic idea is to pre-compute the distance from centroid to the origin (a.k.a. norm $\|c\|$), and further employ the triangle inequality to derive an annular area around the origin which all the candidate centroids are inside:
\vspace{-0.5em}
\begin{equation}
\label{equ:annu}
\mathcal{J}(i)=\{j:|\|c_j\|-\|x_i\| |\le \max\big(ub(i), \|x_i-c_{j^{''}}\|\big)\},
\end{equation}
\vspace{-1.5em}

\noindent where $c_{j^{''}}$ is the second nearest centroid of $x_i$.

\subsubsection{Exponion Algorithm: Tighter Lower Bound}
To further shrink the annular range of candidates \cite{Drake2013} around the origin,
Newling et al. \cite{Newling2016b} proposed to use a circle range around the assigned centroid. 
It filters out the centroids which will not be the nearest, and returns a centroid set $\mathcal{J}^{'}(i)$ which is a subset of $\mathcal{J}(i)$:
\vspace{-0.5em}
\begin{equation}
\small
\label{equ:newling}
\mathcal{J}^{'}(i)=\{j:\|c_j-c_{a^{'}(i)}\|\le2 ub(i)+\|c_{a^{'}(i)}-c^{'}_{a^{'}(i)}\|\},
\end{equation}
\vspace{-1.5em}

\subsubsection{Tighter Centroid Drift Bound}
{
In the \elkan method, the centroid drift (e.g., $\|c^{'}_j-c_j\|$ has to be computed every time when the centroid moves to a new one.
{It is used to update the lower bound of distance from every point to the new cluster $\|x_1-c_1\|$ in every iteration, while using triangle inequality among $x_1$, $c_1$, and $c^{'}$ to update cannot get a tighter bound, and finding a smaller drift is crucial.
By replacing this in \elkan, Rysavy et al. \cite{Rysavy2016a} proposed to use the distance between centroid and the origin point (e.g., [0, 0] in two dimension space), and compute a tighter drift $\delta$.
Note that the distance from $c_{a^{'}(i)}$ to the origin point, i.e., $\|c_{a^{'}(i)}\|$, can be pre-computed.}
Here, we only show the drift computation in a 2-dimension case.
\vspace{-0.5em}
\begin{equation}
\small
\delta(i, j) =2\cdot\frac{c_{a^{'}(i)}[1]\cdot ra - c_{a^{'}(i)}[2]\cdot \sqrt{\|c_{a^{'}(i)}\|^{2}-ra^2}}{\|c_{a^{'}(i)}\|^{2}} 
\end{equation}
\vspace{-1em}

\noindent
It has been proven that $\delta(i, j)<\|c_j^{'}-c_j\|$ \cite{Rysavy2016a}, then we can update $lb(i,j) = lb(i,j)- \delta(i, j)$.
For high-dimensional cases, $c_{a^{'}(i)}[1]$ and $c_{a^{'}(i)}[2]$ can be computed using a more complex conversion in Algorithm 2 of \cite{Rysavy2016a}, and we will not elaborate.

\subsubsection{Block Vector}
Bottesch et al. \cite{Bottesch2016} calculated bounds based on norms and the H\"older’s inequality to have a tighter bound between the centroid and a point.
The idea is to divide each data point into multiple blocks of equal size, similar to \textit{dimensionality reduction}.
By default, the data point is divided into two blocks, i.e., $x_i^{B}=\{\frac{\sum_{z=1}^{d/2}x_i[z]}{d/2}, \frac{\sum_{z=d/2}^{d}x_i[z]}{d/2}\}$.
Then a tighter lower bound can be obtained by using the pre-computed norm $\|x_i\|$ and $\|c_j\|$, and inner product of the block vector $x_i^{B}$ and $c_j^{B}$.

\vspace{-0.5em}
\begin{equation}
\small
lb(i,j)=\sqrt{\|x_i\|^2+\|c_j\|^2-2\cdot\langle x_i^{B}, c_j^{B}\rangle}
\end{equation}
where $\langle x_i^{B}, c_j^{B}\rangle$ denotes the inner vector, i.e., $\langle x_i^{B}, c_j^{B} \rangle=x_i^{B}[1]\cdot c_j^{B}[1]+x_i^{B}[2] \cdot c_j^{B}[2]$.


}

\section{Evaluation Framework}
\label{sec:frame}
{After reviewing the sequential methods, we conclude a common pruning pipeline in Figure~\ref{fig:workflow}.
Five core operators (highlighted in red color) execute in the following order: (1) access the data point, (2) then access the global bound to see whether $lb$ is bigger than $ub$ -- (3) if yes, the point maintains in the current cluster; (4) otherwise, access group and local bounds to determine whether to read centroids and compute distances to find the nearest cluster $c_j$ to assign, \shengnew{then it will be inserted into $c_j$'s list of covered points,}\marginnote{W2@R1 \\\ref{sec:r1}}and (5) then update the bounds.

Next, we will present an evaluation framework (\uni) based on the above pipeline, as shown in Algorithm~\ref{alg:kmeans}.
\uni supports multiple traversal mechanisms to smoothly switch between different index-based and sequential methods.
Note that the above pruning pipeline mainly works for point data, and existing index-based methods need to scan $k$ centroids without bound-related operations. 
Hence, before describing \uni, we will introduce an optimized assignment such that the bound-based prunings on both nodes and points can well fit into \uni in a unified way, followed by an optimized refinement without any more data access.
Our experiments show that \uni with these optimizations can further accelerate \kmeans.}

\subsection{Optimizations in UniK}
\label{sec:opt-unik}


\subsubsection{The Assignment Step}
To scan all nodes and points simultaneously, we define an advanced node that shares exactly the same property with the point, and meanwhile is attached with more information that can help prune via bounds.

\begin{figure}
	\centering
		\scalebox{1}{\includegraphics[height=4.2cm]{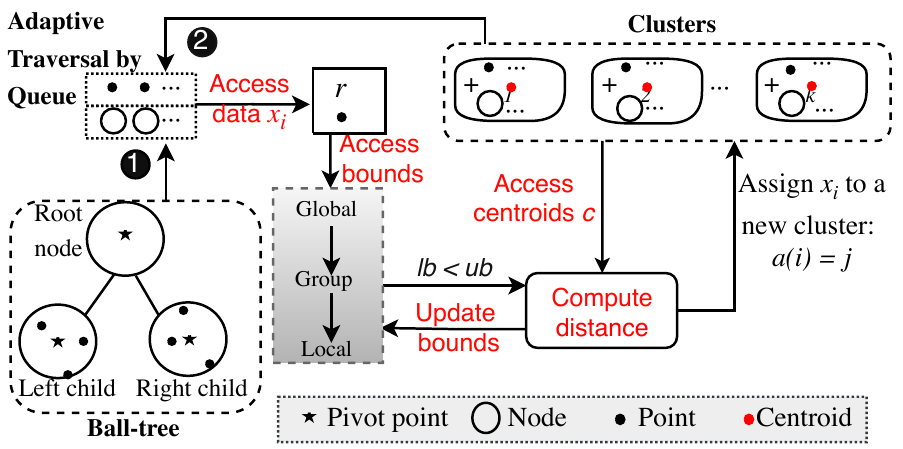}\marginnote{W2@R1 \\ \ref{sec:r1}}}
	\vspace{-2em}
		\caption{\uni framework with a common pruning pipeline.}
		\label{fig:workflow}
	\vspace{-0.8em}
\end{figure}

\myparagraph{Advanced Index}
Given a tree-structured index $T$ built from $D$, there are usually three kinds of nodes: root node $N_{rt}$, leaf node $N_l$, and internal node $N_i$.
To enable the pruning capability for \kmeans, we enrich each node with some extra information as defined below:
\begin{definition1}(\textbf{Node})
	\label{def:node}
	{$N=(p, r, sv, \psi, L_N, L_P, num, h)$ covers its child nodes $L_N$ if $N$ is an internal or root node, or a set of child points $L_P$ if $N$ is a leaf node.}
\end{definition1}

\noindent where pivot point $p$ is the mean of all the points, radius $r$ is the distance from $p$ to the furthest point in $N$, the sum vector $sv$ represents the sum of all the points under $N$, $\psi=\|N^{'}.p-p\|$ is the distance from $p$ to the pivot of its parent node $N^{'}$, $num$ is the number of points covered in the range of $N$, and $h$ is $N$'s height (or depth).
We also analyze the space cost of this index in our \textit{technical report} \cite{wang2020tr} (see Section~\ref{sec:space}).
To accommodate the assigned points and nodes, we further enrich the clusters as below:

\begin{definition1}(\textbf{Cluster})
	$S_j=(c_j, sv, L_N, L_P, num)$ covers a list of nodes $L_N$ and points $L_P$, and $num$ is the number of points in $S_j$, i.e., $num=|L_P|+\sum_{N\in L_N}N.num$.
\end{definition1}

\myparagraph{Node Assignment}
Given a node $N$, we will assign $N$ to a cluster $S_j$ if the gap between the distance from pivot $p$ to the nearest centroid $c_{n_1(p)}$ and the distance from $p$ to the second nearest centroid $c_{j^{''}}$ is larger than $2r$, i.e., 
\vspace{-0.5em}
\begin{equation}
\label{equ:nodeprune}
\|p-c_{j^{''}}\|-r >\|p-c_{n_1(p)}\|+r \rightarrow a(N)=n_1(p),
\end{equation}

\marginnote{W2@R1 \\\ref{sec:r1}\\ Node assignment}\shengnew{This is also a more general rule of Equation~\ref{equ:ball} and Figure~\ref{fig:node}(b) where $c_1$ and $c_2$ are the two nearest centroids.}
To achieve the above condition, a straightforward way is to scan $k$ centroids to find the two nearest centroids for $p$, similar to \cite{Moore2000}.
To further avoid distance computation, we use the idea of group pruning (Equation~\ref{equ:newling}) based on the pivot point $p$ to prune those centroids that cannot be the nearest two.
Before the group pruning, we use a global pruning similar to Equation~\ref{equ:nodeprune} by combining $r$:
\vspace{-0.5em}
\begin{equation}
\label{equ:pruneall}
lb(p)-r > ub(i)+r \rightarrow a(N)=a^{'}(N),
\end{equation}

%
After the global and group pruning, we use the local bounds \shengnew{in Equation~\ref{equ:notassign}} by {combining $r$ again} to avoid the distance computation between the pivot point $p$ and centroids: 
\begin{equation}
\label{equ:prune}
lb(p,j)-r > ub(p)+r \rightarrow a(N)\ne j,
\end{equation}
By comparing the bounds for point and node, we combine them into a single pipeline by setting $r=0$ when the checked object is a point.
Node assignment differs from point assignment in the way that, the node needs to be further split and checked if it is not pruned by Equation~\ref{equ:nodeprune} and \ref{equ:pruneall}.

Before computing the distance when the child node or the point is scanned, we estimate the child's pivot bounds to the centroids based on the parent-to-child distance $\psi$ using triangle inequality, which will be further used to prune in a new round without computing the real distance.
Specifically, let $N_c$ denote one child of $N$, then the upper bound and lower bound of $N$ can be passed to $N_c$ by:
\begin{equation}
\label{equ:child}
\small
\begin{split}
& lb(N_c.p,j)=lb(N.p,j)-N_c.\psi,\\
& lb(N_c.p)=lb(N.p)-N_c.\psi,~~
ub(N_c.p)=ub(N.p)+N_c.\psi.
\end{split}
\end{equation}

\subsubsection{The Incremental Refinement Step}
{To update centroids after all the points are assigned, a traditional refinement will read all the points in each center again, sum them up and get the mean vector. 
Hence, the whole dataset needs to be scanned again in each iteration.
Ding et al. \cite{Ding2015} optimized this by updating the previous centroids with those points that changed clusters only,
	so only a subset of points will be re-accessed.
	However, we can save this cost if maintaining a sum vector in the assignment}. 

More specifically, we maintain a sum vector $sv$ that will be updated when an object (node or point) moves in or out during the assignment step. Such an update will be rare when the iteration goes deeper, as only some points change their clusters across iterations.
Since index node is assigned in batch, then in order to avoid accessing all the points under it,
each node's sum vector $sv$ and the number of points $num$ can be computed in advance by an iterative procedure \cite{Moore2000} once the index is built.
For example, as shown in the right part of Figure~\ref{fig:workflow}, when an object is assigned to a new cluster, we will update the sum vector (the symbol of ``+'') of the two clusters which this object moves in and out accordingly.

\vspace{-0.8em}
\subsection{Detailed Framework Design}
\label{sec:alg}

\begin{algorithm}\small
	\caption{\uni-\kmeans($k$, $D$)}
	\KwIn{$k$: \#clusters, $D$: dataset.}
	\KwOut{$k$ centroids: $c=\{c_1,\dots, c_k\}$.}
	\label{alg:kmeans}

	\texttt{Initialization}($t$, $c$, $\mathcal{Q}$, $S$)\;
	
	\While{$c$ changed or $t<t_{max}$}{
		\For{every cluster $S_j\in S$}{
			{\small\knob{Update $lb(i,j)$ with centroid drifts for $x_i\in S_j$}}\label{line:assstart}\; 
			$Q.add(L_N, L_P)$\label{line:push}\;
			
			\While{$\mathcal{Q}.\var{\textit{isNotEmpty}}()$}{
				$o\leftarrow \mathcal{Q}.poll()$, $r\leftarrow 0 $\;
				\knob Update $ub$ and $lb$ \label{knob2}\;				
				
				\If{\knob o is node \label{knob3}}{$r\leftarrow o.getRadius()$\;}

				\eIf{\knob$lb(i)-r>ub(i)+r$\label{line:1:b:globalbound}}
				{
					{ $o$ stays in current cluster: $a(o)\leftarrow a^{'}(o)$}\label{line:1:b:4}\; 
				}{
					
					
					\knob$gap=$\texttt{GroupLocalPruning}($o, i, j$) \label{knob5}\;
					
					\texttt{AssignObject}($o$, $S$, $\mathcal{Q}$, $gap$)\;
				}
				
			}
		}	
		\texttt{Refinement}($S$)\label{line:refine}\;
		$t\leftarrow t+1$\;
	}

	\KwRet\{$c_1,\dots, c_k$\}\;	
	\vspace{-0.5em}
	\algrule
	\vspace{-0.5em}
	\SetKwFunction{FMain}{Initialization}
	\SetKwProg{Fn}{Function}{:}{}
	\Fn{\FMain{$t$, $c$, $\mathcal{Q}$, $S$}}{
		$t\leftarrow 0$, $\mathcal{Q}\leftarrow \emptyset$,
		initialize centroids $c = \{c_1,\cdots,c_k\}$\label{line:1:init}\;
		
		\eIf{\knob $N_{rt}\ne null$}{
			{$S_1.add(N_{rt}.sv, N_{rt})$\label{line:aroot}}\tcp*[r]{index \cite{Kanungo2002}}\marginnote{W2@R1\\ \ref{sec:r1}}
		}{
			\knob Search on each centroid\tcp*[r]{\small Search \cite{Broder2014}}
			$S_1.add(D)$\label{line:apoint}\;
		}
	}	
	\vspace{-0.5em}
	\algrule
	\vspace{-0.5em}
	\SetKwFunction{FMain}{GroupLocalPruning}
	\SetKwProg{Fn}{Function}{:}{}
	\Fn{\FMain{$o, i, j$}}{
		{\small\knob $\mathcal{J} \leftarrow \mathcal{J}(i)$ or $N_{c_j}$ or $\mathcal{J}^{'}(i)$;\label{line:j}\tcp*[f]{Expo, Pami20, Annu}}
		
		$min \leftarrow +\infty$, $min_2 \leftarrow +\infty$\;		
		\For{every centroid $c_j \in \mathcal{J}$}{
			\knob Pass group pruning \label{knob6}\tcp*[r]{Yinyang, Regroup}
			\If{\knob$ub(i) > lb(i,j)$}{\label{line:local} 
				$lb(i,j) \leftarrow \|o- c_j\|$\label{line:ebd}\;
				\If{ $lb(i,j)<min$}
				{
					$a(i)\leftarrow j$,
					$min\leftarrow lb(i,j)$, $min_2\leftarrow min$\label{line:compute}\;
				}
			}
		}
		\KwRet $min_2-min$\;
	}	
	\vspace{-0.5em}
	\algrule
	\vspace{-0.5em}
	\SetKwFunction{FMain}{AssignObject}
	\SetKwProg{Fn}{Function}{:}{}
	\Fn{\FMain{$o$, $S$, $\mathcal{Q}$, $gap$}}{
		\eIf{\knob o is node and Equation~\ref{equ:nodeprune}: $gap<2r$}{\label{line:nodeagain}
			$S_{a^{'}(i)}.\var{\textit{remove}}(o.v, o)$\;
			
			\For{every child $N_c\in o.L_N$}{
				Update $N_c$'s bound by Equation~\ref{equ:child}\label{line:child1}\;
				
				$S_{a^{'}(i)}.\var{\textit{add}}(N_c.sv, N_c)$\label{line:child2},
				 $\mathcal{Q}.push(N_c)$\;
			}
		}{
			\If{$a^{'}(i)\ne a(i)$}{\label{line:stay}
				$S_{a^{'}(i)}.\var{\textit{remove}}(o.sv, o)$\label{line:update1}\; $S_{a(i)}.\var{\textit{add}}(o.sv, o)$\label{line:update2}\;
			}
		}
	}
	\vspace{-0.5em}
	\algrule
	\vspace{-0.5em}
	\SetKwFunction{FMain}{Refinement}
	\SetKwProg{Fn}{Function}{:}{}
	\Fn{\FMain{$S$}}{
		\For{every cluster $S_j\in S$}{
			$c_j\leftarrow \frac{S_j.sv}{S_j.num}$\label{line:refinement}\;
		}
	}
\end{algorithm}

Algorithm~\ref{alg:kmeans} presents a general form of our unified framework under \uni, which is composed of a queue data structure and four core functions.
The queue $\mathcal{Q}$ is used to hold both the points and the nodes (called as \textit{object} for unity purpose) to be assigned. When a node cannot be assigned by Equation~\ref{equ:nodeprune}, its child nodes are enqueued. The four core functions are elaborated as follows.


\myparagraph{\texttt{Initialization}} After randomly selecting $k$ points as the centroids $c$ (line~\ref{line:1:init}), we assign the root node $N_{rt}$ to the first cluster $S_1$ if an index has been built (line~\ref{line:aroot}); otherwise, we temporally store the whole point set $D$ in {$S_1$ (line~\ref{line:apoint}).}
Then in each cluster, all the nodes and points are pushed into $\mathcal{Q}$.
An object $o$ polled from the queue $\mathcal{Q}$ is pruned in the following manner: the global pruning is first conducted in line~\ref{line:1:b:globalbound}, we {a}ssign $o$ (denoted as $a(o)$) if the upper bound is smaller than the lower bound minus twice of the radius.

\myparagraph{\texttt{GroupLocalPruning}}{If the global pruning fails}, we use the group pruning in line~\ref{line:j} to filter partial centroids.
Then we conduct the local pruning (line \ref{line:local}); if it fails, we compute the distance from the centroid to the pivot point, and update the two nearest neighbors' distances (line~\ref{line:compute}).

\myparagraph{\texttt{AssignObject}}After the scanning of centroids, we obtain the gap between the distance to the two nearest clusters. 
Then, we check whether the object is a node, and compute the gap using Equation~\ref{equ:nodeprune} to see whether we can assign the whole node (line~\ref{line:nodeagain}).
If not, we further split the node and push all its children into the queue (line~\ref{line:child1}--\ref{line:child2}).
If it can be assigned or the object is a point, we will further check whether the object should stay in the current cluster~(line~\ref{line:stay}); otherwise, we update the cluster (line~\ref{line:update1}).


\myparagraph{\texttt{Refinement}}We divide the sum vector $sv$ by the number of points inside to update the centroid in line~\ref{line:refinement}. The refinement can be done without accessing the dataset again, as we have updated the cluster's sum vector in line~\ref{line:update1} when the object cannot remain in the current cluster.

\vspace{-0.5em}
\subsection{Multiple Traversal Mechanisms}
\label{sec:travel}
\marginnote{W2@R1\\ \ref{sec:r1} \\ knobs}
{The symbols \knob~in Algorithm~\ref{alg:kmeans} are knobs that indicate whether to apply pruning techniques used in various index-based and sequential methods which we evaluate. By turning certain knobs on, we can switch to a specific algorithm.
\shengnew{For example, by turning on knobs at Lines~\ref{line:assstart}, \ref{knob2}, \ref{line:1:b:globalbound}, \ref{knob5}, \ref{knob6}, \ref{line:local}, and others off, the algorithm will run as the \yiny \cite{Ding2015}.}
Formally, we formulate the knobs as below:}

\vspace{-0.5em}
\begin{definition1}{\textbf{(Knob Configuration)}}
	A knob \knob = \{0,1\} controls a setting (selection), e.g., whether to use index or not. 
	A knob configuration $\theta \in \Theta$ is a vector of all knobs in Algorithm~\ref{alg:kmeans}, e.g., $[0,1,0,\cdots,1]$, where $\Theta$ denotes the configuration space.
\end{definition1}
\vspace{-0.5em}

\begin{figure*}
	\vspace{-1em}
	\begin{minipage}{0.19\textwidth}
		\centering
		\vspace{1.8em}
			\scalebox{1}{\includegraphics[width=.9\linewidth]{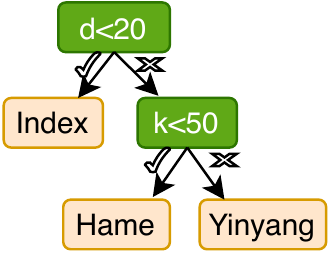}}
		\caption{A basic decision tree (BDT) for simple algorithm tuning.}\label{fig:dt}
	\end{minipage}
	\begin{minipage}{0.8\textwidth}
		\centering
		\vspace{-0.5em}
			\scalebox{1}{\includegraphics[width=0.99\linewidth]{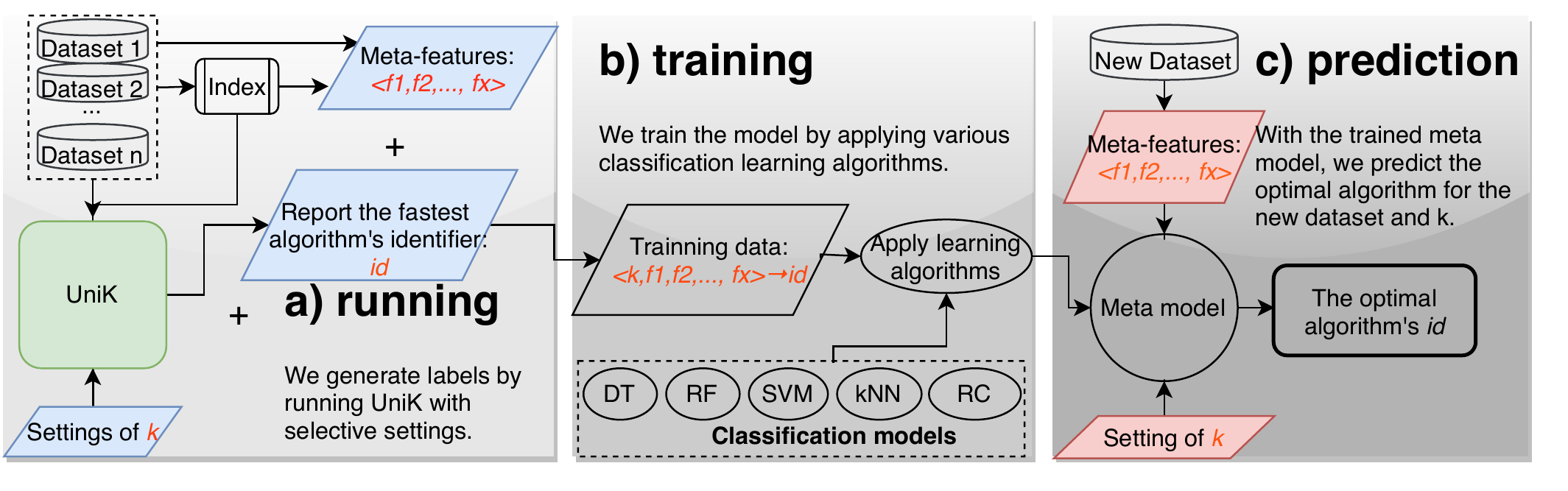}}
		\vspace{-1.8em}
		\caption{Three modules of \utune: a) selective running, b) meta-model training, and c) prediction.}\label{fig:utune}
	\end{minipage}
	\vspace{-1.2em}
\end{figure*}

{Most existing algorithms can fit into \uni by projecting to a specific knob configuration,\footnote{{They run in the same way as presented in the original papers, and will not incur any extra costs. Thus, it is a fair evaluation for any algorithm under comparison.}} and
our optimization proposed in Section~\ref{sec:opt-unik} can be activated by a new knob configuration.}\footnote{{{
By enabling all bound knobs, we will get the \textit{Full} method in Figure~\ref{fig:motivation}.} We also argue that this formulation will be useful to cater for more new configurations in $\Theta$ to be explored by future studies, but it is not the focus of this paper.}}
For example in Figure~\ref{fig:workflow} we present two traversal mechanisms (i.e. parts ``\textcircled{1}'' and ``\textcircled{2}'') based on knobs, where we traverse from the root node by using \textcircled{1} in the first iteration,
but in the following iterations we traverse from the nodes maintained in the current cluster by using \textcircled{2}.
This is because the pruning effect is not always good especially for high dimensional data,
and it will still cost extra time in the tree traversal in next iteration if starting from the root node.
If most nodes in the index can be pruned, traversal from the root (i.e., \textcircled{1}) in each iteration can gain better performance than scanning nodes from clusters.\marginnote{W2@R1 \\\ref{sec:r1} \\ Adaptive traversal}
\shengnew{In Algorithm~\ref{alg:kmeans}, we implement such an adaptive mechanism as scanning both the points and nodes as an object $o$ with a radius $r$, where $r=0$ when $o$ is a point}.

%

To be more adaptive, in the first and second iterations of \uni, we can traverse from the root \textcircled{1} and current nodes in the clusters \textcircled{2}, respectively (as shown in \shengnew{line \ref{line:aroot} of Algorithm~\ref{alg:kmeans}, where we put the root node $N_{rt}$ into $S_1$ before the first iteration)}.
Then we can compare the assignment time of these two iterations.
If the time spent on the first iteration \textcircled{1} is bigger than that on the second iteration \textcircled{2},
we give up traversing from the root in subsequent iterations, and scan the current nodes and points maintained in each cluster;
otherwise, we empty the point and node list of each cluster, and then push the root node at the end of each iteration to keep benefiting from the index in multiple iterations.
We denote these two traversal styles as \textit{index-single} and \textit{index-multiple}.

By default, we set the index as Ball-tree and set the bound configuration same as Yinyang (Section~\ref{sec:yinyang}). 
{Our experiments in Section~\ref{sec:expuni} and \ref{sec:expauto} show the the superiority of such a configuration over both Ball-tree and Yinyang in some datasets.}
When we disable all the bound configurations, it will be a pure index-based method \cite{Moore2000}.
Hence, besides multiple knobs of bound configuration, we also have four configuration knobs on the index traversal: 1) not using index; 2) pure; 3) index-single; 4) index-multiple.
Next, we will further discuss how to choose the right knobs to turn on.

\section{Configuration Auto-Tuning}
\label{sec:auto}


In this section, we study how to select a fast algorithm for a given clustering task. This is critical to both an evaluation framework (like this paper) and practitioners. Our literature review shows that existing studies \cite{Curtin2017,Ding2015} rely on simple yet fuzzy decision rules, e.g., not using index-based method for high-dimensional data or choosing \texttt{Yinyang} when $k$ is big. In Figure~\ref{fig:dt}, we illustrate a basic decision tree (BDT) based on these rules. However, our experiments show this BDT does not work very well. 

Therefore, we choose to train an ML model based on our evaluation data to automatically select a fast algorithm for a given clustering task. The algorithm selection is equivalent to a problem of tuning ``knobs'' in our evaluation framework \uni (Algorithm~\ref{alg:kmeans}), in the sense that each of the existing algorithms corresponds to a unique knob configuration.

\myparagraph{An Overview} We model the knob configuration (algorithm selection) problem as a typical classification problem, where our goal is to predict the best knob configuration for a given clustering task. To this end, we extract meta-features to describe clustering datasets, and generate class labels (i.e., ground truth) from our evaluation logs that contain the records of which configuration performing the best for a particular dataset. We then feed the meta-features and class labels into an off-the-shelf classification algorithm to learn a mapping from a clustering dataset (with certain features) to the best performing algorithm configuration. We can then use the trained model to predict a good configuration for clustering a given dataset.\footnote{{Note that the learning component in this paper will not tune any existing algorithm or parameter; its main task is to predict the best one among existing algorithms and our optimized algorithm. Moreover, our learning model is not limited to some specific learning methods such as deep learning; most classification models (such as decision tree, SVM, and kNN) can be trained to complete this prediction task.}} We name our auto-tuning tool as \utune, and illustrate its three modules in Figure~\ref{fig:utune}. 



\vspace{-0.5em}
\subsection{Generating Training Data}\label{sec:rskc}
\vspace{-0.5em}
\myparagraph{Class Label Generation} 
Since the knob configuration space is large, it is computationally intractable to 
try all possible knob configurations and label the dataset with the best-performing configuration. Thus, we only focus on a few knob configurations corresponding to the high-performing existing methods as our selection pool \cite{VanAken2017}. The pseudocode of our selective running can be found in Algorithm~\ref{alg:ground} of our \textit{technical report} \cite{wang2020tr}, and the main idea is three-fold:
1) we limit the number of iterations $t_{max}$ as the running time of each is similar after several iterations (see Figure~\ref{fig:ite});
2) we exclude those algorithms that have low rank during our evaluation, e.g., {\small\texttt{Search}} \cite{Broder2014}, {and our experiments (see Figure~\ref{fig:lead}) show that only five methods always have high ranks;}
3) we test index optimizations if the pure index-based method outperforms sequential methods. 
Such a selective running enables us to generate more training data within a given amount of time, and thus further improves the prediction accuracy (see Table~\ref{tab:precision}).



\begin{table}
	\centering
	\ra{0.8}
	\caption{A summary of features $\bm{F}=\{f1,f2,\cdots,fx\}$.}
	\vspace{-1.2em}
	\scalebox{0.90}{\begin{tabular}{cccc}
			\toprule
			\textbf{Type} & \textbf{Feature} & \textbf{Description} & \textbf{Normalize} \\ \midrule
			\multirow{3}{*}{Basic} & $n$ & The scale of dataset & - \\
			& $k$ & Number of clusters & - \\
			& $d$ & Dimensionality of dataset & - \\ \midrule
			\multirow{3}{*}{Tree} & $h(T)$ & Height of index tree $T$& $\log_2\frac{n}{f}$ \\
			& $|N_i|, |N_l|$ & \#Internal \& leaf nodes & $\frac{n}{f}$ \\
			& $\mu(h)$, $\sigma(h)$ & Imbalance of tree & $\log_2\frac{n}{f}$\\ \midrule
			\multirow{3}{*}{Leaf} & $\mu(r)$, $\sigma(r)$ & Radius of leaf nodes & $N_{rt}.r$ \\
			& $\mu(\psi)$, $\sigma(\psi)$ & Distance to parent node & $N_{rt}.r$ \\
			& $\mu(|L_p|)$, $\sigma(|L_p|)$ & \#Covered points in $N_i$& $f$\\ \bottomrule
	\end{tabular}}
	\label{tab:features}
	\vspace{-1em}
\end{table}

\myparagraph{Meta-Feature Extraction}
We extract a feature vector $\bm{F}$ to describe (or represent) a dataset, such as dimensionality, scale, and $k$ shown in Table~\ref{tab:features}.
\marginnote{D1@Meta \\ \ref{sec:meta} \& \\D3@R2 \ref{sec:r2}\\ Here we clearly mentioned and considered the importance of data distribution.}
\ul{In addition to the basic features, we also extract novel and more complex features that can capture certain properties of data distribution based on our built index.} 
\ul{Recall the pruning mechanisms in }Section~\ref{sec:frame}, \ul{they work well when data shows good assembling distribution.} \ul{The index construction actually conducts a more in-depth scanning of the data and reveals whether the data assembles well in the space.}
\ul{Specifically, the information we got includes the tree depth $h(T)$, number of leaf nodes $|N_l|$, number of internal nodes $|N_i|$, and imbalance of tree} {(i.e., the mean value and standard deviation of all leaf nodes' heights $h$)}.

\ul{Further, we extract more features of all the leaf nodes, including the radius of nodes $r$, the distance from child to parent nodes $\psi$, and the number of points in the leaf nodes $|L_p|$, during index construction.
Specifically, we select all the leaf nodes $N$ and extract their $(r, \psi, |L_p|)$, then compute their mean $\mu$ and standard deviation $\sigma$, which are all normalized by their maximum values in the tree.}

\vspace{-0.5em}
\subsection{Meta-Model Training and Prediction}

We model the algorithm selection as a multi-label classification problem in terms of using an index or not, and using which one of the five bound configurations.
The prediction needs to be conducted in two parts based on our two ground truth files which shows the rank of various index and bound configurations, respectively.
Firstly, we predict the optimal bound configuration.
Secondly, we predict the index configuration and whether the bound should be combined, {as mentioned in Section~\ref{sec:travel}.}
Then we combine the results from these two predictions and generate the final configuration.
For a new clustering task, we can extract the features based on our built index (e.g., Ball-tree), and use the learned model to predict a high-ranked knob configuration of algorithms.

\section{Experimental Evaluations}
\label{sec:exp}
\begin{figure*}[h]
	\centering
	\vspace{-1em}
	\includegraphics[width=12cm]{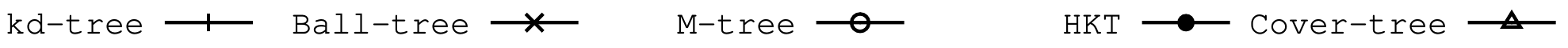}\\
	\hspace{-2em}
	\includegraphics[width=3.57cm,height=0.10\textheight]{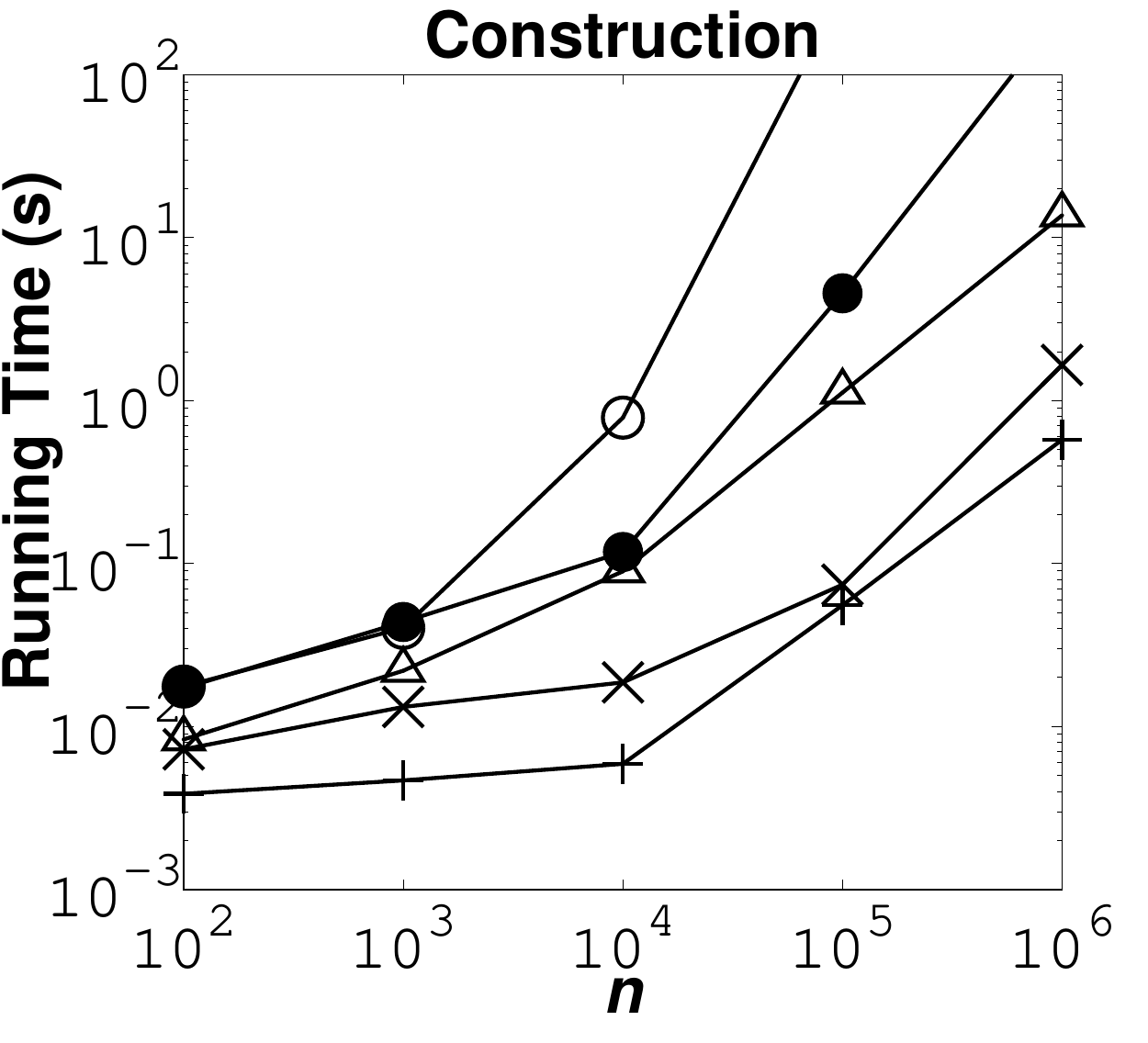}
	\includegraphics[width=3.57cm,height=0.10\textheight]{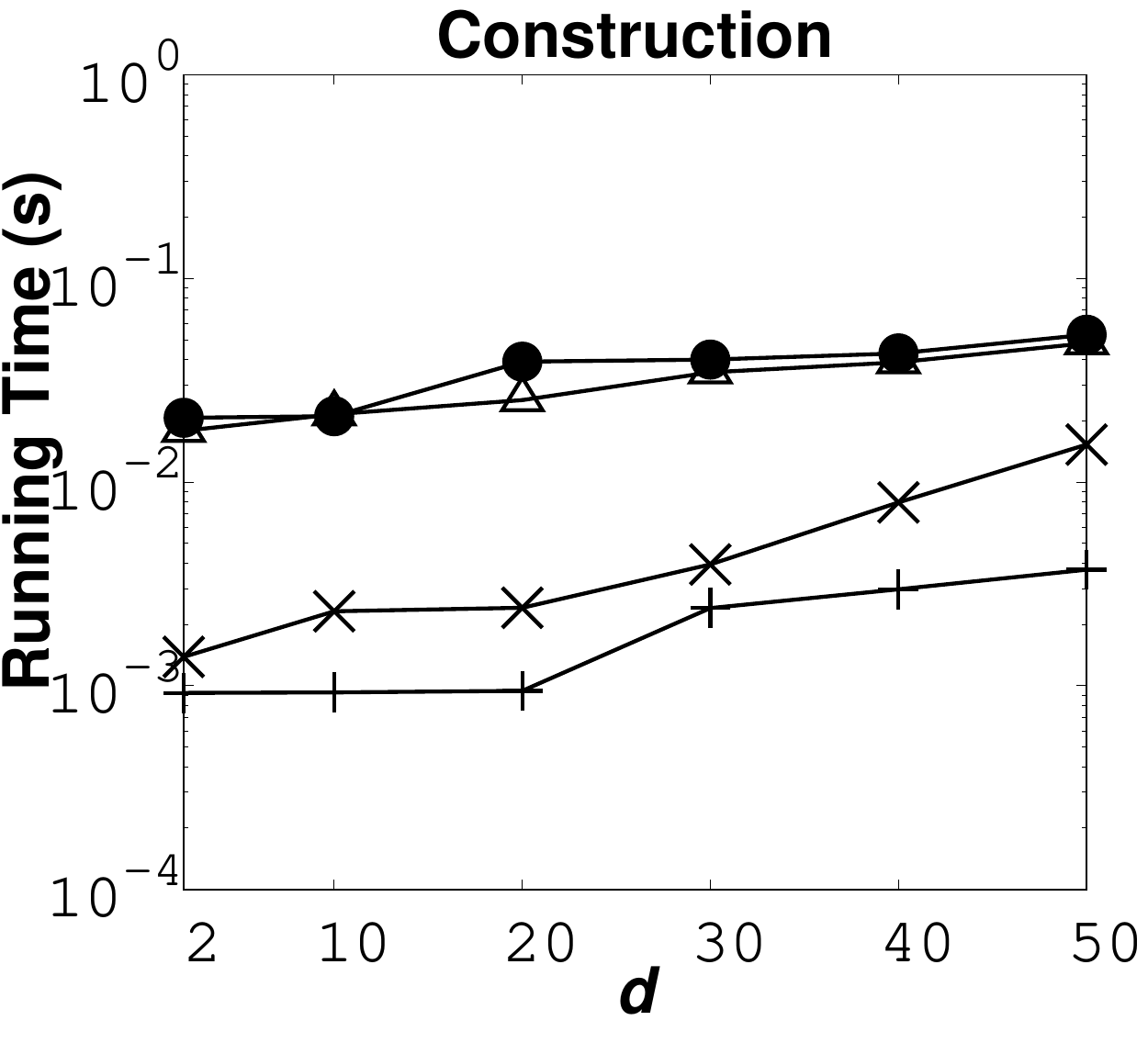}
	\includegraphics[width=3.57cm,height=0.10\textheight]{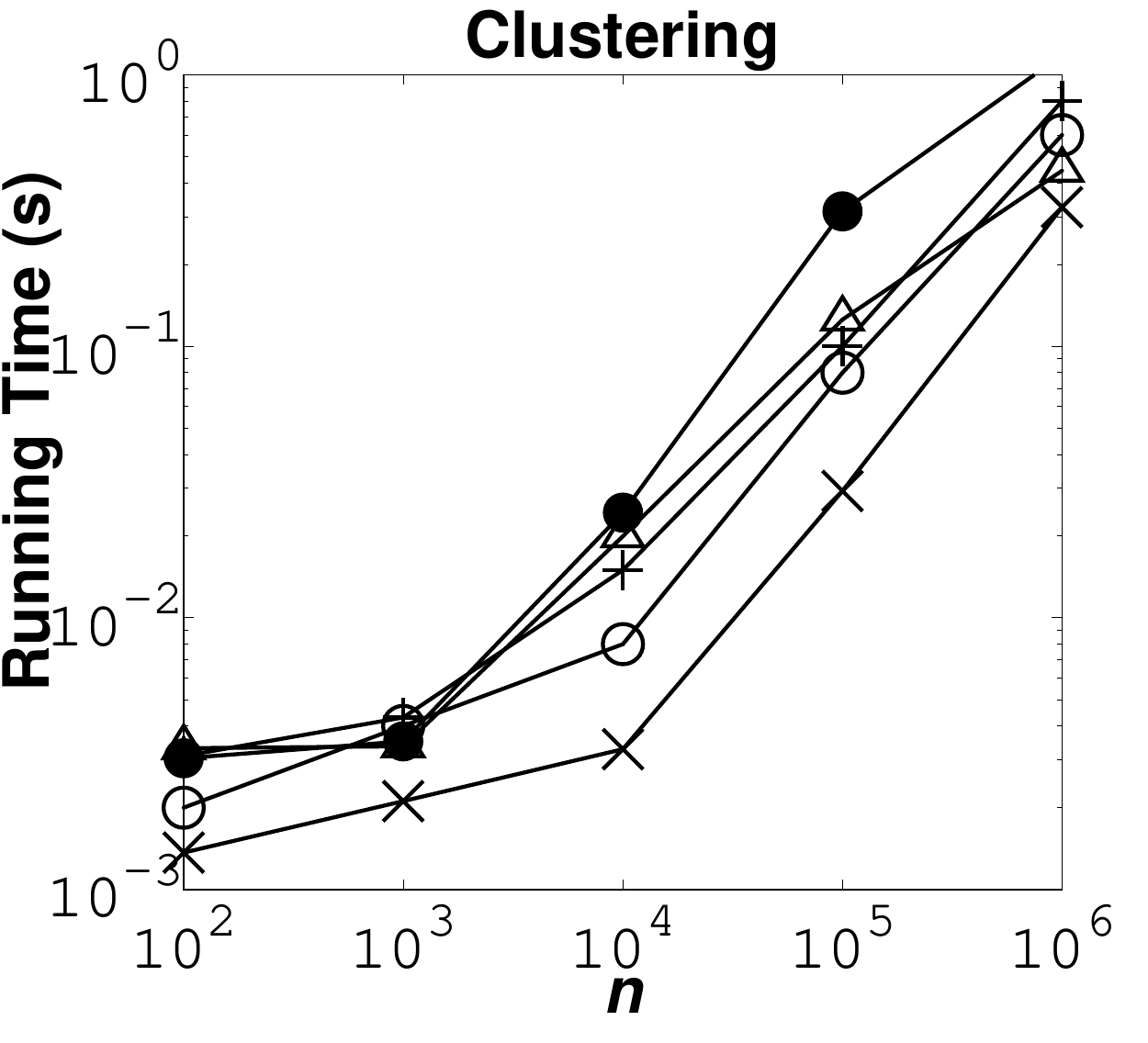}
	\includegraphics[width=3.57cm,height=0.10\textheight]{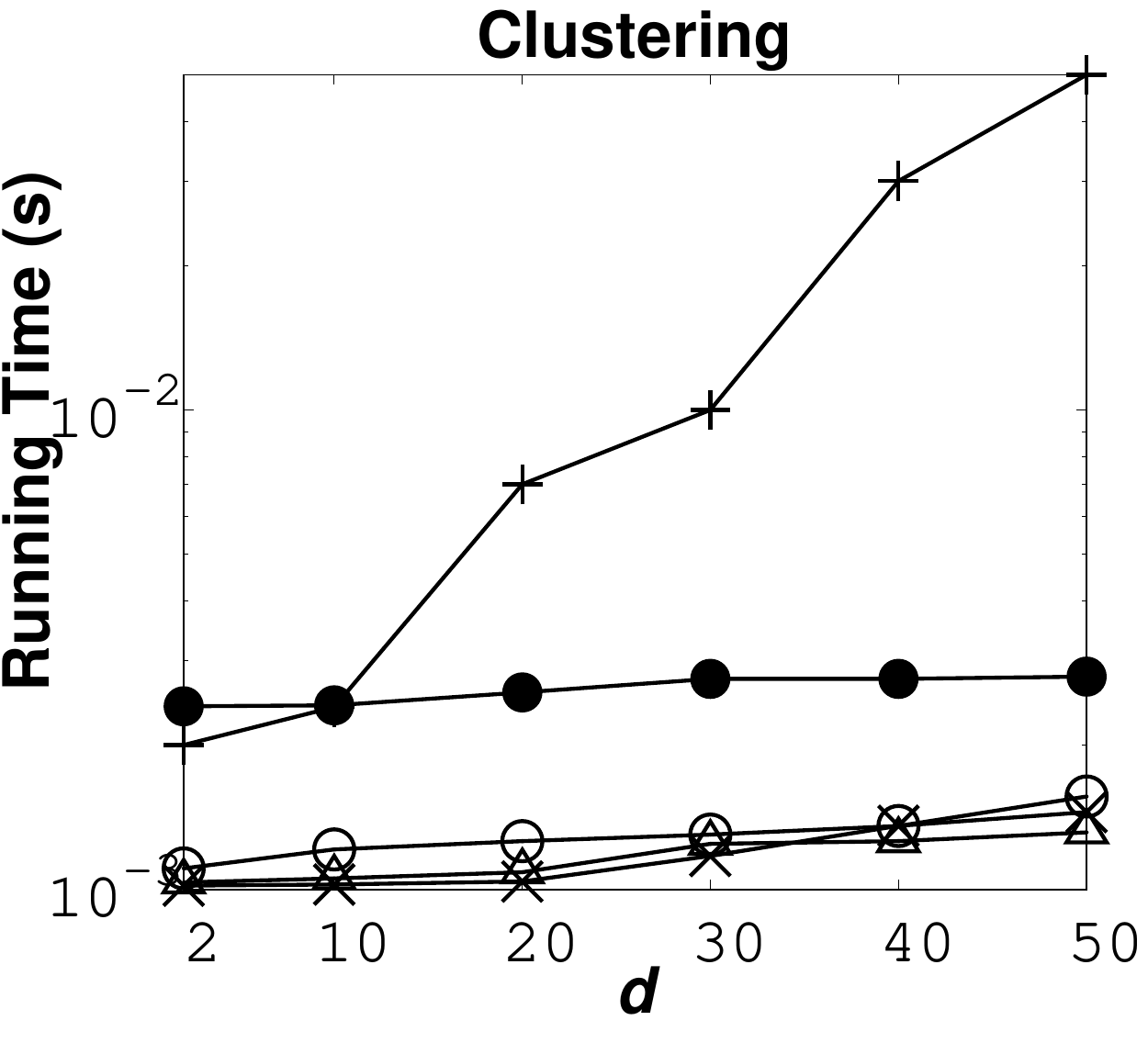}
	\includegraphics[width=3.57cm,height=0.10\textheight]{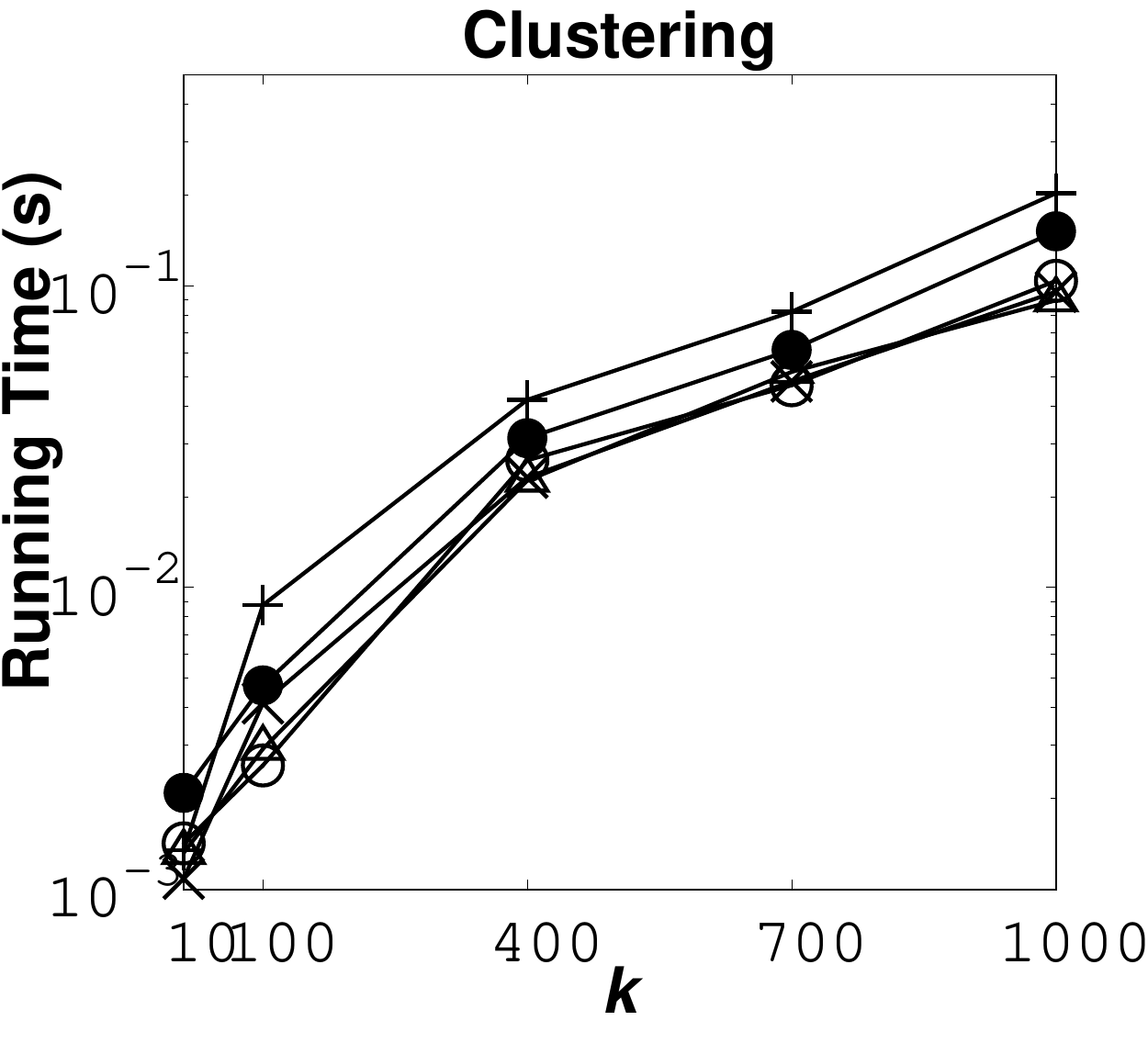}
	\vspace{-1.3em}
	\caption{{Construction and clustering performance of five index structures.}}
	\vspace{-1em}
	\label{fig:index}
\end{figure*}

\begin{table}
	\centering
	\ra{0.8}
	\caption{{An overview of datasets, the index construction time (second), and \#nodes of Ball-tree.}}
	\vspace{-1.2em}
	\scalebox{0.85}{\begin{tabular}{ccccccc}
			\toprule
			\multirow{1}{*}{ \textbf{Ref.}} & \multirow{1}{*}{ \textbf{Name}} & \multirow{1}{*}{\textbf{$n$}} & \multirow{1}{*}{\textbf{$d$}} & \textbf{Time} & \textbf{\#Nodes} & \textbf{Used by} \\ \midrule
			\cite{II} & Big\underline{Cross} & 1.16M & 57 & 10.8 & 183k & \cite{Shindlera} \\ 
			\cite{VI} & \underline{Conf}long & 165k & 3 & 0.26 & 21.8k & \cite{Newling2016b}\\ 
			\cite{VII} & \underline{Covt}ype & 581k & 55 & 3.87 & 88.3k &\cite{Hamerly2010,Hamerly2015a,Elkan2003} \\ 
			\cite{VIII} & \underline{Euro}pe & 169k & 2 & 0.27 & 11.2k & \cite{Newling2016b}\\ 
			\cite{III} & \underline{KeggD}irect & 53.4k & 24 & 0.17 & 2.8k & \cite{Newling2016b,Ding2015} \\ 
			\cite{IV} & \underline{Kegg}Undirect & 65.5k & 29 & 0.31 & 4.5k & \cite{Newling2016b,Ding2015} \\ 
						\cite{XI} & \underline{NYC}-Taxi & 3.5M & 2 & 8.7 & 228k & - \\ 
			\cite{IX} & \underline{Skin} & 245k & 4 & 0.33 & 21.2k&\cite{Newling2016b} \\ 
			\cite{X} & \underline{Power} & 2.07M & 9 & 4.3 & 43.7k & \cite{Curtin2017} \\ 
			\cite{V} & \underline{Road}Network & 434k & 4 & 0.55 & 6.9k & \cite{Ding2015} \\ 
				\cite{I} & US-\underline{Census} & 2.45M & 68 &204 & 135k& \cite{Newling2016b}\\
				\cite{Minist} & \underline{Mnist} & 60k & 784 & 4.8 & 7.3k &	\cite{Newling2016b,Hamerly2010,Ding2015} 
				 \\ 
\bottomrule
	\end{tabular}}
	\label{tab:dataset}
	\vspace{-1em}
\end{table}


%

Our evaluation seeks to answer the following questions:
\begin{itemize}
	\item Which index structure is proper for index-based methods?
	\item How does the performance of sequential methods vary?
	\item Can our evaluation framework \uni enable further improvement on the performance of existing clustering methods?
	\item Can \utune predict a fast algorithm through learning?
\end{itemize}

\vspace{-0.5em}
\subsection{Experimental Setup}
\label{sec:expset}
\vspace{-0.5em}

\myparagraph{Implementation}We implemented all the algorithms in Java 1.8.0\_201, and used Scikit-learn 0.22.2 \cite{scikit,Pedregosa2011} to train our classification models.
All experiments were performed on a server using an Intel Xeon E5 CPU with 256 GB RAM running RHEL v6.3 Linux.
Our code is available at \cite{unik-git} for reproducibility. 

{\myparagraph{Parameter Settings} The performance of index-based methods and three sequential algorithms (i.e., {\small\texttt{Yinyang}} \cite{Ding2015}, {\small\texttt{Drak}} \cite{Drake2012}, and {\small\texttt{Vector}} \cite{Bottesch2016}) will be affected by parameters. To be fair, we follow the suggestions of those sequential methods and set fixed parameters, detailed settings can be found in their description in Section~\ref{sec:seq}.
The effects on index-based methods will be studied in Section~\ref{sec:exp_im}.}



\myparagraph{Measurement} 
Same as the traditional methods \cite{Ding2015,Newling2016}, we measure the running time and the percentage of pruned distance computation (i.e., pruning power). 
In order to run more rounds of experiments, we record the total running time of the first ten iterations (after which the running time usually becomes stable as shown in Figure~\ref{fig:ite}). 
Moreover, we measure \#data access, bound access, bound updates, and footprint.
For each measurement above, we report the average value across ten sets of randomly initialized centroids using $k$-means++ \cite{Arthur2007}.

\myparagraph{Datasets}We select a range of real-world datasets (Table~\ref{tab:dataset}), most of them are from the UCI repositories \cite{Bache2013}, and also used in the state-of-the-art such as \cite{Ding2015,Newling2016b}.
Moreover, we introduce several recent new datasets, including pick-up locations of NYC taxi trips.
The datasets' hyperlinks can be found in the reference.


\vspace{-0.7em}
\subsection{Evaluation of Existing Methods in {UniK}}
\label{sec:expunik}

\begin{figure*}
	\centering
	\begin{minipage}{1\textwidth}
		\centering
		\hspace{-4.3em}\includegraphics[width=1.06\textwidth]{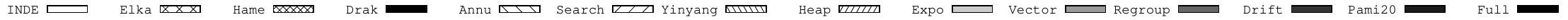}\\
		\hspace{-2.1em}\includegraphics[width=0.52\textwidth,height=0.11\textheight]{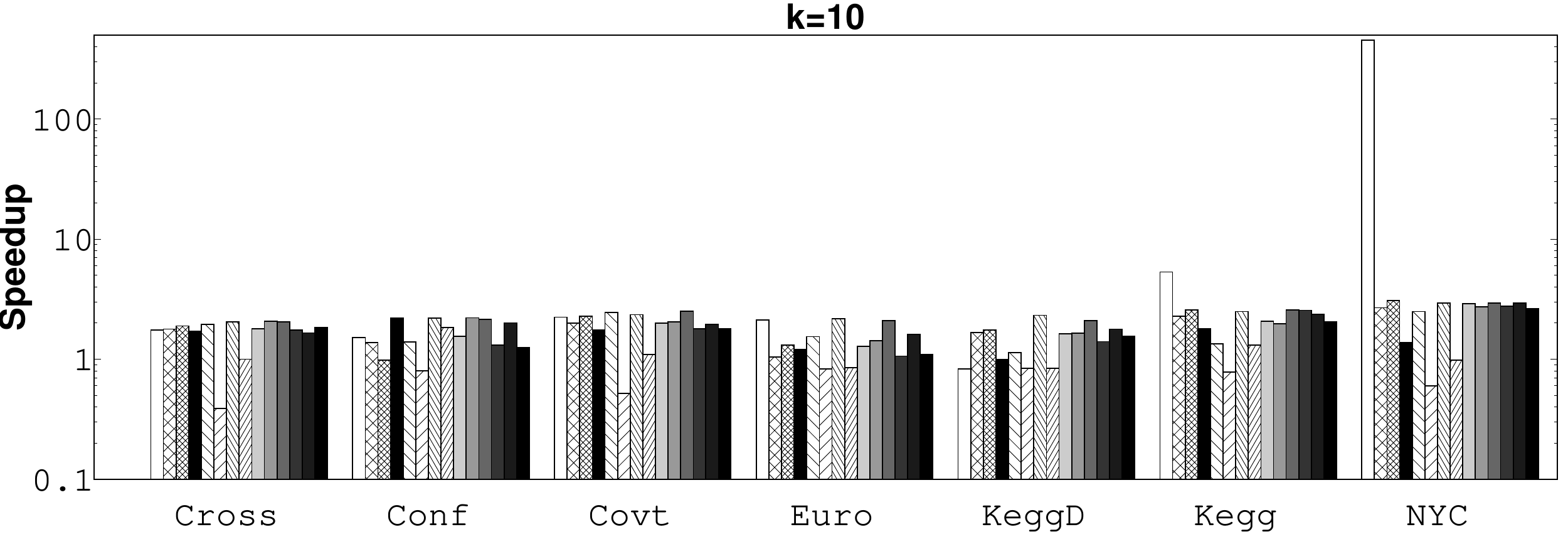}\hspace{-0.4em}
		\includegraphics[width=0.52\textwidth,height=0.11\textheight]{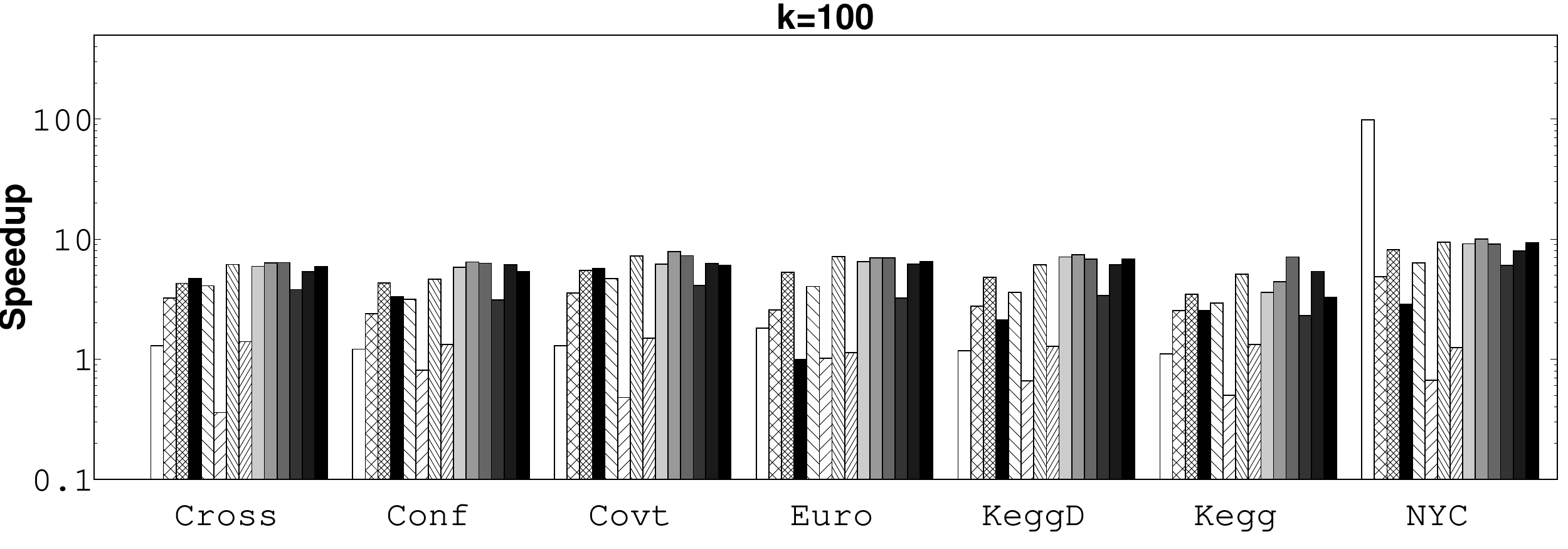}\\
		\vspace{-1.3em}
		\captionof{figure}{Overall speedup in various datasets when setting $k$ as 10 and 100, respectively.}
		\vspace{-0.5em}
		\label{fig:seq}
	\end{minipage}\\
	\begin{minipage}{1\textwidth}
		\centering
		\hspace{-2.1em}\includegraphics[width=0.52\textwidth,height=0.11\textheight]{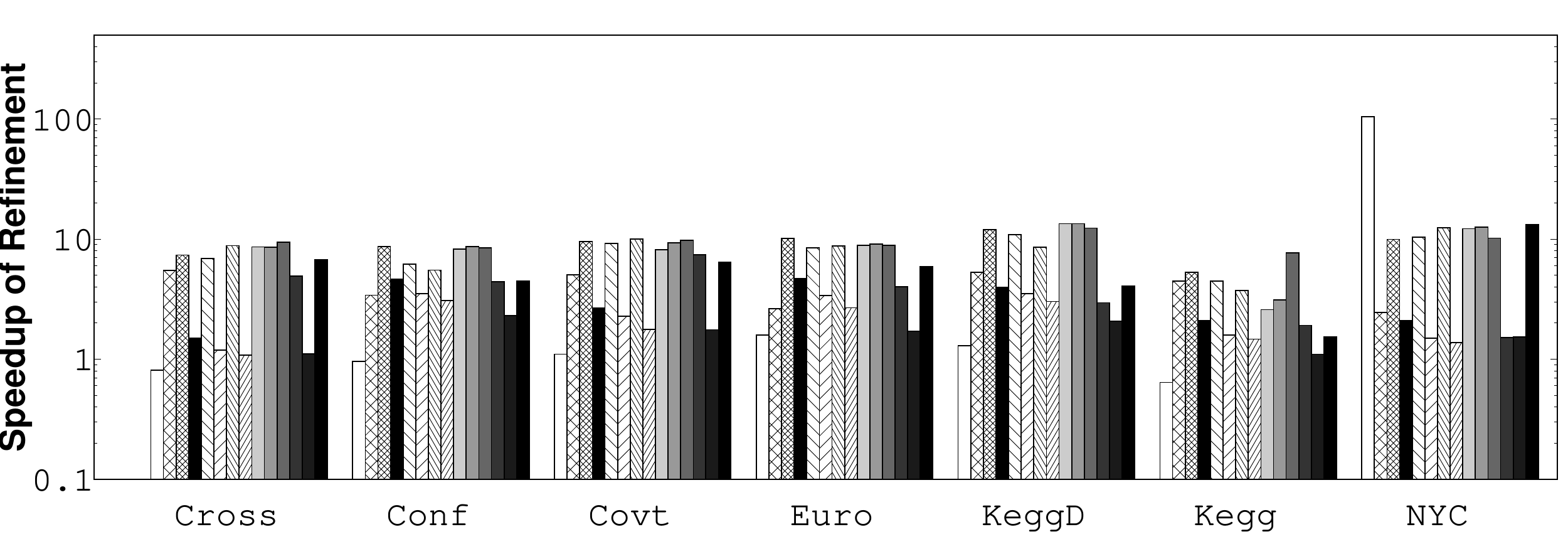}\hspace{-0.4em}\includegraphics[width=0.52\textwidth,height=0.11\textheight]{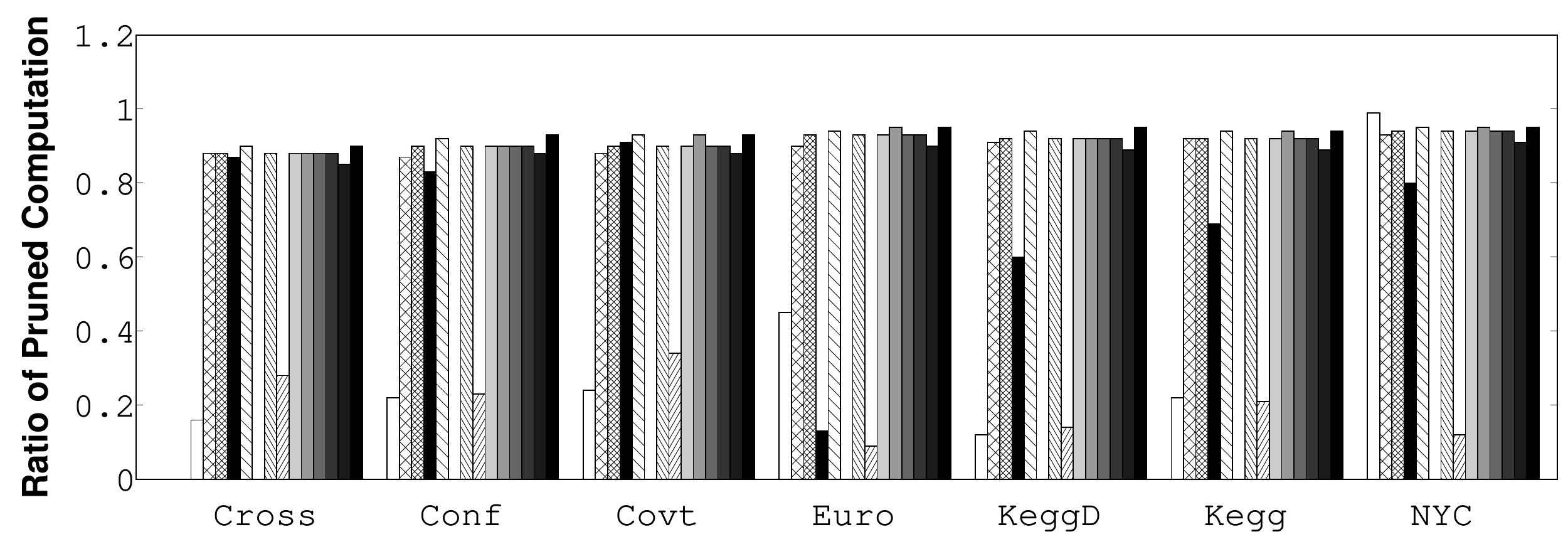}
		\vspace{-1.3em}
		\captionof{figure}{Refinement's speedup and distance computation pruning ratio ($k=100$).}
		\vspace{-0.5em}
		\label{fig:seq-refine}
	\end{minipage}
	\begin{minipage}{1\textwidth}
		\centering
		\hspace{-2.1em}\includegraphics[width=0.52\textwidth,height=0.11\textheight]{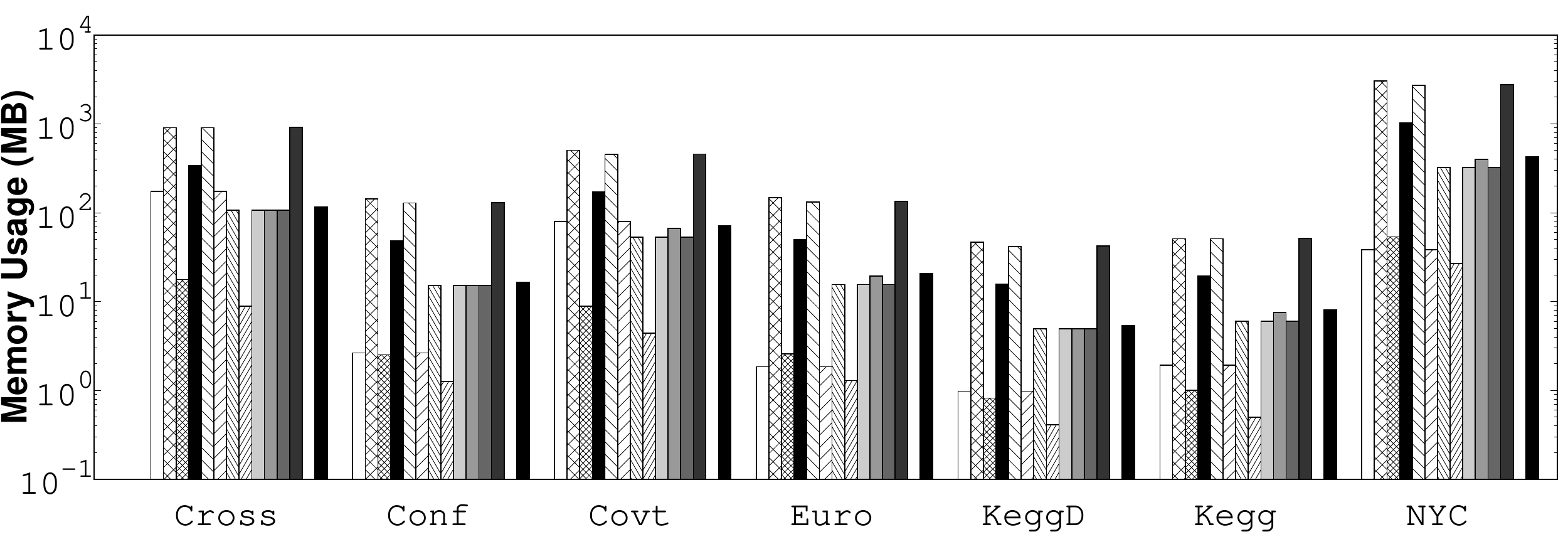}\hspace{-0.4em}
		\includegraphics[width=0.52\textwidth,height=0.11\textheight]{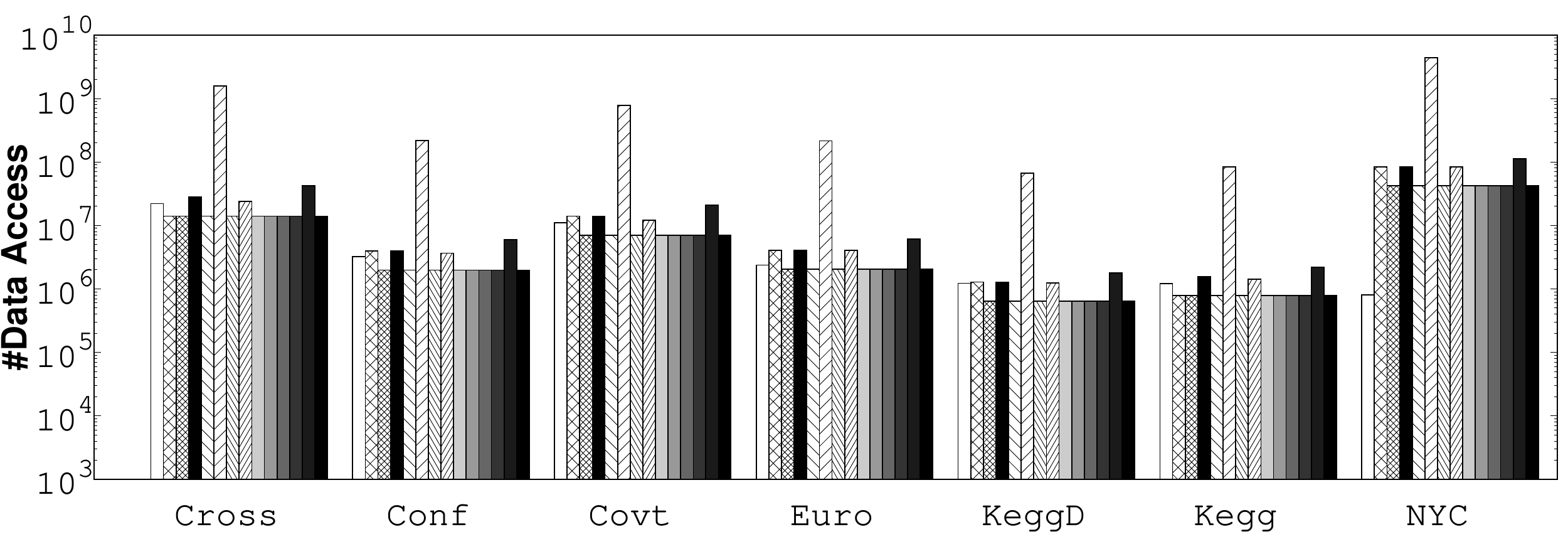}
		\vspace{-2.5em}
		\captionof{figure}{Statistics on the footprint of bound (index) and data accesses ($k=100$).}
		\vspace{-0.5em}
		\label{fig:seq1}
	\end{minipage}
	\begin{minipage}{1\textwidth}
		\centering
		\hspace{-2.1em}\includegraphics[width=0.52\textwidth,height=0.11\textheight]{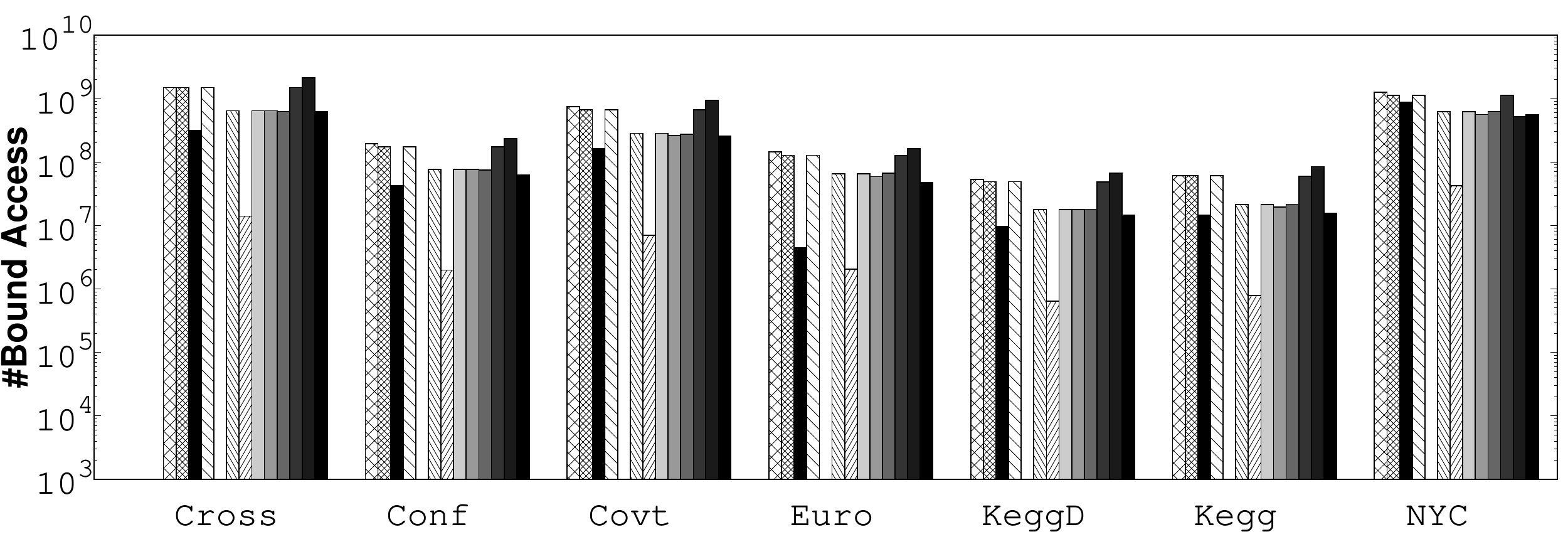}\hspace{-0.4em}
		\includegraphics[width=0.52\textwidth,height=0.11\textheight]{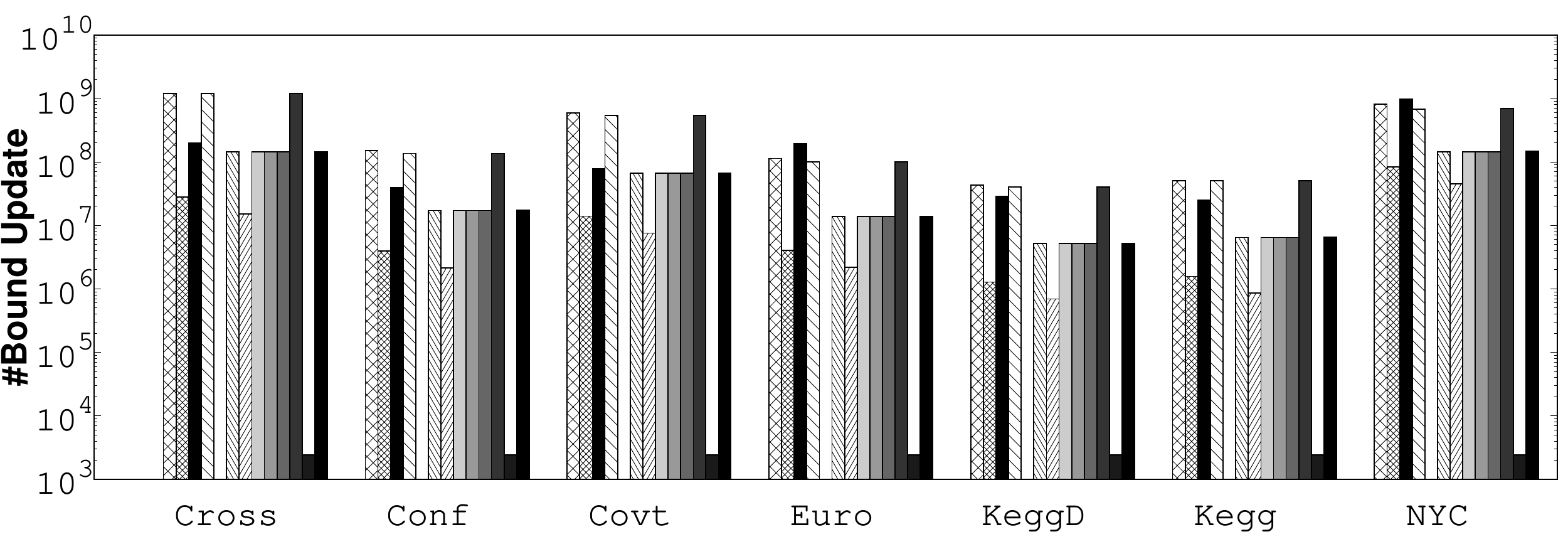}
		\vspace{-2.5em}
		\captionof{figure}{Statistics on the bound accesses and updates ($k=100$).}
		\vspace{-1em}
		\label{fig:seq2}
	\end{minipage}
\end{figure*}

\vspace{-0.3em}
\subsubsection{Index-based Methods}\label{sec:exp_im}
We implemented five indices: kd-tree, Hierarchical \kmeans tree (HKT) \cite{Ross1975}, Ball-tree \cite{Omohundro1989}, M-tree \cite{Ciaccia1997a}, and Cover-tree \cite{Beygelzimer2006} covered in Section~\ref{sec:kd}. The latter four that bound points by a radius can be easily extended to support \uni.

\myparagraph{Index Construction}{Figure~\ref{fig:index} compares these five indices over BigCross, and shows the construction time and clustering time w.r.t. the dimension $d$ and the data scale $n$, respectively.}
Here, we set $n=10,000$ when varying $d$, and M-tree is quite slow for a bigger $n$; that also explains why we ignore its performance.

\underline{\textit{Observations}}.~(1) With the increase of $d$ and $n$, the construction time increases and is more sensitive to $n$, but it is still tolerable for Ball-tree, Cover-tree, and kd-tree.
{(2) In average, Ball-tree is the fastest in clustering and 2nd fastest in index construction.
(3) Even though kd-tree is the fastest in index construction, its leaf-nodes can only cover one point whereas Ball-tree can cover multiple points. Thus, kd-tree has many more nodes than Ball-tree (the ratio is around the capacity $f=30$ according to our experiments), and Ball-tree is more applicable for large-scale clustering.} 
(4) Columns 5 and 6 of Table~\ref{tab:dataset} also show the index construction time and the number of nodes of Ball-tree. We find that the index of most datasets can be built within ten seconds, which means the delay of index building before clustering is tolerable.

\myparagraph{Clustering Efficiency}
\underline{\textit{Observations}}.~(1) With the increase of data scale, the cost of clustering rises dramatically for every index, but Ball-tree still beats other indices.
(2) When the dimensionality increases, kd-tree's performance degrades the most due to its complex pruning mechanism by using hyperplane; other indices are not impacted much as they use a radius to prune without any extra cost.
(3) A bigger $k$ makes the clustering slower, in a linear manner.

\begin{figure}[t]
	\centering
	\includegraphics[width=4.16cm,height=0.10\textheight]{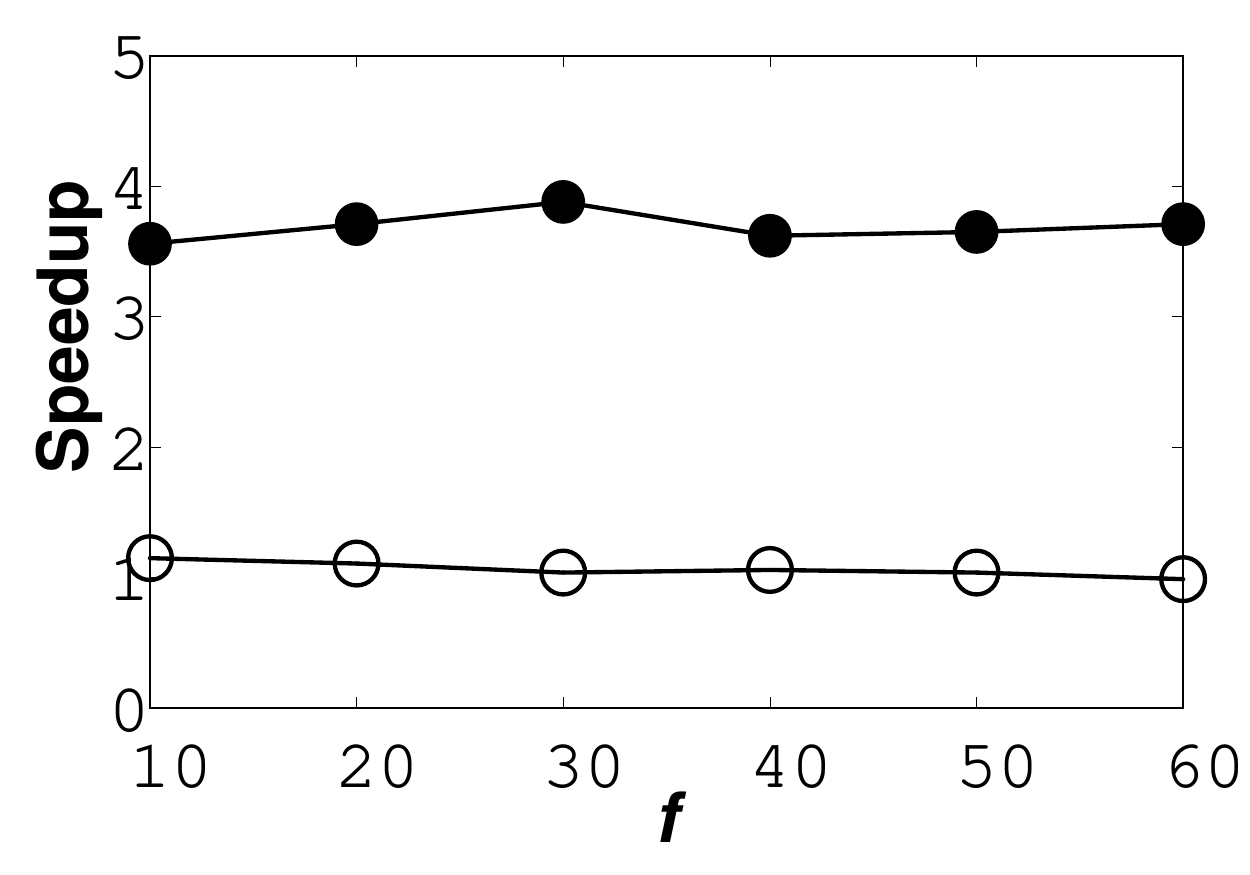}
	\includegraphics[width=4.16cm,height=0.10\textheight]{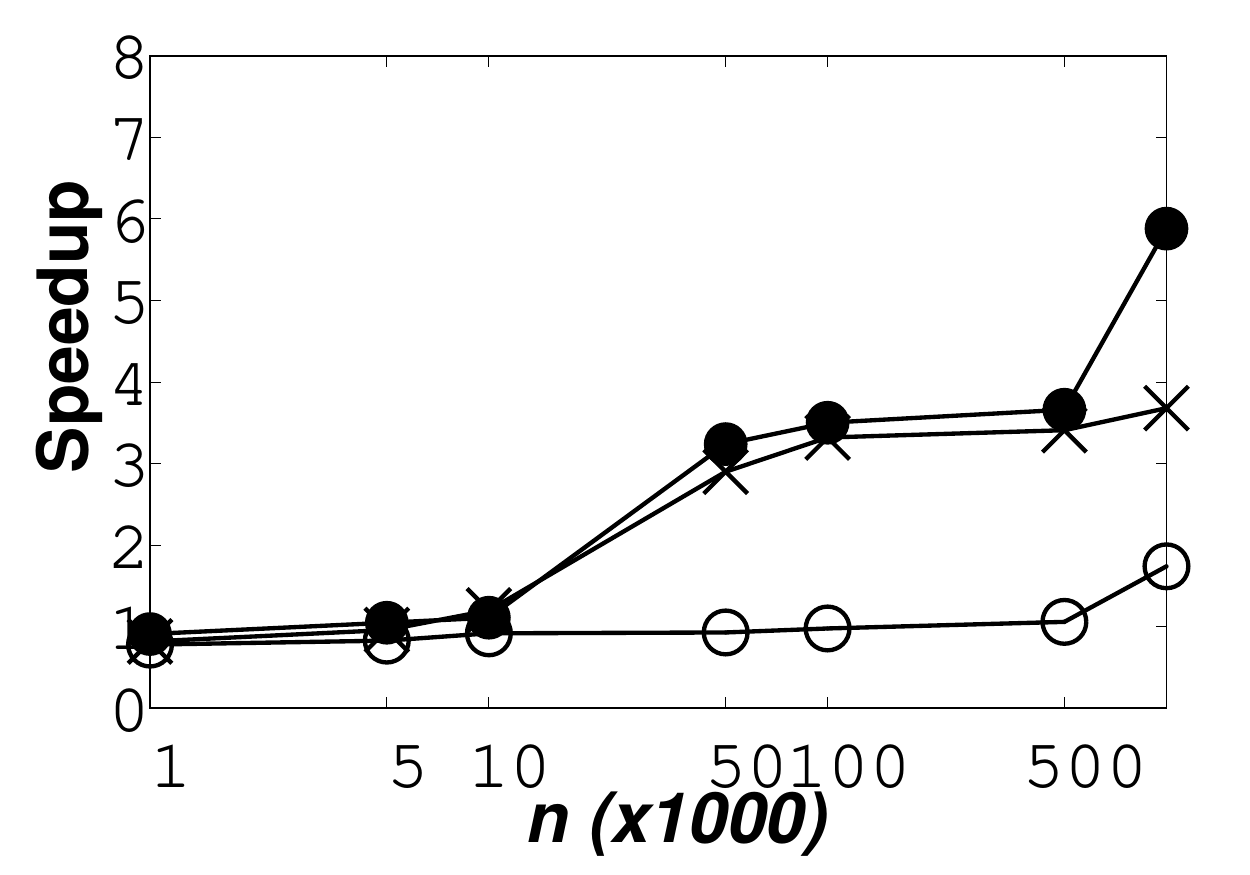}\\ \vspace{-0.5em}
	\includegraphics[width=4.16cm,height=0.10\textheight]{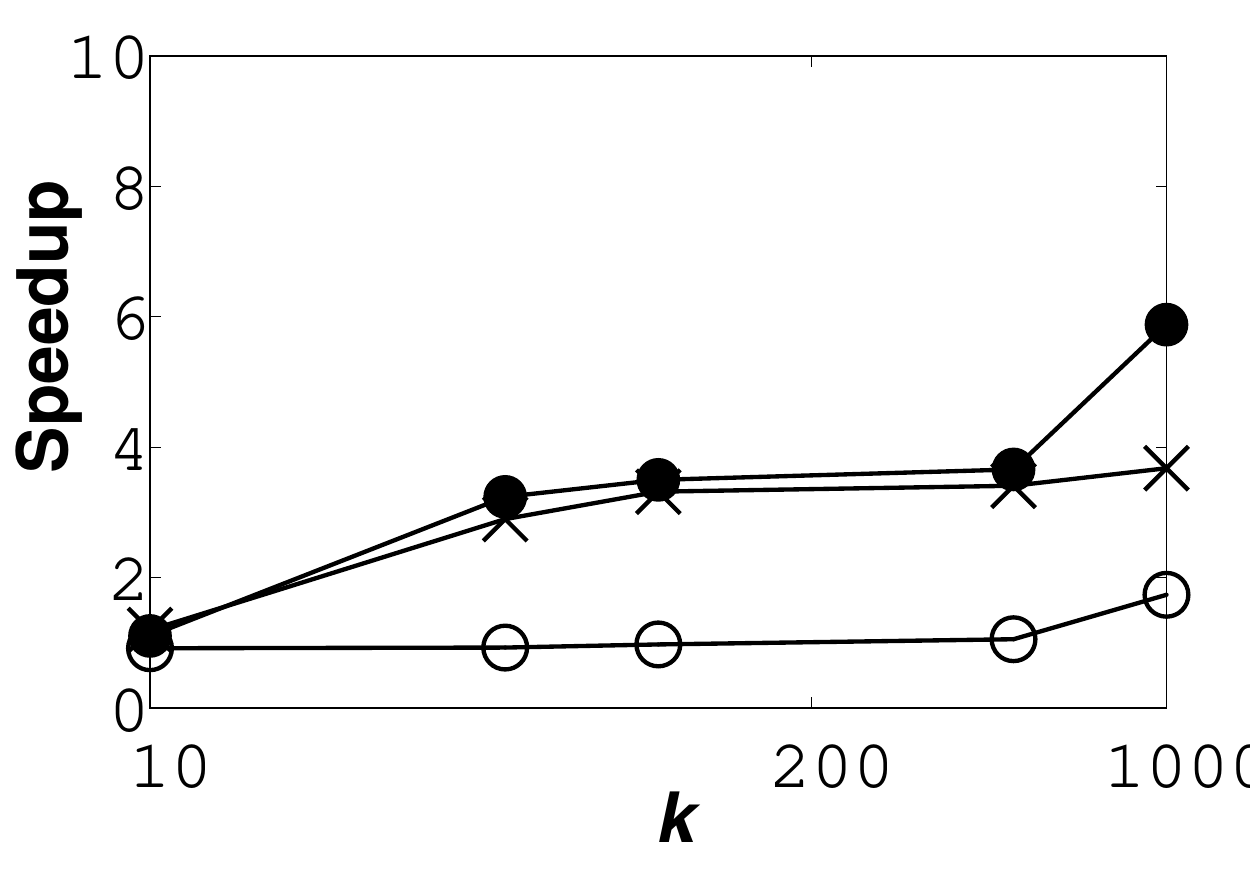}
	\includegraphics[width=4.16cm,height=0.10\textheight]{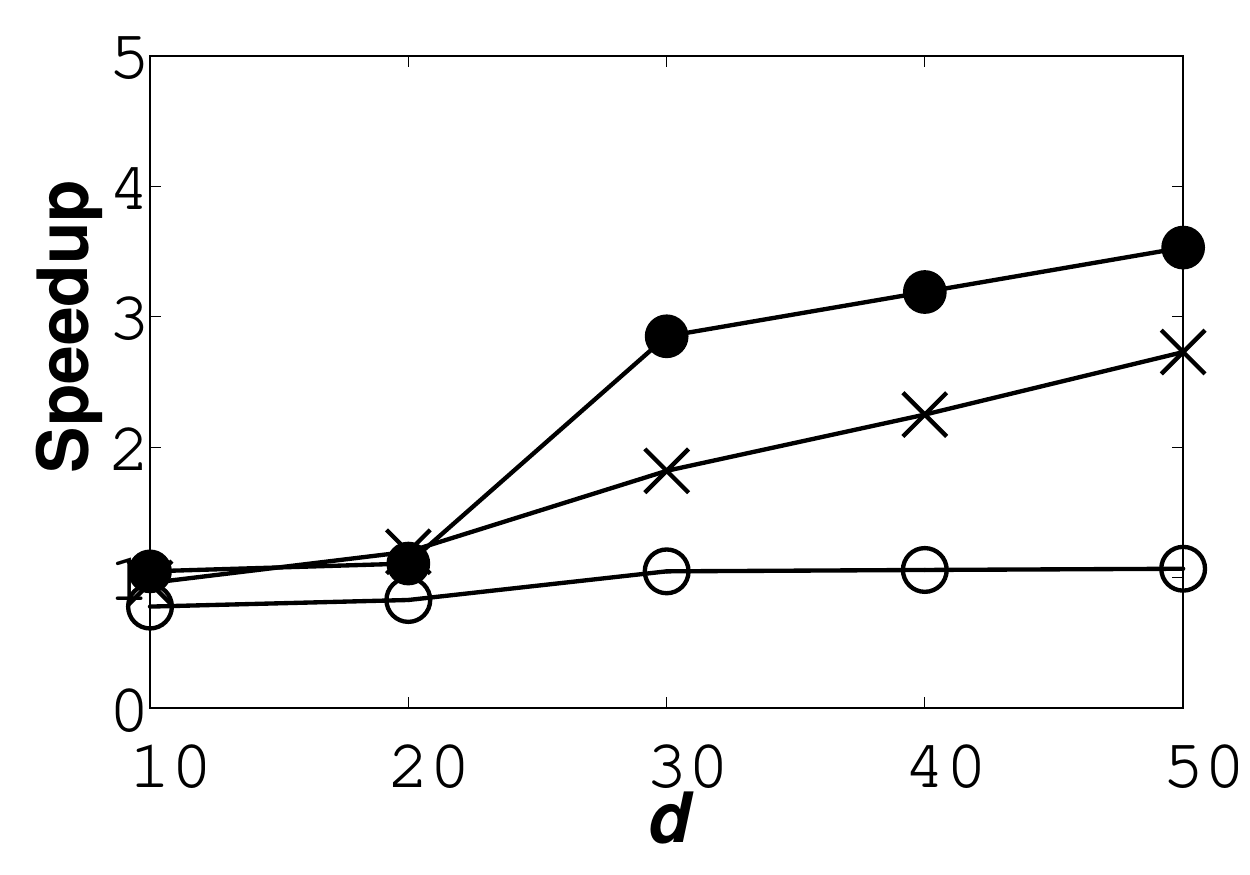}
	\vspace{-1.5em}
	\caption{{Sensitivity test on capacity $f$, data scale $n$, $k$, and dimension $d$ over the BigCross dataset.}}
	\vspace{-1em}
	\label{fig:increasing}
\end{figure}

\noindent \underline{\textit{Our Choice}}.~Thus, we choose Ball-tree \cite{Omohundro1989} as \uni's default index structure and conduct a further comparison with sequential methods and our proposed \uni and \utune.
The space cost of the index is not high compared with most sequential methods, which will be presented in Figure~\ref{fig:seq1}.
Moreover, setting a proper capacity $f$ for leaf nodes in M-tree, HKT, and Ball-tree is crucial,\footnote{Cover-tree and kd-tree do not have a parameter on capacity.} and a larger $f$ can save more space as the number of nodes will decrease.
To balance the space and efficiency, we set a relatively small capacity (i.e. $30$), and a sensitivity test of capacity will be conducted later in Figure~\ref{fig:increasing}, together with \uni. It shows that the clustering efficiency is not affected much by the capacity.

\vspace{-0.8em}
\subsubsection{Sequential Methods}
\label{sec:seqexp}
\myparagraph{Learderboard}
\begin{figure}
	\centering
	\includegraphics[height=0.16\textwidth]{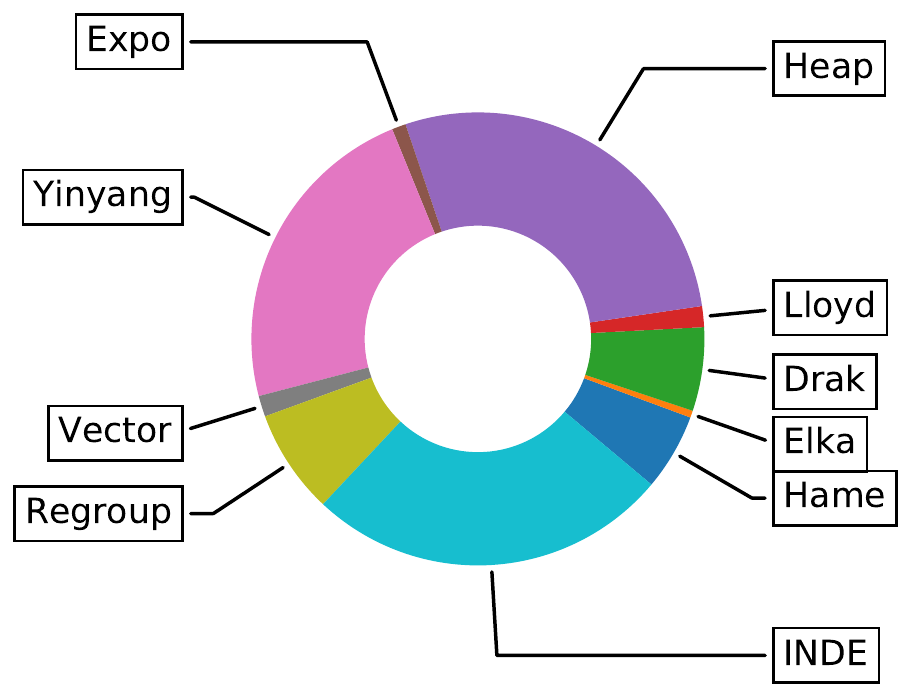}
	\includegraphics[height=0.16\textwidth]{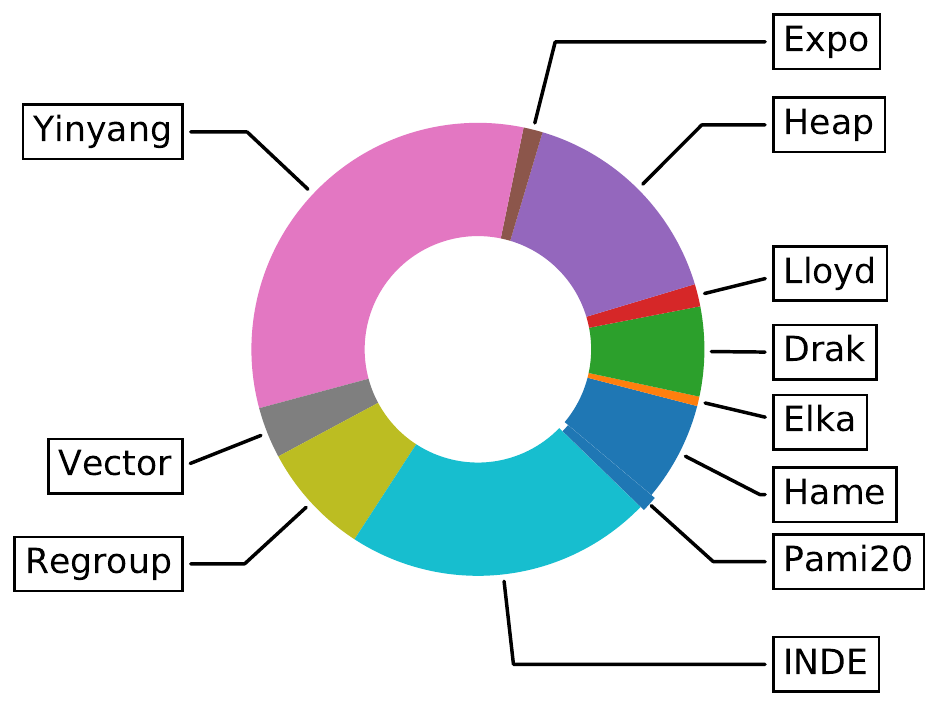}
	\vspace{-1.2em}
	\caption{\shengnew{Leaderboard of sequential methods as top 1 \& 3.}}\marginnote{D5@R2 \ref{sec:r2}}
	\vspace{-1em}
	\label{fig:lead}
\end{figure}
{
{Figure \ref{fig:lead} shows a leaderboard of various sequential methods by setting different parameter settings ($k$, $n$, and $d$) across all the datasets in Table~\ref{tab:dataset}.}
We can observe that five sequential methods have competitive performance: {\small\texttt{Hame}}, {\small\texttt{Drak}}, {\small\texttt{Heap}}, {\small\texttt{Yinyang}}, and {\small\texttt{Regroup}}. 

\noindent\underline{\textit{Our Choice}}.~Thus, we will use these five sequential algorithms as the selection pool in our auto-tuning model \utune.} 

\myparagraph{Speedup and Pruning Ratio}
We first investigate the speedup and the pruning ratio in distance computation. Figure~\ref{fig:seq} shows the overall speedup over the Lloyd's algorithm, compared with the representative index-based method ({\small\texttt{INDE}}): Ball-tree.
Since assignment occupies most time of the clustering time and it shows a similar trend with the overall speedup, we ignore it here, {and it can be found in our \textit{technical report}} \cite{wang2020tr} (see Section~\ref{sec:moreexp}).
Interestingly, the improvement of refinement (Figure~\ref{fig:seq-refine}) using our incremental method significantly improves the efficiency for all algorithms.


On the low-dimensional NYC dataset, we observe that the index-based method can beat all existing sequential methods in term of running time. This also occurs in several relatively high-dimensional datasets when $k=10$, such as KeggD and Kegg.
Among all the sequential methods, the {\small\texttt{Regroup}} and {\small\texttt{Yinyang}} are two fastest methods on most datasets.
We also observe that the speedup is not consistent with the pruning ratio, e.g., the index is almost 150 times (it is 400 when $k=10$) faster on NYC, while its pruning ratio is only 10\% (30\% when $k=10$) more than others. 
A reverse trend happens in the fully optimized algorithm ({\small\texttt{Full}}) which has the highest pruning ratio but very low efficiency.

The above observations verify our claims that \textbf{index-based method can be very fast, and a higher pruning ratio cannot simply guarantee better performance}. 

\myparagraph{Role of Data Access \& Bound Access/Update} 
Figure~\ref{fig:seq1} presents the memory usage of each algorithm. We find: among all the space-saving sequential methods, \texttt{Heap} can save the most, and index-based method's overhead is also much less than those sequential methods, and it will not increase with the rise of $k$.
Moreover in Figure~\ref{fig:seq2}, the index-based method has much less data access, which can explain why it is much faster; {\small\texttt{Yinyang}} has much less bound access and update which can explain why it is faster than other methods that have very similar pruning ratio. 
This also reveals that \textbf{data access, bound access and bound updates are three crucial factors that should also be considered in the future algorithm design and evaluation}.
{A similar performance breakdown of setting $k=10$ and more analysis can be found in our \textit{technical report} \cite{wang2020tr} (see Section~\ref{sec:moreexp}).}

\begin{figure}
	\centering
	\includegraphics[width=5.585cm]{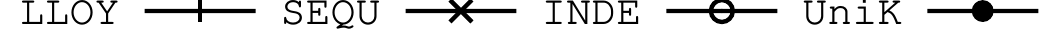}\\
	\includegraphics[width=4.16cm,height=0.11\textheight]{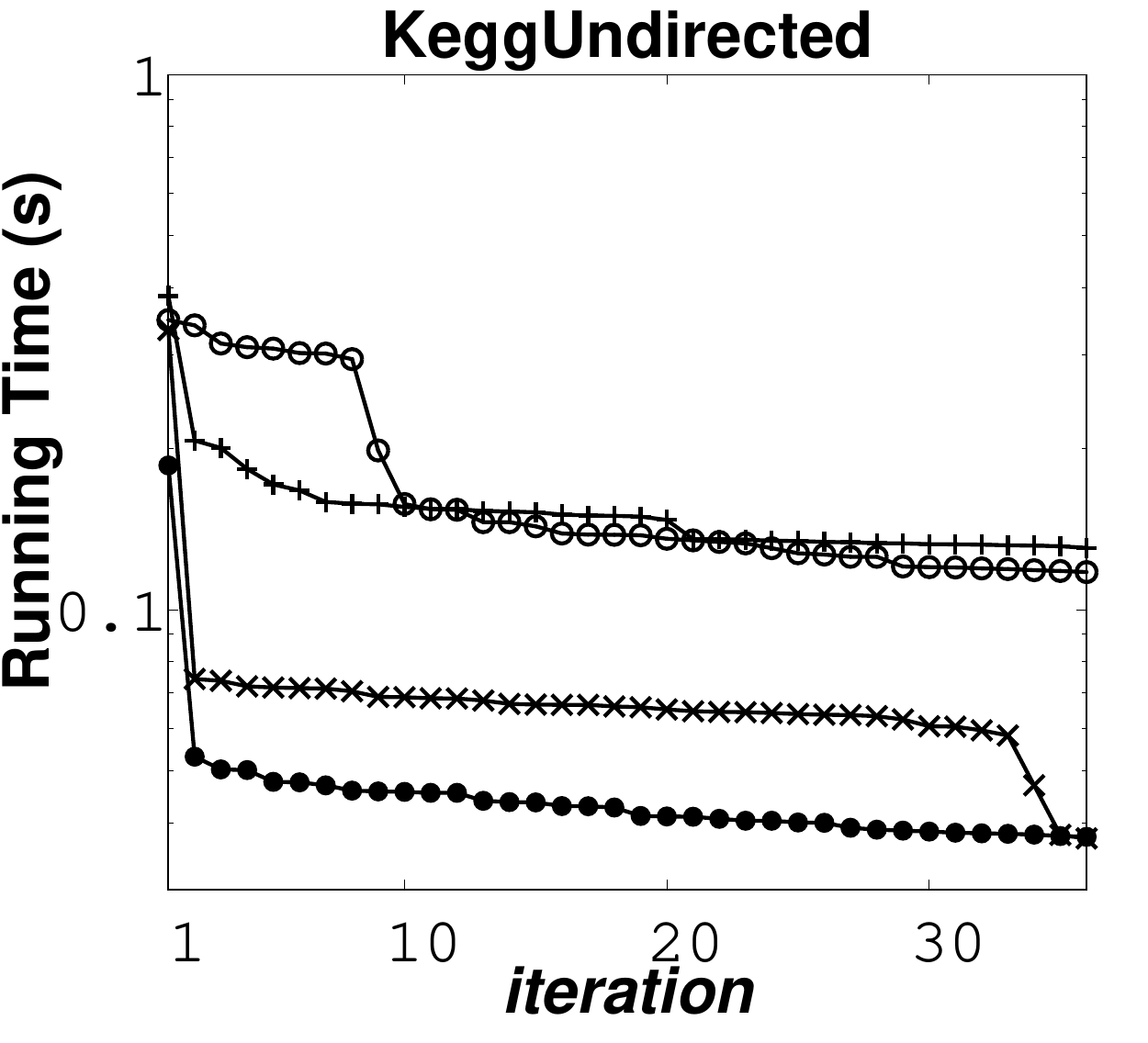}
	\includegraphics[width=4.16cm,height=0.11\textheight]{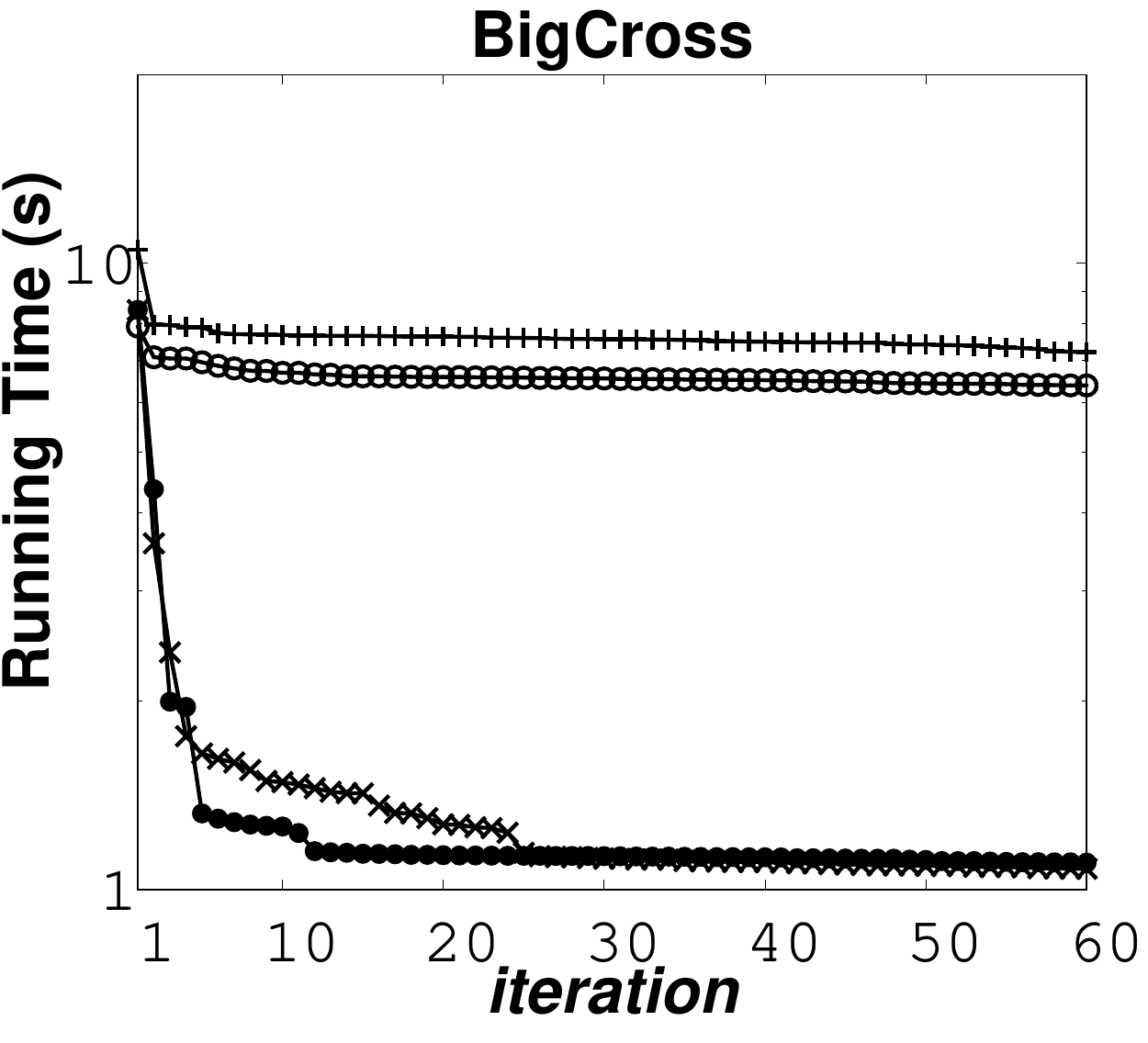}
	\vspace{-1.5em}
	\caption{{Running time of each iteration over the KeggUndirected ($d=29$) and BigCross dataset ($d=57$).}}
	\vspace{-1.5em}
	\label{fig:ite}
\end{figure}

\begin{table}
	\centering
	\ra{0.8}
	\caption{{Bound and data accesses in the first iteration.}}
	\vspace{-1.2em}
	\scalebox{0.95}{\begin{tabular}{c|c|c|c|c|c}
			\toprule
			\textbf{\small Dataset} & \textbf{\small Criteria} & \lyd & \seq & \ind & \uni \\ \midrule
			\multirow{5}{*}{\small \makecell{Cross\\$k=100$}} & Time (s) & 96.0 & 33.1 & 55.3 & 16.3 \\
			& Pruned & 0 & 84\% & 45\% & 91\% \\
			& Bound & 0 & 1.5B & 0 & 0.9B \\
			& Point & 100M & 30M & 15.8M & 9.8M \\
			& Node & 0 & 0 & 931k & 108k \\ \bottomrule
		\end{tabular}
		\label{tab:accessing}}
	\vspace{-1em}
\end{table}

\subsubsection{Our Unified Method \uni}
\label{sec:expuni}
Next, we compare \uni with (arguably the best) index-based method (\ind): {\small\texttt{Ball-tree}} and sequential method (\seq): {\small\texttt{Yinyang}}, and conduct a series of fine-grained evaluations on specific datasets.
Detailed results on each dataset are in Table~\ref{tab:overview} when comparing with \utune later.

\myparagraph{Running Time Per Iteration}
Figure~\ref{fig:ite} shows the running time on each iteration. 
We observe that the time spent per iteration decreases sharply in the first few iterations and then becomes stable.
\uni \ul{is the fastest because of our adaptive traversal mechanism}, and \ind and \seq dominate each other in two datasets respectively.

\myparagraph{Access on Data and Bound}
Specifically, we count the number of bound accesses, data accesses, and distance computations for BigCross, as shown in Table \ref{tab:accessing}.
The number of bound accesses and data accesses of \seq is much higher than \ind and \uni,
and \uni has the minimum number of accesses. 
For example, \seq needs 1.5 billion number of bound accesses, and \uni only needs 0.9 billion, as most points have been pruned by index.

\myparagraph{Robustness to Parameters}We test the robustness of \uni to various parameters in Figure~\ref{fig:increasing}, especially the capacity of the leaf nodes. 
By increasing the capacity, the performance decreases slightly but does not fluctuate much.
Like other two methods, \uni's speedup rises slightly when increasing $n$, $d$, and $k$.

\vspace{-0.5em}
\subsubsection{Summary of Lessons Learned}
We first summarize some key insights that might differ from existing studies:
	\begin{itemize}
		\item Ball-tree is very fast in both construction and clustering. 
		For low-dimensional spatial datasets such as NYC, Ball-tree can beat all sequential methods to a great extent.
		Moreover, Ball-tree also wins in high-dimensional datasets such as BigCross and Kegg when $k$ equals 10. This is because the data points assemble well, and having a small $r$ helps the pruning in batch. 
		Hence, by choosing the right index structure (Ball-tree), the index-based method can also perform well in high-dimensional data, which breaks the traditional conclusions towards the index-based methods \cite{Newling2016b}.
		
		\item Among all the sequential methods, there are five methods alternately being the fastest in various clustering tasks. They are: {\small\texttt{Hame}}, {\small\texttt{Drak}}, {\small\texttt{Heap}}, {\small\texttt{Yinyang}}, and {\small\texttt{Regroup}}. As a common target on reducing space on bounds, they also reduce the overhead to maintain and update those bounds, which further leads to faster performance.
		
		\item Several tight bounds proposed do not work well and their space consumption is high (e.g. \cite{Bottesch2016,Rysavy2016a}), probably because frequent bound access, comparison, and expensive updates are needed, and large amount of bounds have to be stored which further increases maintenance costs.
		Recall Section~\ref{sec:tightbound}, we can also see that it is complex to compute and update these bounds.
		\end{itemize}
	
	We also have several new and deeper insights:
		\begin{itemize}
		\item {By integrating the index-based and sequential methods, \uni achieves {an average of 53\% performance improvement} in our tested clustering tasks, and even up to 73\% when $k\ge100$ (details in Table~\ref{tab:overview}}). Compared with index-based methods, it avoids many distance computations when assigning nodes by bound-based pruning. Compared with sequential methods, it prevents intensive data accesses and distance computations simultaneously using batch pruning.
		
		\item {We have shown the aggregate rank of each method over all datasets in term of efficiency using pie charts in Figure~\ref{fig:lead}.} Further, we rate them in Table~\ref{tab:com} based on multiple metrics we used, and more experiments results are not shown due to space limit. Here we do not list other index-based methods as Ball-tree clearly dominates the rest. {We divided the metrics into two groups: beginners and researchers, where the former is for general data mining users, and the latter is for research purposes on accelerating \kmeans over existing algorithms. With our rates, users can choose a proper algorithm based on their demands.}
		
		\item Besides the metrics we have evaluated, Table~\ref{tab:com} also includes the robustness to various clustering tasks and ease of implementation that system engineers can refer to.
		For domain-specific analysts, we highly suggest spatial analysts (\earth) to adapt Ball-tree as it {can bring significant acceleration for low-dimensional spatial data}.
		Based on users' computing devices, we recommend that analysts with servers (\server) can choose {\small\texttt{Yinyang}} or our \uni, and {\small\texttt{Hame}} and {\small\texttt{Pami20}} will be proper for laptops (\laptop) with limited memory.
	\end{itemize}
\begin{table}
	\centering
	\ra{0.5}
		\caption{Evaluation Summary of Section~\ref{sec:expunik}. The darker the circle $\large \circ$, the higher degree of its corresponding criteria.}
		\vspace{-1em}
		\label{tab:com}
	\scalebox{0.85}{\begin{tabular}{c|ccc|ccc|ccc}
		\toprule
		 \textbf{Users} & \multicolumn{3}{|c|}{\textbf{Beginners}} & \multicolumn{3}{|c|}{\textbf{Researchers}} & \multicolumn{3}{|c}{\textbf{Domain}} \\ \midrule
		 	\makecell{\hspace{2em}\rotatebox[origin=c]{-45}{\textbf{Criteria}} \\ \\ \hspace{-2em} \rotatebox[origin=c]{0}{\textbf{Algorithm}}} & \rotatebox[origin=c]{90}{\small Leaderboard} & \rotatebox[origin=c]{90}{\small Space-saving} & \rotatebox[origin=c]{90}{\small Parameter-free} & \rotatebox[origin=c]{90}{\small \makecell{Fewer \\data access}} & \rotatebox[origin=c]{90}{\small \makecell{Fewer \\bound access}} & \rotatebox[origin=c]{90}{\small \makecell{Fewer \\distance}} & \rotatebox[origin=c]{90}{\small Robustness} & \rotatebox[origin=c]{90}{\small \makecell{Ease of \\implementation}} & \rotatebox[origin=c]{90}{\small Analysts} \\ \midrule
		 {\small \texttt{Ball-tree}} \cite{Moore2000} & \like{10} & \like{8} & \like{8} & \like{10} & \like{10} & \like{5} & \like{0} & \like{6} & \earth \\
		 \elkan \cite{Elkan2003} & \like{0} & \like{0} & \like{10} & \like{2} & \like{2} & \like{2} & \like{2} & \like{5} & \\
		 {\small\texttt{Hame}} \cite{Hamerly2010} & \like{5} & \like{8} & \like{10} & \like{2} & \like{6} & \like{8} & \like{2} & \like{8} & \\
		 {\small\texttt{Drak}} \cite{Drake2012} & \like{5} & \like{9} & \like{5} & \like{9} & \like{3} & \like{5} & \like{5} & \like{2} & \\
		 {\small\texttt{Annu}} \cite{Drake2013} & \like{0} & \like{5} & \like{10} & \like{5} & \like{2} & \like{2} & \like{8} & \like{8} & \\
		 {\small\texttt{Heap}} \cite{Hamerly2015a} & \like{10} & \like{10} & \like{10} & \like{6} & \like{6} & \like{6} & \like{5} & \like{4} & \laptop \\
		 \yiny \cite{Ding2015} & \like{10} & \like{7} & \like{5} & \like{8} & \like{8} & \like{8} & \like{8} & \like{4} & \server \\
		 {\small\texttt{Expo}} \cite{Newling2016b} & \like{0} & \like{6} & \like{10} & \like{2} & \like{2} & \like{2} & \like{2} & \like{5} & \\
		 {\small\texttt{Drift}} \cite{Rysavy2016a} & \like{2} & \like{2} & \like{10} & \like{2} & \like{2} & \like{2} & \like{2} & \like{2} & \\
		 {\small\texttt{Vector}} \cite{Bottesch2016} & \like{2} & \like{2} & \like{5} & \like{2} & \like{2} & \like{2} & \like{2} & \like{4} & \\
		 {\small\texttt{Regroup}} \cite{Rysavy2016a} & \like{5} & \like{6} & \like{8} & \like{8} & \like{8} & \like{8} & \like{8} & \like{4} & \\
		 {\small\texttt{Pami20}} \cite{Xia2020} & \like{3} & \like{10} & \like{10} & \like{5} & \like{8} & \like{5} & \like{6} & \like{9} & \laptop \\
		 \uni & \like{10} & \like{5} & \like{7} & \like{8} & \like{7} & \like{7} & \like{10} & \like{4} & \server \\ \bottomrule
	\end{tabular}}
	\vspace{-1em}
\end{table}
%

\vspace{-1em}
\subsection{Evaluation of Auto-tuning}
\label{sec:expauto}

\begin{table*}
	\centering
	\ra{1.1}
	\vspace{-1em}
	\caption{\ul{Evaluation of knob configuration in terms of MRR prediction accuracy.}}
	\vspace{-1.2em}
	\scalebox{0.85}{\begin{tabular}{ccccccccccccccccccc}
			\toprule
			\multirow{2}{*}{\textbf{\makecell{Accuracy}}}	&\multirow{2}{*}{BDT} &\multicolumn{5}{c}{\textbf{Basic features}} & & \multicolumn{5}{c}{\textbf{+ Tree-features}} & & \multicolumn{5}{c}{\textbf{+ Leaf-features}} \\ 	\cmidrule{3-7} \cmidrule{9-13} \cmidrule{15-19} 
		&	& DT &RF & SVM & kNN & RC & & DT & RF & SVM & kNN & RC&&DT & RF& SVM & kNN & RC \\ \midrule
			Bound@MRR &0.41&0.70 & 0.68 & 0.62 & 0.63 & 0.57
			&&0.67 & 0.70 & 0.64 & 0.63 & 0.59&&0.69 & 0.68 & 0.63 & 0.63 & 0.60 \\ 
			Index@MRR&0.37&0.80 & 0.82 & 0.84 & 0.74 & 0.68
			&&0.83 & 0.77 & 0.83 & 0.74 & 0.70&&0.74 & 0.77 & 0.83 & 0.74 & 0.74 \\ 
						S-Bound@MRR &{0.42}& 0.84 & 0.83 & 0.81 & 0.82 & 0.74 & & 0.86 & 0.87 & 0.81 & 0.82 & 0.74
			&&0.89 & 0.87 & 0.88 & 0.88 & 0.80 \\ 
			S-Index@MRR&\textbf{0.43}&0.89 & 0.87 & 0.86 & 0.83 & 0.83& &0.91 & 0.90 & 0.87 & 0.83 & 0.85&&\textbf{0.92} & 0.92 & 0.92 & 0.86 & 0.84
			\\ \bottomrule
		\end{tabular}
		\label{tab:precision}}
	\vspace{-0.5em}
\end{table*}
\begin{table*}[!ht]
	\ra{0.85}
	\centering
	\caption{{Overall speedup over the running time (second) of Lloyd's algorithm (the gray column) on various datasets.}}
	\vspace{-1em}
	\scalebox{0.95}{\begin{tabular}{|c|c|cccc|c|cccc|c|cccc|}
		\hline
		\multirow{3}{*}{ \textbf{Data}} & \multicolumn{5}{c}{$k=10$} & \multicolumn{5}{c}{$k=100$} & \multicolumn{5}{c|}{$k=1000$} \\ \cline{2-16}
& \multirow{2}{*}{\lyd} & \multicolumn{4}{c|}{$\times${Speedup}} & \multirow{2}{*}{\lyd} & \multicolumn{4}{c|}{$\times${Speedup}} & \multirow{2}{*}{\lyd} & \multicolumn{4}{c|}{$\times${Speedup}} \\ \cline{3-6}\cline{8-11}\cline{13-16}
& & \seq & \ind & \uni & \utune & & \seq & \ind & \uni & \utune & & \seq& \ind & \uni & \utune \\ \hline
		 Cross & \cg 262 & \makecell{1.64\\71\%} & \makecell{1.76\\67\%} & \makecell{1.36 \\78\%} & \makecell{1.76\\84\%} & \cg 1463 & \makecell{2.83\\86\%} & \makecell{2.16\\59\%} & \makecell{\textbf{3.24}\\90\%} & \makecell{\textbf{4.70}\\90\%} & \cg13530 & \makecell{3.31\\90\%} & \makecell{1.81\\46\%} & \makecell{\textbf{4.04}\\94\%} & \makecell{\textbf{7.73}\\93\%}\\
		 Conf & \cellcolor{lightgray}2.45 & \makecell{1.32\\68\%} & \makecell{1.30\\61\%} & \makecell{1.30\\74\%} & \makecell{1.32\\68\%} & \cellcolor{lightgray}9.00 & \makecell{1.53\\90\%} & \makecell{1.63\\25\%} & \makecell{2.22 \\90\%} & \makecell{2.50\\90\%} & \cellcolor{lightgray}50.75 & \makecell{2.83\\90\%} & \makecell{1.47\\6\%} & \makecell{2.88 \\93\%} & \makecell{2.93\\93\%} \\ 
		 Covt & \cellcolor{lightgray}0.65 & \makecell{1.89\\74\%} & \makecell{2.18\\72\%} & \makecell{1.68\\87\%} &\makecell{2.18\\72\%} & \cellcolor{lightgray}2.53 & \makecell{5.62\\90\%} & \makecell{1.43\\23\%} & \makecell{5.61\\94\%} & \makecell{5.67\\93\%}& \cellcolor{lightgray}10.39 & \makecell{7.47\\92\%} & \makecell{1.04\\4\%}& \makecell{6.66\\92\%} & \makecell{7.47\\92\%} \\ 
		 Euro & \cellcolor{lightgray}15.3 & \makecell{1.38\\75\%} & \makecell{1.42\\67\%} & \makecell{1.39\\84\%} & \makecell{1.48\\35\%} & \cellcolor{lightgray}111 & \makecell{3.24\\92\%} & \makecell{2.53\\45\%} & \makecell{3.79\\ 90\%} & \makecell{4.03\\ 95\%}& \cg381.9 & \makecell{2.65\\94\%} & \makecell{0.63\\11\%} & \makecell{3.12\\95\%} & \makecell{3.13\\95\%} \\ 
			 KeggD & \cellcolor{lightgray}0.45 & \makecell{2.93\\83\%} & \makecell{3.59\\79\%} & \makecell{\textbf{4.22}\\84\%} & \makecell{\textbf{4.3}\\ 95\%}& \cg1.16 & \makecell{2.61\\92\%} & \makecell{ 1.21\\11\% }& \makecell{4.00\\71\%} & \makecell{5.8 \\95\%}& \cg 8.50 & \makecell{6.58\\93\%} & \makecell{1.23\\11\% }& \makecell{\textbf{7.01}\\89\%} & \makecell{\textbf{7.57}\\95\%} \\ 
			Kegg & {\cellcolor{lightgray}0.49} & \makecell{1.98\\78\%} & \makecell{2.83\\ 83\%} & \makecell{2.40\\94\%} & \makecell{2.83\\ 83\%} & \cellcolor{lightgray}2.49 & \makecell{4.69\\93\%} & \makecell{1.79\\ 31\%} & \makecell{\textbf{5.87}\\95\%} &\makecell{\textbf{6.15}\\ 96\%} & \cellcolor{lightgray}18.64 & \makecell{6.67\\93\%} & \makecell{0.94\\51\%} & \makecell{6.52\\95\%} &\makecell{6.67 \\ 93\%} \\ 

		 NYC & \cg15.3 & \makecell{ 1.39\\84\%} & \makecell{389 \\ 99\%} & \makecell{{31.4}\\99\%} & \makecell{\textbf{389} \\ 99\%} & \cg 75.6 & \makecell{4.19 \\94\%}& \makecell{153\\ 99\%} & \makecell{{55.6}\\99\%} &\makecell{153\\ 99\%} & \cg 229.8 & \makecell{1.69\\93\%} & \makecell{11.05\\93\%} & \makecell{7.53\\95\%} & \makecell{13.3\\96\%} \\ 
		 		 Skin & \cg0.56 & \makecell{1.30\\79\%} & \makecell{2.54\\87\%} & \makecell{2.40 \\88\%} & \makecell{2.54\\87\%} & \cg2.92 & \makecell{2.35\\92\%} & \makecell{2.60\\56\%} & \makecell{\textbf{4.09}\\96\%} & \makecell{\textbf{4.13}\\96\% }& \cg 21.41 & \makecell{ 2.70\\93\% }& \makecell{1.38\\27\%} & \makecell{\textbf{3.28}\\94\%} & \makecell{\textbf{3.54}\\95\%} \\ 
		 Power & \cg 6.38 & \makecell{1.43\\78\%} & \makecell{0.77\\53\% }& \makecell{0.87\\82\%} & \makecell{1.43\\78\%} & \cg32.9 & \makecell{2.39\\91\%} & \makecell{1.02\\18\%} & \makecell{2.53\\93\% } & \makecell{2.60\\91\%}& \cg 223.9 & \makecell{2.17\\92\%} & \makecell{0.96\\2\%} & \makecell{2.26\\92\%} & \makecell{2.5\\92\%} \\ 
		 Road & \cellcolor{lightgray}6.02 & \makecell{1.36 \\84\%} & \makecell{8.64\\96\%} & \makecell{{8.19}\\98\%} & \makecell{8.64\\96\%} & \cellcolor{lightgray}21.2 & \makecell{2.57\\93\%} & \makecell{3.68\\ 69\%} &\makecell{\textbf{4.60}\\93\%} & \makecell{\textbf{4.93}\\97\%}& \cellcolor{lightgray}132.8 & \makecell{2.40\\94\%} & \makecell{1.58\\27\% }& \makecell{2.69\\ 93\%} & \makecell{2.87\\95\%} \\ 
		 Census & \cellcolor{lightgray}11.9 & \makecell{1.31\\62\%} & \makecell{0.82\\26\%} & \makecell{1.14\\67\%} & \makecell{1.55 \\ 69\%} & \cellcolor{lightgray}94.7 & \makecell{3.65\\84\%} & \makecell{1.14\\15\%} & \makecell{3.51\\85\%} & \makecell{3.67 \\ 84\%} & \cellcolor{lightgray}791 & \makecell{5.85\\91\%} & \makecell{1.05\\9\%} & \makecell{5.87\\91\%} & \makecell{5.87\\91\%} \\ 
		 Mnist & \cellcolor{lightgray}7.44 & \makecell{1.13\\1\%} & \makecell{0.91\\0\%} & \makecell{0.98\\1\%} & \makecell{1.36 \\ 27\%} & \cellcolor{lightgray}67.3 & \makecell{1.21\\17\%} & \makecell{0.98\\15\%} & \makecell{1.22\\18\%} & \makecell{3.94 \\ 77\%} & \cellcolor{lightgray}709 & \makecell{1.69\\37\%} & \makecell{1.04\\2\%} & \makecell{1.54\\38\%} & \makecell{5.13\\83\%} \\ 	

		 \hline	 
		 Spam & \cg0.12 & \makecell{1.13} & \makecell{1.42} & \makecell{1.15} & \makecell{1.62} & \cg0.69 & \makecell{5.80} & \makecell{2.12} & \makecell{12.59} & \makecell{12.59 }& \cg 4.87 & \makecell{ 4.35 }& \makecell{2.87} & \makecell{8.87} & \makecell{8.87} \\ 
		 
		 Shuttle & \cg 0.20 & \makecell{3.65} & \makecell{0.72}& \makecell{0.57} & \makecell{3.65} & \cg1.15 & \makecell{5.62} & \makecell{3.67} & \makecell{5.47 } & \makecell{6.53}& \cg 4.61 & \makecell{4.85} & \makecell{1.94} & \makecell{5.17} & \makecell{5.17} \\ 
		 MSD & \cellcolor{lightgray}8.93 & \makecell{1.17 } & \makecell{0.72} & \makecell{{0.92}} & \makecell{1.17} & \cellcolor{lightgray}21.2 & \makecell{2.04} & \makecell{1.21} &\makecell{2.17} & \makecell{2.17} & \cellcolor{lightgray}592 & \makecell{2.33} & \makecell{1.17}& \makecell{2.57} & \makecell{2.57} \\ 
		 \hline
	\end{tabular}}
	\label{tab:overview}
	\vspace{-0.5em}
\end{table*}

Based on the above comprehensive evaluations on index-based and sequential methods, we choose two representative algorithms, one for each class, denoted as \ind (Ball-tree) and \seq ({\small\texttt{Yinyang}}).\footnote{{These two algorithms also compose the BDT in Figure~\ref{fig:dt}, where {\texttt{Yinyang}} is same as {\texttt{Hame}} by setting $t=1$ when $k<50$.}}
Next we compare them with \uni (with our default index and bound configurations), and our \utune with various learning models.

\vspace{-0.5em}
\subsubsection{Model Training}
\myparagraph{Ground Truth Generation}
Since available public datasets for clustering are not that many, 
when running clustering methods to generate the ground truth, we alter $k=\{10, 100, 200, 400, 600, 800, 1000\}$, $n=\{10^3, 10^4, 10^5, 10^6\}$, and $d=\{10, 20, 30, 40, 50\}$ over all the datasets in Table~\ref{tab:dataset}.
{In the appendix of \textit{technical report} \cite{wang2020tr} (see Figure~\ref{fig:train-time}), we present the efficiency of the full running and selective running (proposed in Section~\ref{sec:rskc} and we used five methods) on the datasets chosen based on our leaderboards (see Figure~\ref{fig:lead}).
We can observe that selective running is much faster, as it skips those slow methods and saves much time to run more parameter settings and get more training data, which can further improve the precision, as later exhibited in Table~\ref{tab:precision}.}

\myparagraph{Adopted ML Models}
With the ground truth obtained, we adopt most classical classification models \cite{Kotsiantis2007} for training and prediction: decision trees (\textbf{DT}), random forests (\textbf{RF}), $k$ nearest neighbor (\textbf{kNN}), support vector machine (\textbf{SVM}), and Ridge linear classifier (\textbf{RC}).\footnote{We divide the ground truth into two parts: 70\% for training and 30\% for testing.} %

\myparagraph{Prediction Accuracy}
We adopt a rank-aware quality metric called \textit{mean reciprocal rank} (\textbf{MRR}) \cite{craswell2009mean} to measure the precision. 
Given a set of testing records $R$, our loss function is defined as below:

\vspace{-0.8em}
\begin{equation}
\var{MRR}(R) = \frac{1}{|R|}\cdot\sum_{l\in R}{\frac{1}{rank(p_l)}}
\end{equation}
\vspace{-0.8em}

\noindent where $rank(p_l)$ denotes the ranking of prediction $p_l$ in the labeled ground truth. 
%

Table~\ref{tab:precision} shows the MRR of different models (the {BDT} in Figure~\ref{fig:dt}, DT, RF, SVM, kNN, RC) in predicting the index configuration (i.e. Index@MRR) and the choice of bound to be used (i.e. Bound@MRR). The MRR result can be further interpreted from two dimensions: (i) For each model, we distinguish the MRR precision over the training data obtained from the full running and the selective running (highlighted as a prefix ``S-''), respectively; (ii) For each model, we keep adding three groups of features in the training phase, namely \textit{basic features}, index features (\textit{Tree}), and advanced features on leaf level (\textit{Leaf}), to verify their effectiveness.
Details on these features are in Table~\ref{tab:features}. 
For completeness purpose, we also report the training and prediction time in our \textit{technical report} \cite{wang2020tr} (see Table~\ref{tab:precision-prediction}).

\underline{\textit{Observations}}. (1) Within the same limited time, selective running has higher precision and can achieve 92\% if using decision tree (with a depth of 10) or SVM, while BDT that relies on fuzzy rules only achieves 43\%. This is because selective running manages to generate more training records than full running (e.g., 1600 vs. 436 in this case). (2) With more index and leaf features, we can have a higher precision than using basic features only when using the selective running ground-truth. (3) Among all the classifier models, the decision tree has the highest precision and it is also very fast for both training and prediction.

\subsubsection{Verification}
Among all the prediction models of high accuracy, we select the decision tree (DT) to support \utune. Then, we compare \utune with the representatives: \ind, \seq, \uni, to verify whether our predicted configuration works well. 
Table~\ref{tab:overview} presents the running time of Lloyd's algorithm, and the speedup brought by \ind, \seq, \uni, and \utune over Lloyd's.
The percentage of the pruned distance computations is shown below the speedup.
{In the appendix our of \textit{technical report} \cite{wang2020tr}, we also show the corresponding assignment and refinement time.}

\textit{\underline{Observations.}}~(1) On average, both \uni and \utune \ul{outperform the index-based and sequential methods in multiple cases, especially when $k$ is big (see bold numbers for significant improvements), and the integration of bound and indexing further improves the pruning ratio.}\marginnote{W3@R1\\ \ref{sec:r1} \\ Bold numbers}
(2) \uni cannot always be fast over high-dimensional data such as Power, which is also consistent with the \textit{no free lunch theorem} \cite{Wolpert1997}.
The main reason is that the index cannot prune well in the first iteration for the datasets that do not assemble well, while \uni can alter to sequential method and avoid being slow in the following iterations. 
(3) By applying an auto-tuning in predicting the optimal configuration, \utune achieves the best performance across all datasets. The performance gap is even larger on low-dimensional datasets such as NYC (up to $389$ times speedup), where
\utune rightly predicts the configuration and improves the pruning percentage over \seq.

{
To investigate whether our {adopted ML models} can generalize well to datasets that have not been seen during training, we further test \utune on three new datasets: Spam \cite{Spam}, Shuttle \cite{Shuttle}, and MSD \cite{msd}.
The results are consistent with our earlier finding, i.e., \utune beats other methods in most times, and is always in the leaderboard.
This confirms a good generalization capability of our ML model. 
}

\vspace{-0.6em}
\subsubsection{Summary of Lessons Learned}
	Through an evaluation on multiple learning models for auto-configuration, we further learn:
	\begin{itemize}
		\item Automatic algorithm selection for fast \kmeans is feasible by using a meta-learning model. Through \utune, we can predict an algorithm configuration that leads to better performance than state-of-the-art methods which we have reported in last section, also including our \uni.
		Moreover, the learning cost is low if using the offline evaluation logs.
		
		\item There are several ways to improve the precision of models. Firstly, our selective running based on evaluation can save much time to generate more logs for training. Secondly, building indexes can provide more features that help improve the prediction accuracy.
		
		\item By using very basic machine learning models without fine-tuning, our algorithm selection approach already achieves 92\% prediction accuracy. {Note that finding the best choice of learning models is orthogonal to this work, though.}
	\end{itemize}

\vspace{-1em}
\section{Conclusions}
We evaluated existing accelerating algorithms for fast \kmeans, in a unified framework which enables a fine-grained performance breakdown. 
To auto-configure the pipeline for the optimal performance, we 
trained a meta-model to predict the optimal configuration. 
Experiments on real datasets showed that our unified framework with autotuning can accelerate \kmeans effectively.
In the future, we will work on learning with a rank-aware loss function like MRR, i.e., designing specific machine learning models to further improve precision.
{Recalling the parameter settings in Section~\ref{sec:expset}, it will also be interesting to study the tuning of three sequential methods' parameters.}
{More discussions on these future opportunities can be found in our \textit{technical report} \cite{wang2020tr} (see Section~\ref{sec:future}).}

{\myparagraph{Acknowledgment} Zhifeng Bao is supported by ARC DP200102611, DP180102050 and a Google
Faculty Award.}

\balance
{
\bibliographystyle{abbrv}
\bibliography{url,library} 
}
\newpage
\appendix
\section{Appendix}
	


\begin{table*}
	\centering
	\ra{1.2}
	\caption{{Evaluation of knob configuration: training time and prediction time.}}
	\vspace{-1em}
	\scalebox{0.9}{\begin{tabular}{ccccccccccccccccccc}
			\toprule
			\multirow{2}{*}{\textbf{\makecell{Efficiency}}}	&\multirow{2}{*}{BDT} &\multicolumn{5}{c}{\textbf{Basic features}} & & \multicolumn{5}{c}{\textbf{+ Tree-features}} & & \multicolumn{5}{c}{\textbf{+ Leaf-features}} \\ 	\cmidrule{3-7} \cmidrule{9-13} \cmidrule{15-19} 
			&	& DT &RF & SVM & kNN & RC & & DT & RF & SVM & kNN & RC&&DT & RF& SVM & kNN & RC \\ \midrule
			Training (ms)&-&1.63 & 577 & 3.74 & 1.73 & 4.40
			&&1.89 & 573 & 4.35 & 1.39 & 4.54&&2.53 & 598 & 5.42 & 1.43 & 4.36 \\ 
			Prediction ($\mu$s)&-&3.98 & 251 & 6.28 & 36.4 & 5.33
			&&6.09 & 259 & 7.14 & 39 & 6.72&&7.15 & 271 & 10.2 & 37.7 & 8.57 
			\\ 
			S-Training (ms)&-&1.98 & 660 & 29 & 1.85 & 4.90 &
			& 2.81 & 764 & 53.3 & 2.33 & 4.55&&4.75 & 897 & 54.3 & 1.87 & 4.68
			\\ 
			S-Prediction ($\mu$s)&-&1.53 & 97 & 10.7 & 27.9 & 1.39 &
			& 1.24 & 105 & 13.3 & 32.00 & 1.82&&\textbf{1.35} & 97.2 & 14.6 & 34.5 & 1.90\\ \bottomrule
		\end{tabular}
		\label{tab:precision-prediction}}
\end{table*}

\begin{table*}
	\ra{1.2}
	\centering
	\caption{{The assignment speedup over the running time (second) of Lloyd's algorithm (the gray column).}}
	\vspace{-1em}
	\hspace{-2em}
	\scalebox{1}{\begin{tabular}{|c|c|cccc|c|cccc|c|cccc|}
			\hline
			\multirow{3}{*}{ \textbf{Data}} & \multicolumn{5}{c}{$k=10$} & \multicolumn{5}{c}{$k=100$} & \multicolumn{5}{c|}{$k=1000$} \\ \cline{2-16}
			& \multirow{2}{*}{\lyd} & \multicolumn{4}{c|}{$\times${Speedup}} & \multirow{2}{*}{\lyd} & \multicolumn{4}{c|}{$\times${Speedup}} & \multirow{2}{*}{\lyd} & \multicolumn{4}{c|}{$\times${Speedup}} \\ \cline{3-6}\cline{8-11}\cline{13-16}
			& & \seq & \ind & \uni & \utune & & \seq & \ind & \uni & \utune & & \seq& \ind & \uni & \utune \\ \hline
			Cross & \cg 219.7 &1.45 & 1.86 & 1.16 & 1.56& \cg 1381 & 2.73 & 2.20 & 3.10 & 4.54& \cg13325 & 3.30 & 1.82 & 4.00 & 7.68 \\
			Conf & \cg1.73 & 1.09 & 1.44 & 0.62 &1.09 & \cg7.42 & 1.37& 1.69 & 2 & 2.26 & \cg48.17 & 2.76 & 1.46 & 2.79 & 2.86 \\
			Covt & \cg0.48 & 1.72 & 2.35& 1.53 & 2.35 & \cg 2.09 &5.55 & 1.44 & 5.50 & 5.60& \cg10.13 &7.46& 1.04 & 6.65 &7.46 \\
			Euro & \cg12.72 & 1.15 & 1.46 & 0.32 & 1.31 & \cg103.82 & 2.94 & 2.61 & 1.67	 &3.64	 & \cg379.46 & 2.56& 0.61 & 0.50 & 2.99 \\
			KeggD & \cg 0.35 & 2.57 & 4.11 & 2.69 & 3.74 & \cg1.04 & 2.47 & 1.21 & 2.00 & 5.52 & \cg 8.29 & 6.56 & 1.23 & 2.06 & 7.51 \\
			Kegg & \cg0.37 & 1.69 & 2.89 & 1.2 & 2.89 & \cg2.25 & 2.51 & 3.38 & 5.38 &5.80 & \cg18.23 & 6.69 & 0.94 & 5.81 & 6.69 \\
			NYC & \cg 12.26 & 1.16 &397.3 & 25.1 & 397.3 & \cg 75.6 & 3.83 & 158.8 & 50.10 &158.8 & \cg221.7 & 1.65 & 10.9 & 7.36 & 13.0 \\
			Skin & \cg0.44 & 1.03	 & 2.68 & 2.10 &2.68 & \cg2.63 & 4.13 & 2.73 & 2.39 & 3.8 & \cg 20.86 & 2.65 & 1.38 & 2.53 & 3.45 \\ 
			Power & \cg 5.31 & 1.26 & 0.90 & 0.77 & 1.26 & \cg 29.79 & 2.26& 1.06 & 2.38 &2.46 & \cg 220.19 & 2.15 & 0.97 & 2.24 & 2.4 \\
			Road & \cg4.28 & 1.09 & 9.01 & 2.57 &9.01& \cg17.54 & 2.38 & 3.92 & 3.11 & 4.52& \cg126.84 & 2.32& 1.58 & 1.90 &2.75 \\
			\hline

	\end{tabular}}
	\label{tab:assignment}
\end{table*}

\begin{table*}
	\ra{1.2}
	\centering
	\caption{{The refinement speedup over the running time (second) of Lloyd's algorithm (the gray column).}}
	\vspace{-1em}
	\hspace{-2em}
	\scalebox{1}{\begin{tabular}{|c|c|cccc|c|cccc|c|cccc|}
			\hline
			\multirow{3}{*}{ \textbf{Data}} & \multicolumn{5}{c}{$k=10$} & \multicolumn{5}{c}{$k=100$} & \multicolumn{5}{c|}{$k=1000$} \\ \cline{2-16}
			& \multirow{2}{*}{\lyd} & \multicolumn{4}{c|}{$\times${Speedup}} & \multirow{2}{*}{\lyd} & \multicolumn{4}{c|}{$\times${Speedup}} & \multirow{2}{*}{\lyd} & \multicolumn{4}{c|}{$\times${Speedup}} \\ \cline{3-6}\cline{8-11}\cline{13-16}
			& & \seq & \ind & \uni & \utune & & \seq & \ind & \uni & \utune & & \seq& \ind & \uni & \utune \\ \hline
			Cross & \cg 42.6 &8.03&1.34 & 5.56 & 8.74& \cg 83.0 & 7.34 & 1.68 & 12.8 &16.6 & \cg205 & 4.77 & 1.26 & 11.3 & 18.8 \\
			Conf & \cg0.72 & 6.60 & 6.60 & 3.54 & 6.60 & \cg1.58 &1.37& 1.27 & 11.5 & 10.1 & \cg2.58 & 7.59& 1.55 & 11.1 & 8.40 \\
			Covt & \cg0.17 & 7.46 & 1.39 & 7.54 & 1.39 & \cg0.44 & 1.17 & 1.17 & 12.76 & 8.90 & \cg0.25 &8.13 & 0.97 & 9.85 & 8.13 \\
			Euro & \cg2.62 &7.90 & 1.25 & 2.69 &3.19 & \cg6.79 & 11.41 & 2.08 & 3.49	 & 16.9 & \cg2.44 & 6.87& 1.65 & 2.48 & 15.3 \\
			KeggD & \cg0.10 & 9.95 & 2.28 & 10.7 & 17.1 & \cg 0.12 & 8.21 & 1.16 & 2.01 & 15.2 & \cg0.21 & 8.08 & 1.63 & 1.97 & 15.6 \\
			Kegg & \cg0.12 & 10.2 & 2.57 & 6.76 & 2.57 & \cg0.25 & 15.8 & 1.78 & 34.6 & 22.0 & \cg0.41 & 5.15& 0.83 & 15.4 & 5.15 \\
			NYC & \cg 3.06 & 6.87 & 361 & 348 & 361 & \cg 8.76 & 14.2 & 122 & 162.7 & 122 & \cg 8.05 & 4.61 & 15.3 & 21.1 & 24.2 \\
			Skin & \cg 0.12 & 6.79 & 2.19 & 8.90 & 2.19 & \cg 0.29 & 8.81 & 1.88 & 9.70 & 15.2 & \cg0.55 &8.10& 1.34 & 8.71 & 13.4 \\ 
			Power & \cg1.07 & 5.60 & 0.44 & 3.52 & 5.60 & \cg3.12 & 6.12 & 0.74 & 7.43 & 6.31 & \cg3.71 & 3.92 & 0.55 & 6.00 & 6.4 \\
			Road & \cg1.74 & 5.88 & 7.65 & 13.0 & 7.65 & \cg3.63 & 6.58 & 2.50 & 8.76 & 16.2 & \cg5.96 & 7.48 &1.61 & 15.2 & 15.3 \\
			\hline
	\end{tabular}}
	\label{tab:refinement}
\end{table*}

\subsection{Common Notations}
Table~\ref{tab:notations} lists the common notations used in this paper.
\begin{table}[h]
	\centering
	\ra{1}
		\caption{A summary of common notations.}
				\vspace{-1em}
		\begin{tabular}{cc}
			\toprule
			\textbf{Symbol} & \textbf{Description} \\\midrule
			$D$, $d$ & The dataset, and its dimensionality\\
			$x_i$ & A data point in $D$\\
			$c_j$ & The centroid of cluster $S_j$\\
			$lb(i,j)$ & The lower bound of $x_i$ to cluster $c_j$\\
			$lb(i)$ & \makecell{The Lower bound of $x_i$ to its \\second nearest cluster}\\
			$ub(i)$ & The Upper bound of $x_i$ to its nearest cluster\\
			$a(i)$, $a^{'}(i)$ & The indices of $x_i$'s new and previous cluster\\
			$p$ & The pivot point of node $N$\\
			$\|c_j\|$ & The L2-norm of centroid $c_j$\\
			$\delta(j)$ & The centroid drift of $c_j$\\
			\bottomrule
		\end{tabular}
		\label{tab:notations}
\end{table}

\subsection{Space Analysis of Index-based Methods}
\label{sec:space}
Recall Definition~\ref{def:node}, building an extra data structure will increase the space overhead, the size of a node $N$ is comparably small -- 
each node is composed of two vectors ($p$, $sv$), four floats ($r$, $\psi$, $num$, $h$), and two pointers to child nodes (left and right) or a set of points in leaf nodes (the number is limited as the capacity $f$).
{Thus, we estimate the space of leaf nodes as $2d+4+f$, and internal nodes as $2d+6$.
Then the overall space cost (number of floats) of all the nodes will be $\frac{n}{f}\cdot(2d+4+f)+q\cdot(2d+6)=n+\frac{n}{f}\cdot(2d+4)+q\cdot(2d+6)$, {where $\frac{n}{f}$ and $q =\frac{n}{f}\cdot(1-2^{1-log_2\frac{n}{f}})$ are the numbers of leaf nodes and internal nodes}. 

{Such an estimation is based on an ideal case where each leaf node has $f$ points and the index is a balanced Ball-tree with a height $log_2\frac{n}{f}$, and we skip the summation for geometric sequence here.}}
In Figure~\ref{fig:seq1} and \ref{fig:seq13}, we can find that the footprint of Ball-tree is comparable with those space-saving sequential methods in the most datasets. 
For the datasets that have high footprint, most of them are high-dimension, which means that they will have more leaf nodes that cover less points than those datasets with lower dimensions.
Note that one advantage of Ball-tree is that its footprint is fixed once it is built, and will not change with the increasing of $k$, while most sequential methods will need much more space.

\begin{algorithm}[h]
	\small
	\caption{\small Selective Running of Knob Configurations}
	\label{alg:ground}
	\KwOut{two ground truth files: $g_1$ and $g_2$}
	\For{every test dataset D}{
		$\bm{F} \leftarrow MetaFeatureExtraction(D, k, T)$\;
		Run partial sequential methods\;
		write($\bm{F}$, identifier of the optimal knob $m_o$, $g_1$)\;
		Test the index-based method $m_i$ with Ball-tree\;
		\eIf{$m_i$ is slower than $m_o$}
		{write($\bm{F}$, not using index \underline{1}, $g_2$)\;}
		{
			Test index-single and index-multiple with $m_o$\;
			\eIf{index-based method is the fastest}
			{write($\bm{F}$, point traveseal \underline{2}, $g_2$)\;}
			{	
				\eIf{index-single is the fastest}
				{write($\bm{F}$, single traveseal \underline{3}, $g_2$)\;}
				{write($\bm{F}$, multiple traveseal \underline{4}, $g_2$)\;}
			}
		}
	}
	\textbf{return} $g_1$ and $g_2$\;
\end{algorithm}

\subsection{More Experiment Results}
\label{sec:moreexp}

\begin{figure*}
	\centering
	\begin{minipage}{1\textwidth}
		\centering
		\hspace{-4.3em}\includegraphics[width=1.06\textwidth]{graph/legend-eps-converted-to.pdf}\\
		\hspace{-2.1em}\includegraphics[width=0.52\textwidth,height=0.11\textheight]{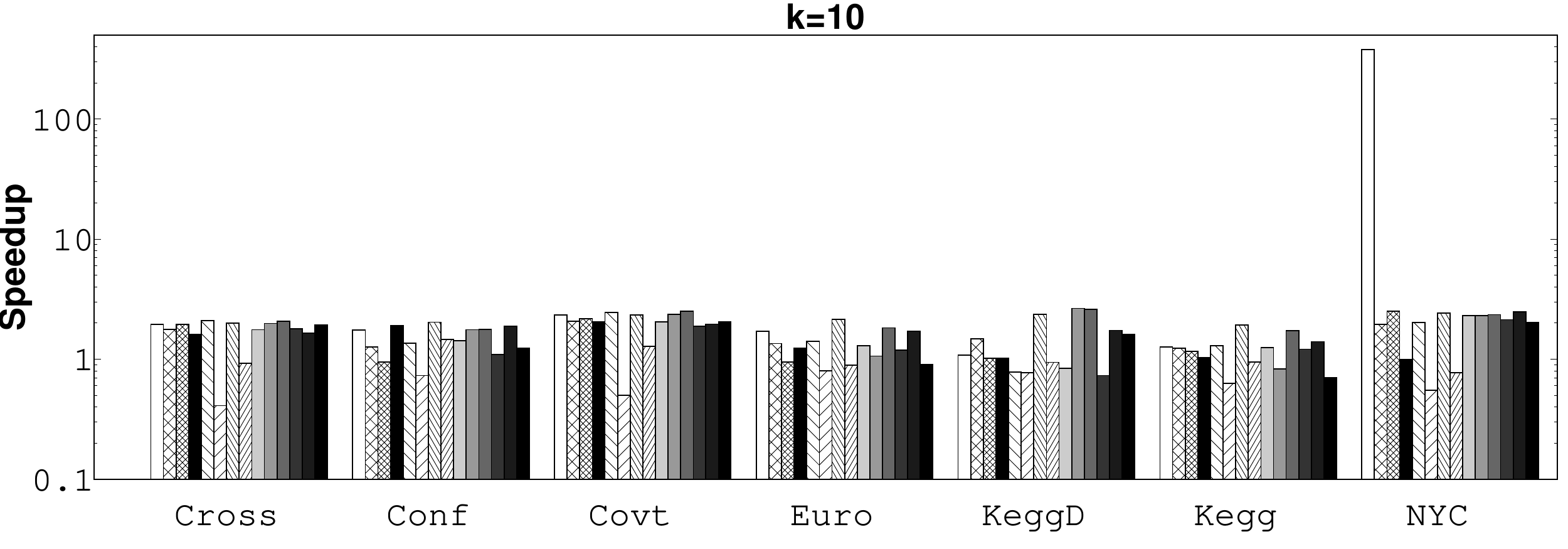}\hspace{-0.4em}
		\includegraphics[width=0.52\textwidth,height=0.11\textheight]{graph/NewPlotS/figure/all-dataset-speedup-eps-converted-to.pdf}
		\vspace{-2.3em}
		\captionof{figure}{Overall speedup in various datasets when using random (left) and $k$-means++ initialization (right), respectively.}
		\label{fig:random}
	\end{minipage}\\
\end{figure*}

\shengnew{\myparagraph{Effect of Initialization Methods}We conduct experiments to observe the effect of initialization methods on the performance. Since the convergence will be different due to various initialization, a fair way is to compare the time spent in the first fixed number of iterations. Here, we record the running time spent for the first ten iterations, and the result is shown in Figure~\ref{fig:random}. We can observe that the performance of using random and $k$-means++ initialization do not affect the efficiency much.}

\begin{figure}
	\includegraphics[width=1\linewidth]{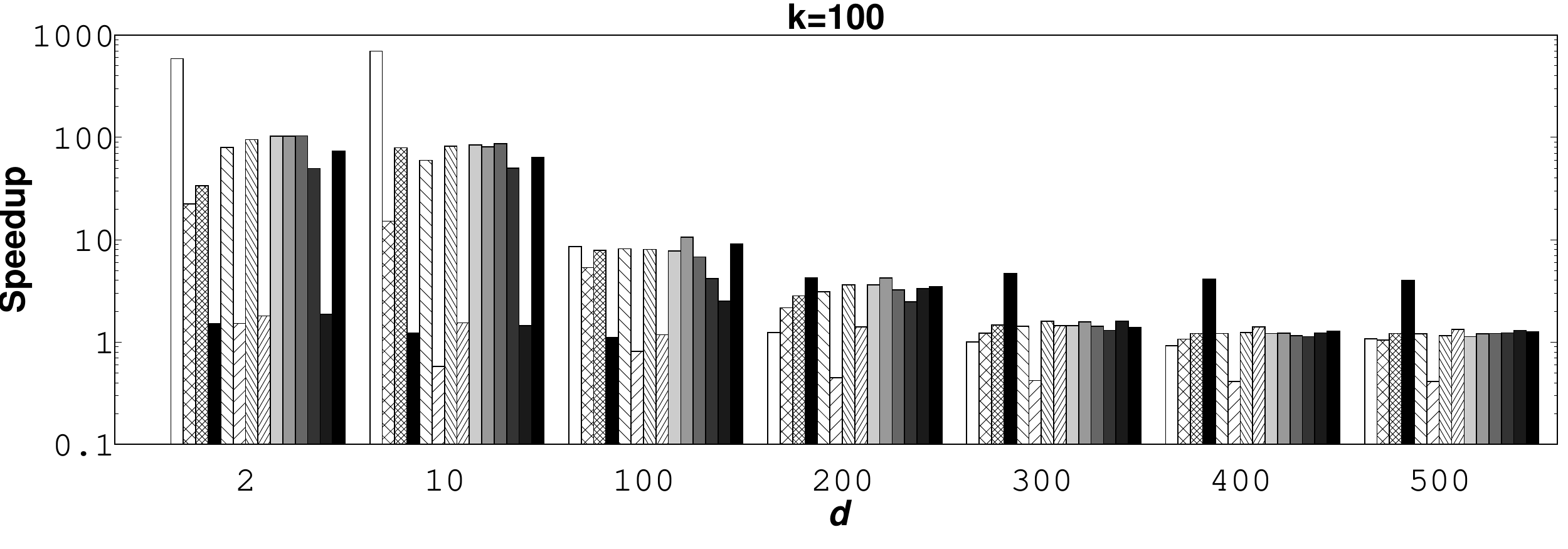}
	\vspace{-2.9em}
	\caption{The effect of increasing dimensionality on Mnist.}
	\vspace{-2em}
	\label{fig:dimension}
\end{figure}

\shengnew{\myparagraph{Effect of Increasing Dimensionality}We also conduct experiments on a specific high-dimensional dataset to observe the effect of increasing dimensionality on sequential methods' performance. Here we choose the Mnist dataset since it has 784 dimensions, and we fix k as 100. The result is shown in Figure~\ref{fig:dimension}. We find that most algorithms can have a good performance in low dimension, and it is hard to maintain when tested with high-dimensional dataset. In contrast, {\small\texttt{Drak}}'s better performance in high-dimensional space also brings itself to the leader board in Figure~\ref{fig:lead}. 
}

\begin{figure*}
	\centering
	\begin{minipage}{0.9\textwidth}
		\centering
		\vspace{-2em}
		\hspace{-3.3em}\includegraphics[width=1.06\textwidth]{graph/legend-eps-converted-to.pdf}\\
		\includegraphics[width=0.9\textwidth,height=0.13\textheight]{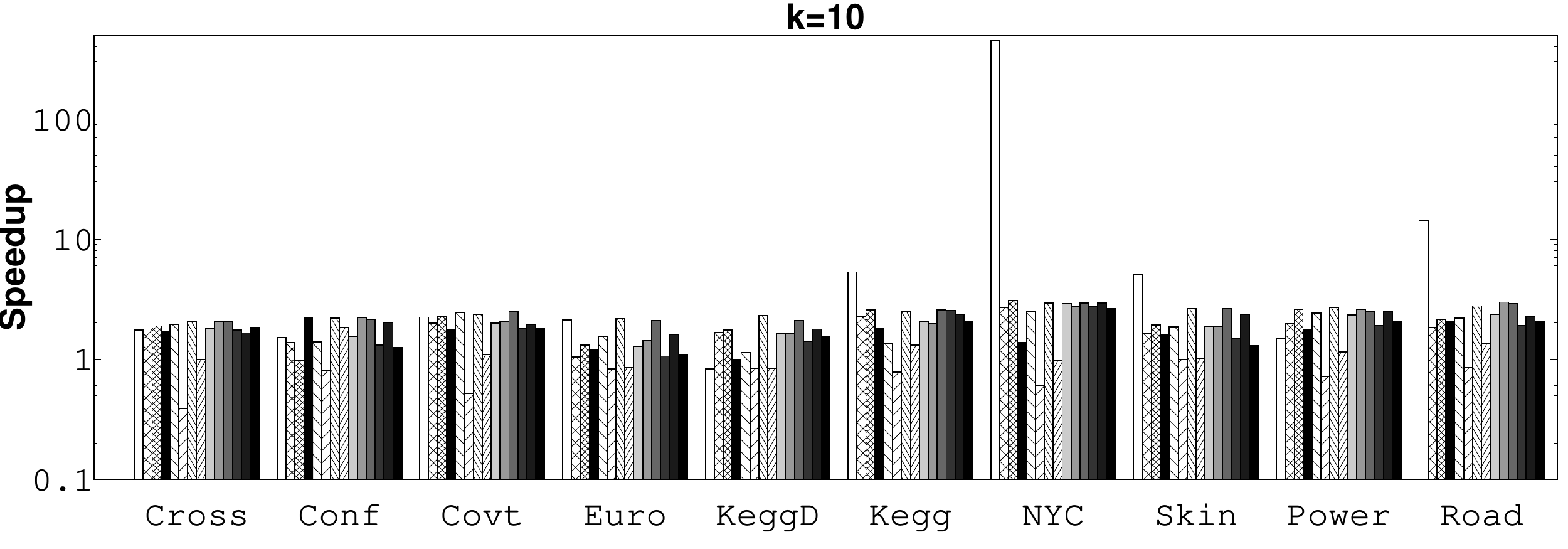}\\\vspace{-0.3em}
		\includegraphics[width=0.9\textwidth,height=0.13\textheight]{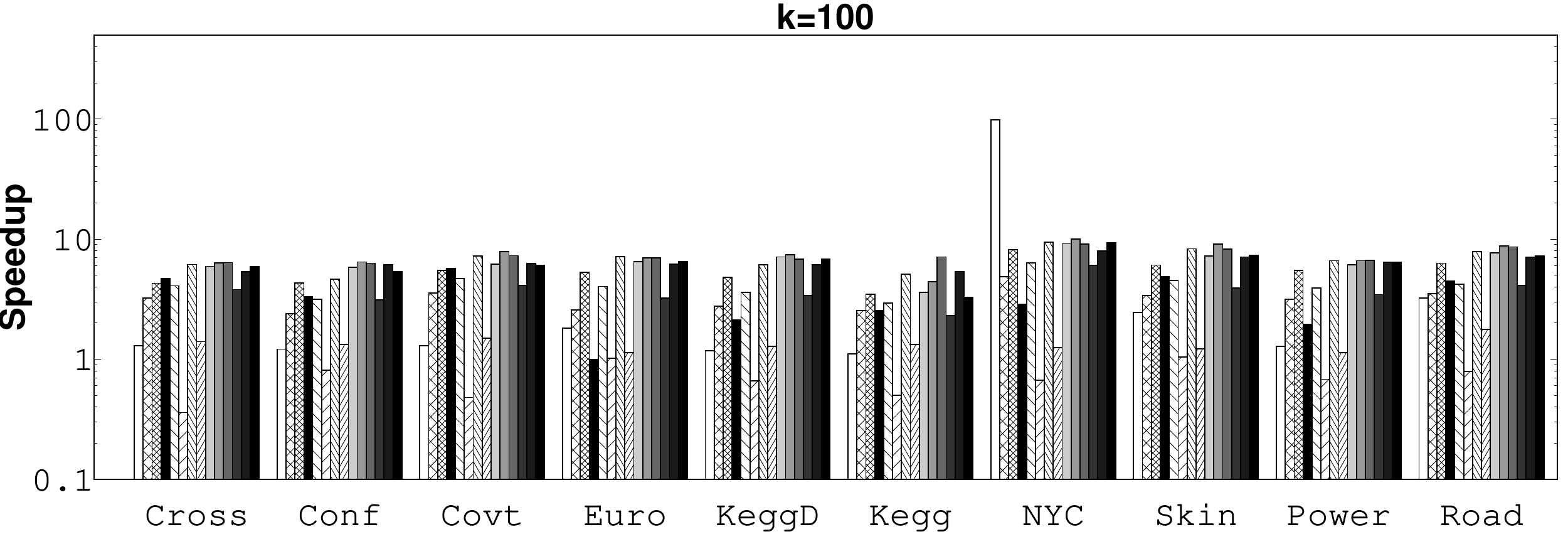}
		\vspace{-1.3em}
		\captionof{figure}{Overall speedup in various datasets when setting $k$ as 10 and 100, respectively.}
		\vspace{-0.5em}
		\label{fig:seqf}
	\end{minipage}\\
	\begin{minipage}{0.9\textwidth}
		\centering
		\includegraphics[width=0.9\textwidth,height=0.13\textheight]{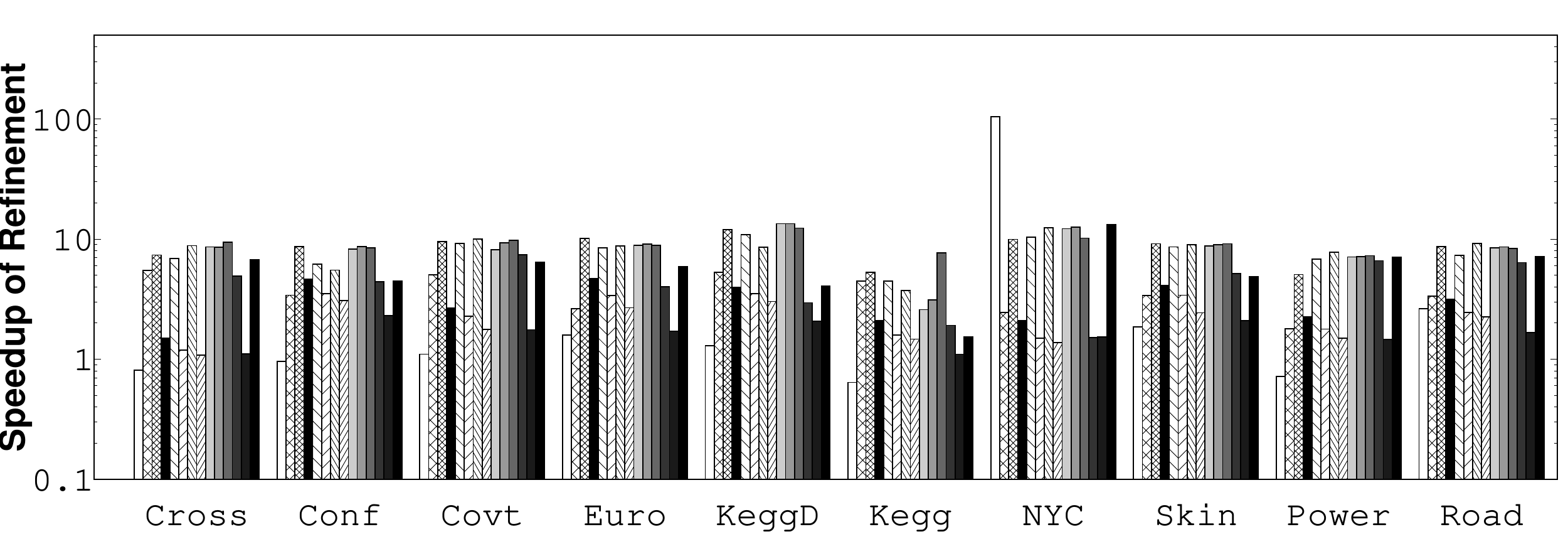}\\\vspace{-0.5em}
		\includegraphics[width=0.9\textwidth,height=0.13\textheight]{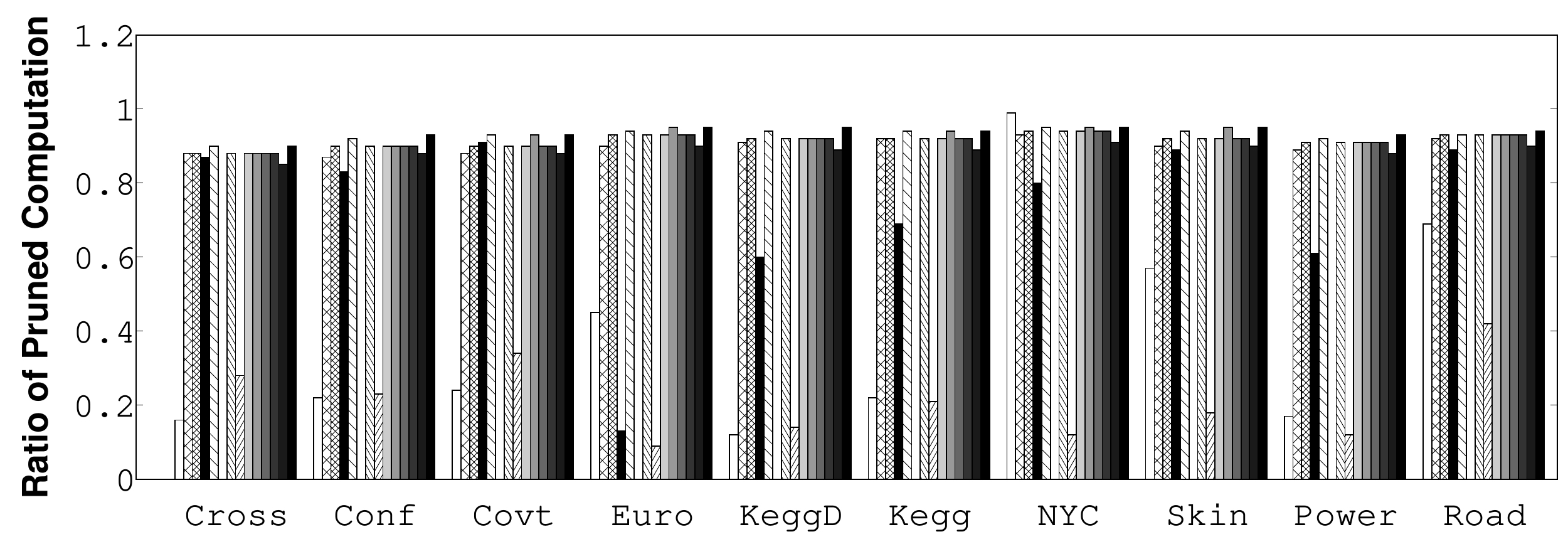}
		\vspace{-1.3em}
		\captionof{figure}{Refinement's speedup and distance computation pruning ratio ($k=100$).}
		\vspace{-0.5em}
		\label{fig:seq-refinef}
	\end{minipage}
	\begin{minipage}{0.9\textwidth}
		\centering
		\includegraphics[width=0.9\textwidth,height=0.13\textheight]{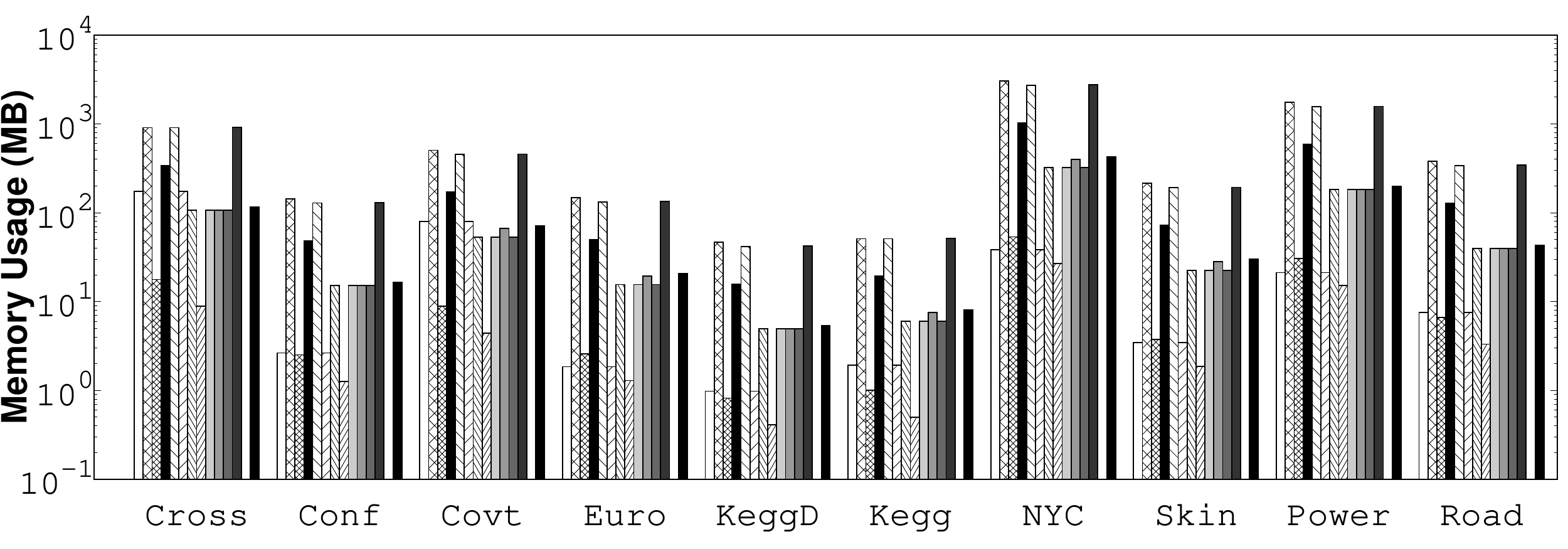}\\\vspace{-0.5em}
		\includegraphics[width=0.9\textwidth,height=0.13\textheight]{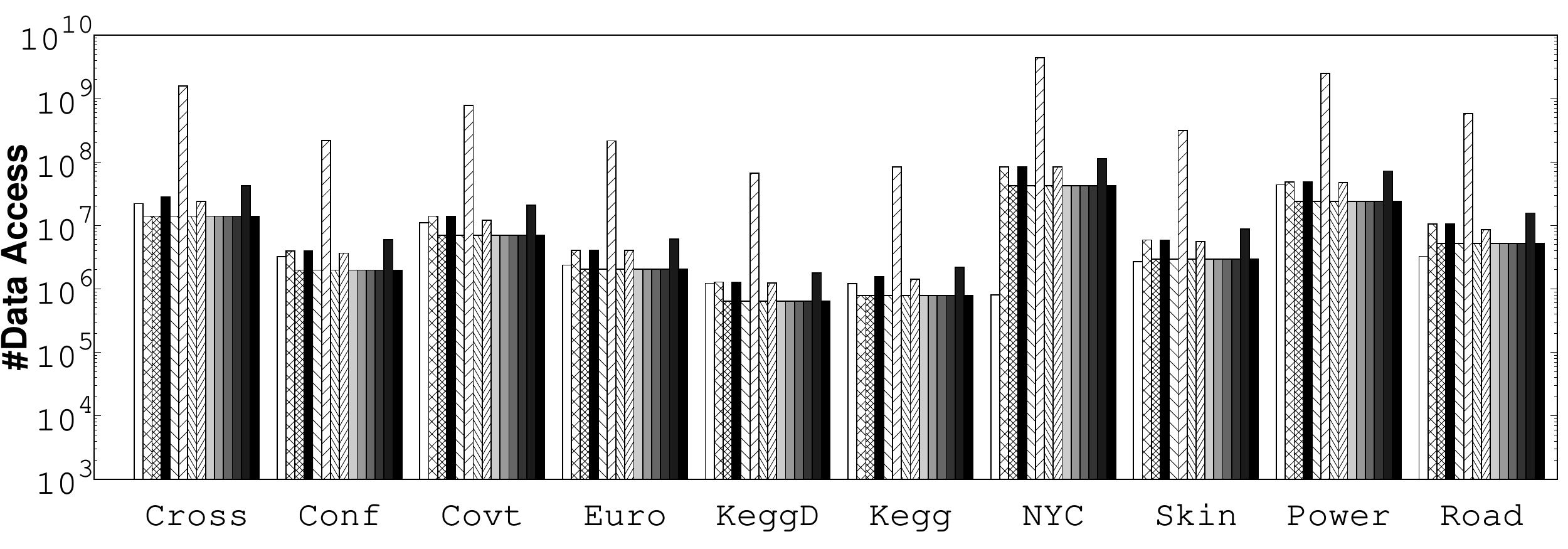}
		\vspace{-1.5em}
		\captionof{figure}{Statistics on the footprint of bound (index) and data accesses ($k=100$).}
		\vspace{-0.5em}
		\label{fig:seq1f}
	\end{minipage}
	\begin{minipage}{0.9\textwidth}
		\centering
		\includegraphics[width=0.9\textwidth,height=0.13\textheight]{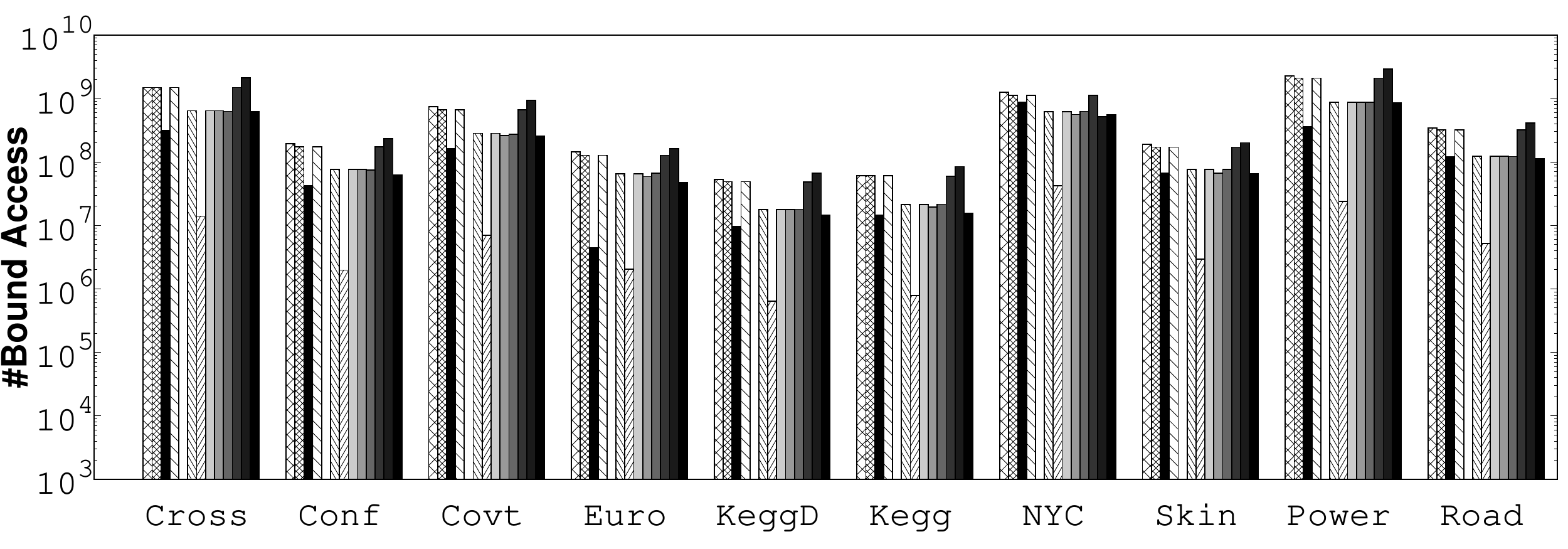}\\\vspace{-0.5em}
		\includegraphics[width=0.9\textwidth,height=0.13\textheight]{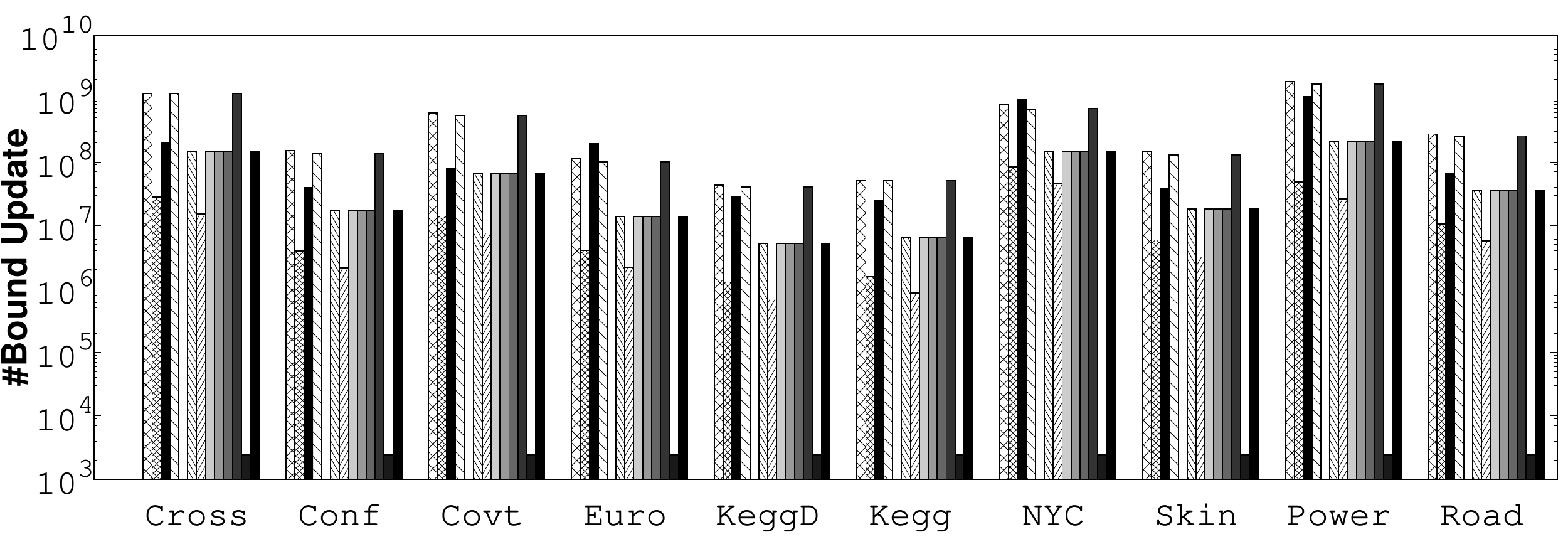}
		\vspace{-1.5em}
		\captionof{figure}{Statistics on the bound accesses and updates ($k=100$).}
		\vspace{-2em}
		\label{fig:seq2f}
	\end{minipage}
\end{figure*}

Figures~\ref{fig:seqf} to \ref{fig:seq2f} are the full version of Figures~\ref{fig:seq} to \ref{fig:seq2}.
Figures~\ref{fig:s11} to \ref{fig:seq14} show a detailed comparison between sequential methods when setting $k=10$, which mainly enhances Section~\ref{sec:seqexp}.
We can observe similar trends with $k=100$, and the main difference is that smaller $k$ has fewer data accesses and lower pruning ratio.
Table~\ref{tab:assignment} and \ref{tab:refinement} present the time of assignment and refinement which partially decompose the results the overall speedup in Table~\ref{tab:overview}.

\begin{figure}
	\includegraphics[width=1\linewidth]{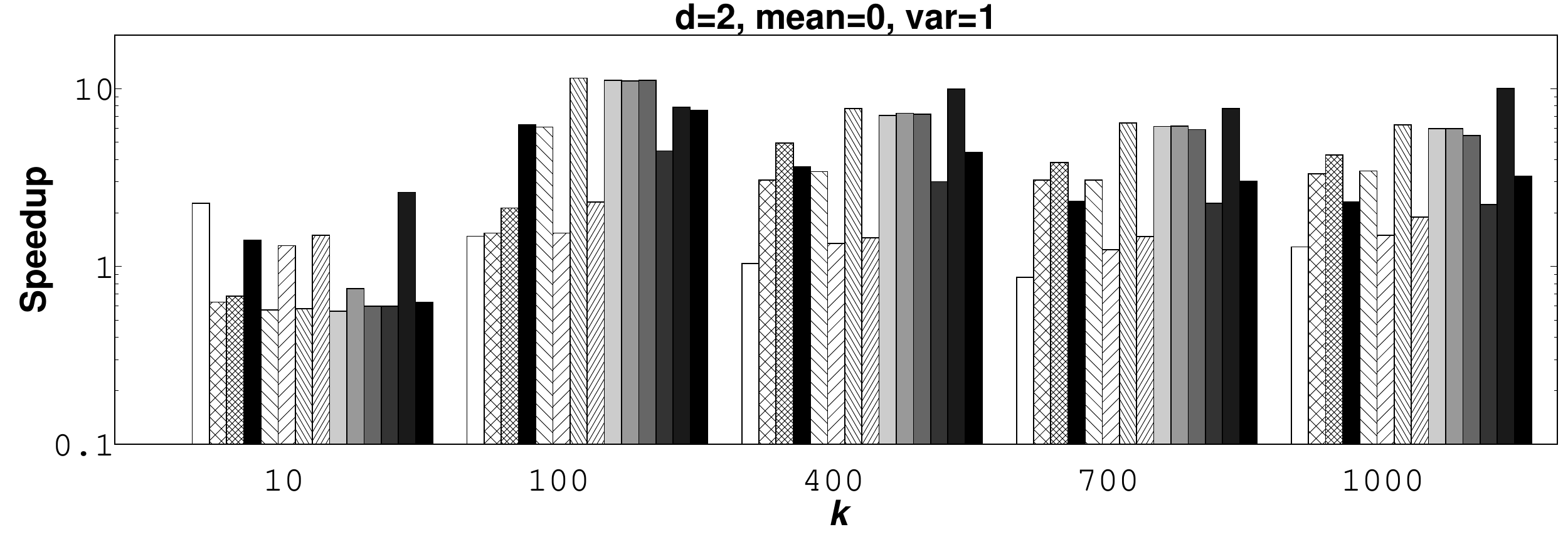}
	\includegraphics[width=1\linewidth]{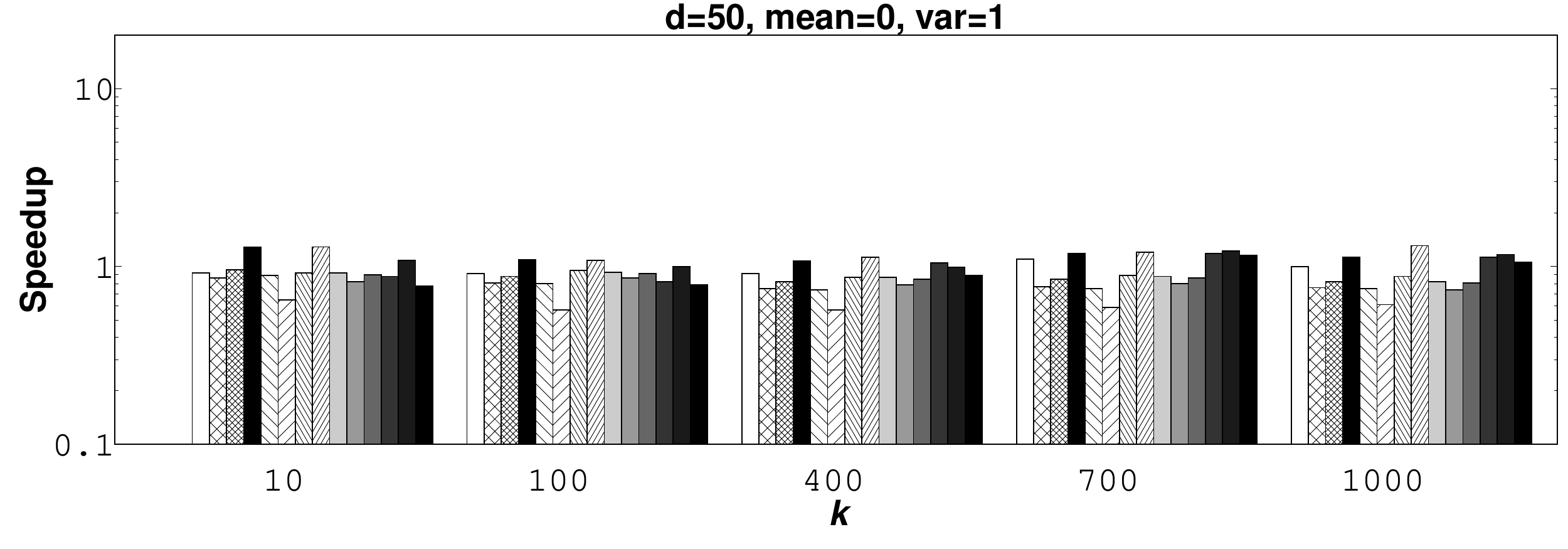}
	\includegraphics[width=1\linewidth]{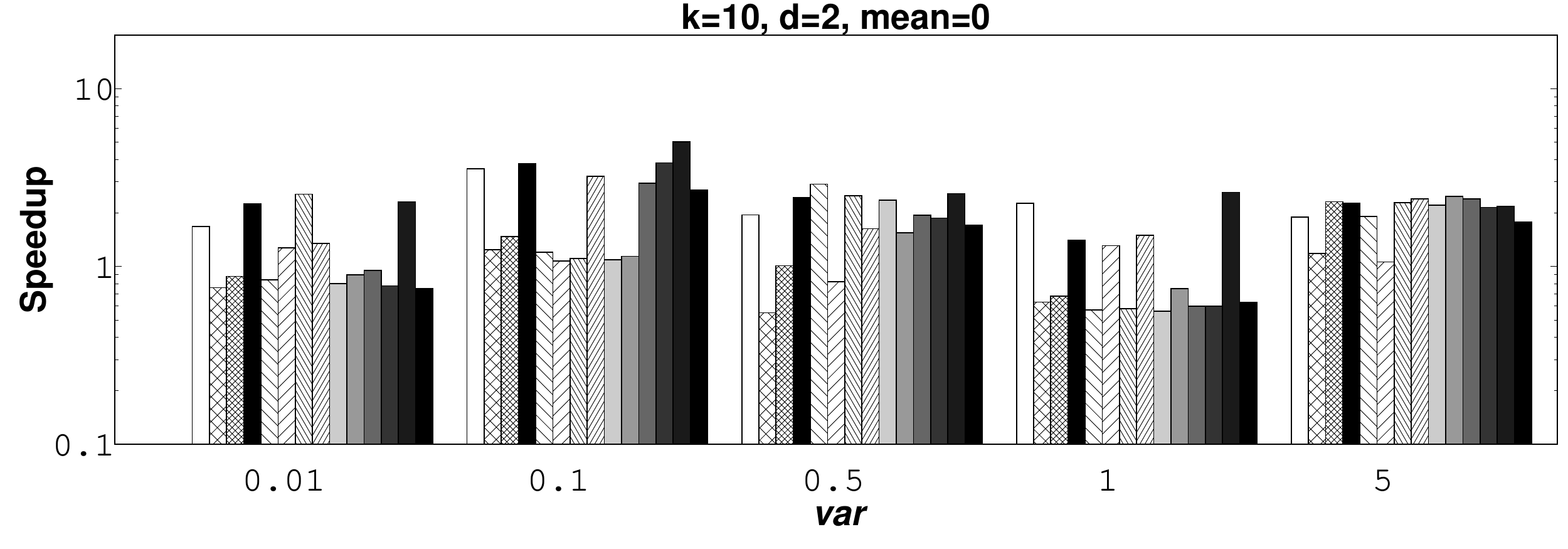}
	\includegraphics[width=1\linewidth]{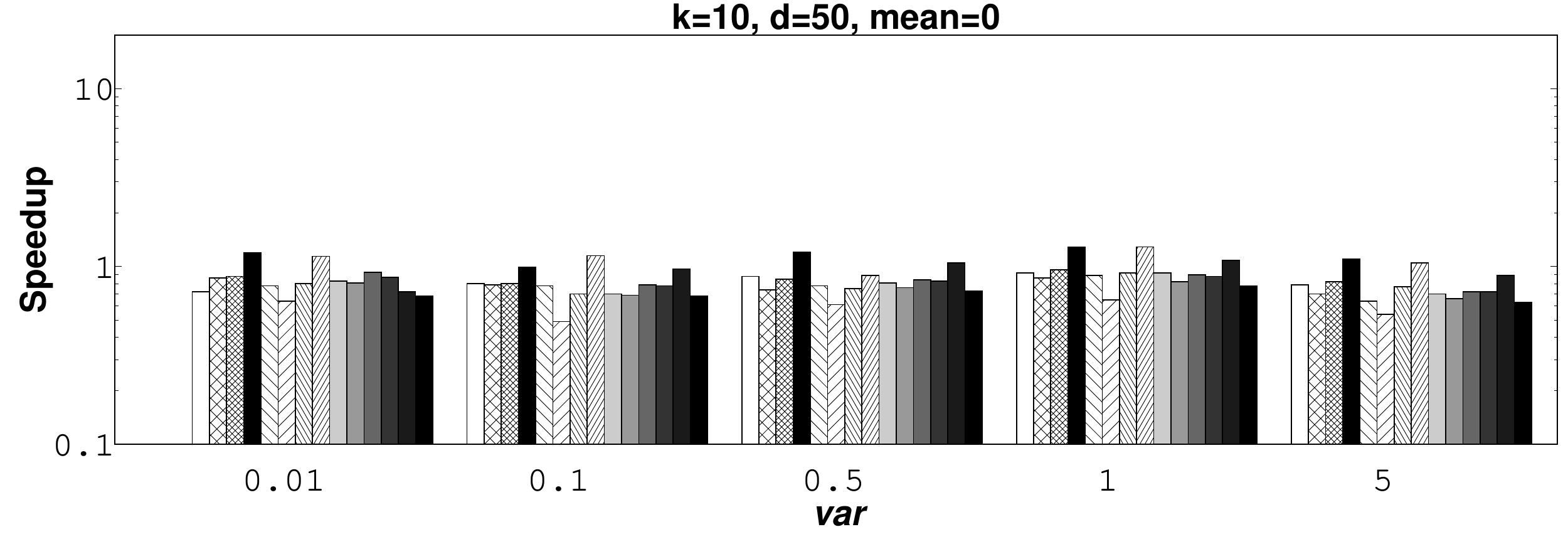}
	\vspace{-2em}
	\caption{The effect of cluster distribution on synthetic datasets.}
	\vspace{-1em}
	\label{fig:distribution}
\end{figure}

{\myparagraph{Effect of Data Distribution} To investigate the effect of data distribution, we generate synthetic datasets with the common normal (or Gaussian) distribution with various parameters. Apart from the \textit{variance} parameter $\var{var}$, we also alter the number of clusters $k$ distributed in the datasets. We employ the scikit-learn library\footnote{\url{https://scikit-learn.org/stable/modules/generated/sklearn.datasets.make_gaussian_quantiles.html}} to generate these datasets. 
We set the fixed scale $|D|$ as 10,000 and set dimension $d$ as 2 and 50 respectively. Then we change the value for parameters $var=\{0.01, 0.1, \underline{0.5}, 1, 5\}$ and $k=\{\underline{10}, 100, 400, 700, 1000\}$, where the default number is underlined when testing other parameters.

From Figure~\ref{fig:distribution}, we find that the performance of most methods increases slightly with the rise of number of clusters in low-dimensional datasets, especially the index-based methods. However, this trend is not that clear when increasing the variance. In contrast, the effects of $k$ and $var$ are slight in high-dimensional datasets, where the pruning ratio is low for both index-based and sequential methods. Overall, these evaluations verify that there are more factors of the datasets that affect the performance in a complex way.}

\begin{figure}
	\centering
	\includegraphics[width=0.42\textwidth]{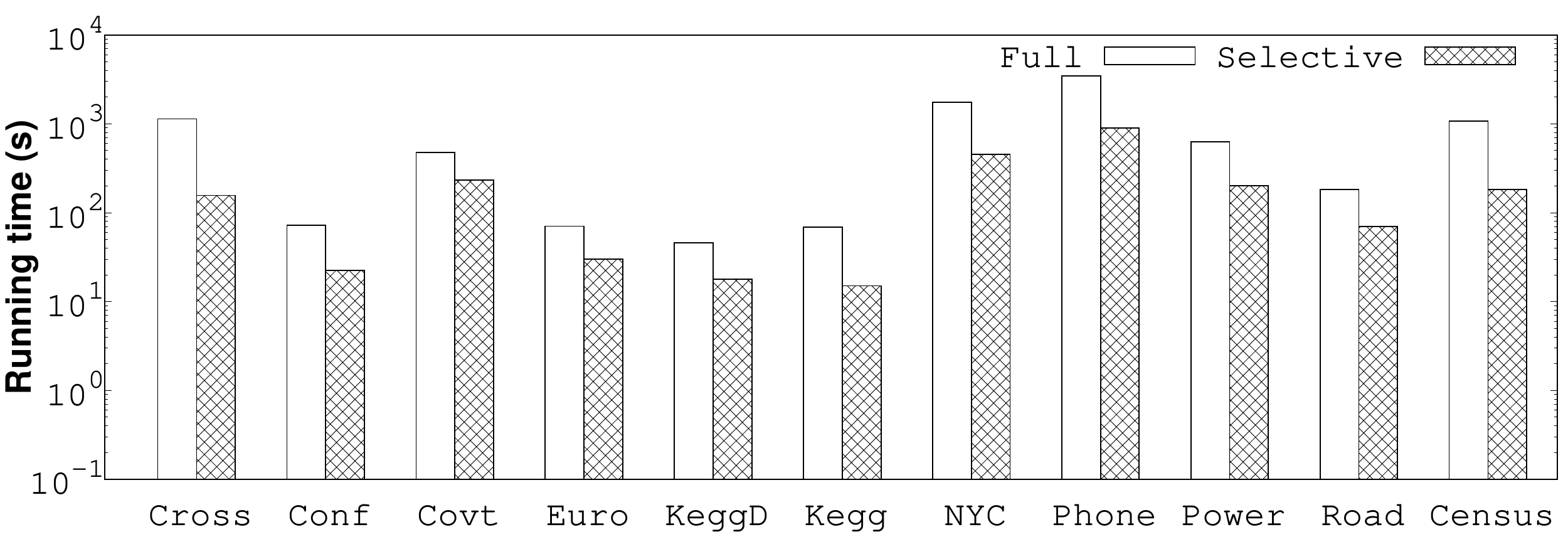}
	\vspace{-1.5em}
	\caption{{Efficiency of ground truth generations.}}
	\vspace{-1em}
	\label{fig:train-time}
\end{figure}

\subsection{Reproducibility}
\label{sec:repro}
Our code is available online.\footnote{\url{https://github.com/tgbnhy/fast-kmeans}} We show how to run our experiments with simple commands and operations in the description of the repository.
Readers can download the datasets in Table~\ref{tab:dataset} online by clicking the links shown there, and put into local folders, then compile the code using Maven (\url{http://maven.apache.org/}).
Clustering tasks can be conducted locally by executing the given exemplar commands by terminals. All the algorithms will be run, and the results will be shown in the terminal to monitor the running progress.
After all the algorithms end, detailed results on each comparison metric, such as speedup and the number of data accesses, will be written into the log files for further plotting.
We also show readers how to interpret the evaluation results based on the terminal and logs files.

\subsection{Future Opportunities}
\label{sec:future}
\myparagraph{More Meta-Features}
In our evaluations, we used multiple measures without extra costs.
The main reason that we do not use other features that can be extracted using data profiling \cite{Abedjan2015} or meta-feature extraction \cite{Rivolli2018} is that, they need much time to extract, such as the model-based method by training decision trees. 
Since a higher precision will recommend a better algorithm and may greatly reduce the clustering time, it is promising to apply these features to improve the precision with fast meta-feature extraction.

\myparagraph{New ML Models to be Adopted}
We have tested multiple classical models to predict the optimal algorithm, while the loss function generally computes exact matches and does not consider the importance of ranked lists of algorithms.
Hence, designing a specific machine learning model with a loss function like MRR, which we used to evaluate the accuracy, will be crucial to further improve the prediction accuracy.
Furthermore, deep neural networks (DNN) have also shown significant successes in various classification tasks, and it will be interesting to design a specific network structure to input a sampled dataset as features directly to predict the optimal algorithm. {It is worth noting that this is out of the scope of this work and our evaluation framework is orthogonal to the choice of ML model adopted in the learning part.}

\myparagraph{Discovering New Configuration Knobs}
In this evaluation, we have compared all existing fast \kmeans clustering algorithms and \uni with our new settings.
It is worth mentioning that the predicted algorithm is only the fastest among the group that we tested, and \kmeans can be further accelerated in the future.
As shown in Algorithm~\ref{alg:kmeans}, we have listed multiple knobs \knob, and the knob configuration space $\Theta$ is large, which means our tested algorithms are only a tiny proportion.
The discovery of new configurations based on Algorithm~\ref{alg:kmeans} will enable us to combine various optimizations that have never been tried and tested. Such new configurations will form new algorithms that can be potentially fast for a certain group of clustering tasks.

\myparagraph{Applying ML to Other Database Algorithm Selection Problems}
In the database field, efficiency evaluations of existing algorithms have been an essential guide for users due to the diversity of tasks and datasets.
Even though several clear insights are given for choosing the right decision in the end of evaluations, they essentially form an elementary and inaccurate decision tree for referring.
Auto-tuning based on the evaluation logs and ML models can interpret the superiority of algorithms in a more accurate way than humans, and can directly predict the optimal algorithm. 
Hence, it will be interesting to apply the auto-tuning methodology presented in this paper to those evaluation studies that have made a good comparison but still manually selects algorithms.

\begin{table*}[t]
		\begin{minipage}{0.9\textwidth}
			\end{minipage}
			\begin{minipage}{0.9\textwidth}
			\centering
			\vspace{-2em}
			\includegraphics[width=1.06\textwidth]{graph/legend-eps-converted-to.pdf}\\\vspace{-0.3em}
			\includegraphics[width=0.9\textwidth,height=0.13\textheight]{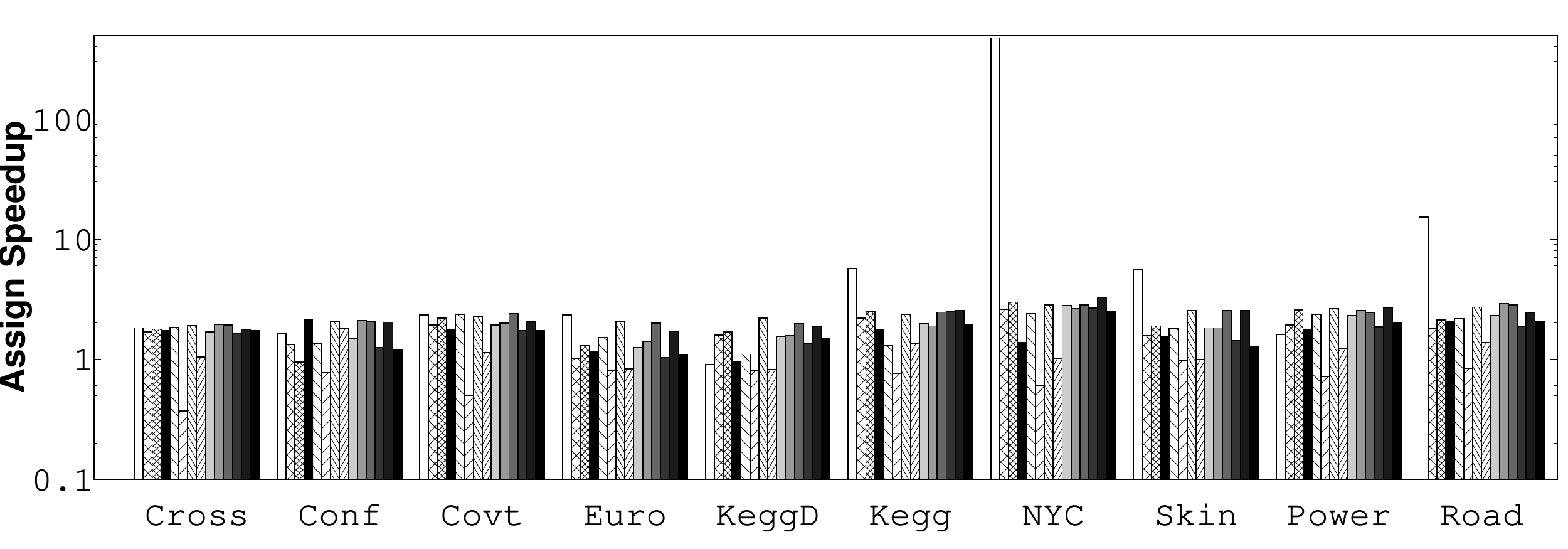}\\\vspace{-0.4em}
			\includegraphics[width=0.9\textwidth,height=0.13\textheight]{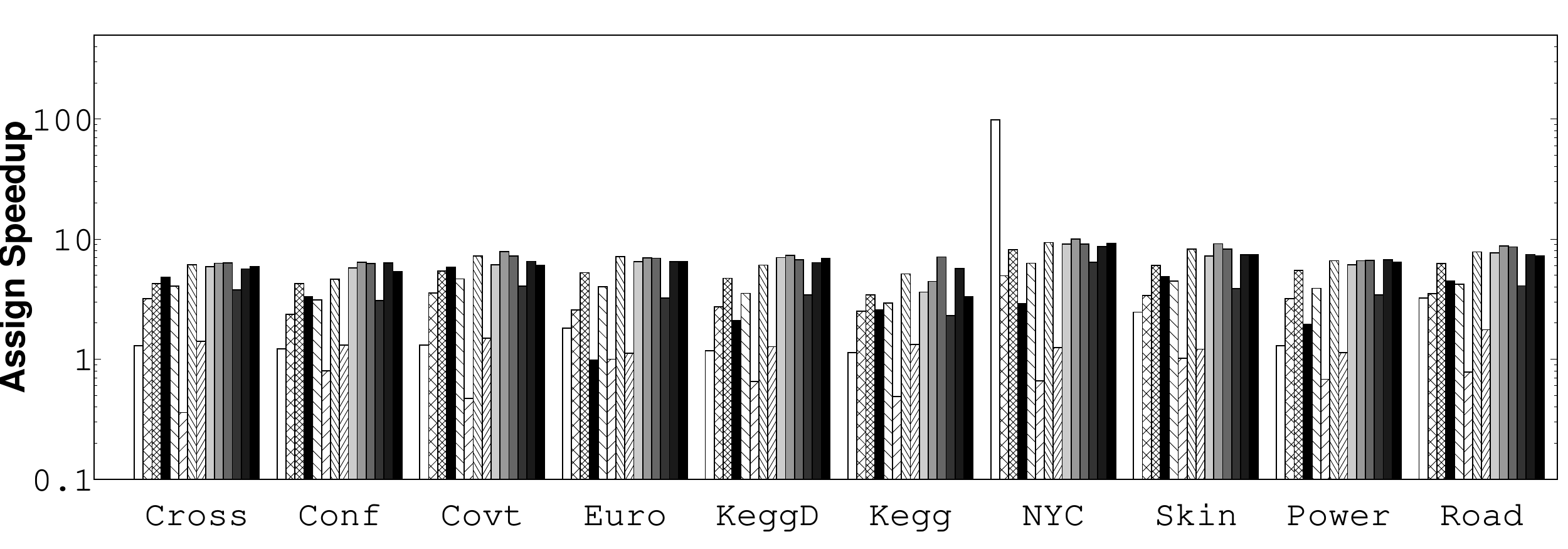}
			\vspace{-1.5em}
			\captionof{figure}{Overall assignment speedup in various datasets when setting $k$ as 10 and 100, respectively.}
			\vspace{-0.5em}
			\label{fig:s11}
		\end{minipage}\\
		\begin{minipage}{0.9\textwidth}
			\centering
			\includegraphics[width=0.9\textwidth,height=0.13\textheight]{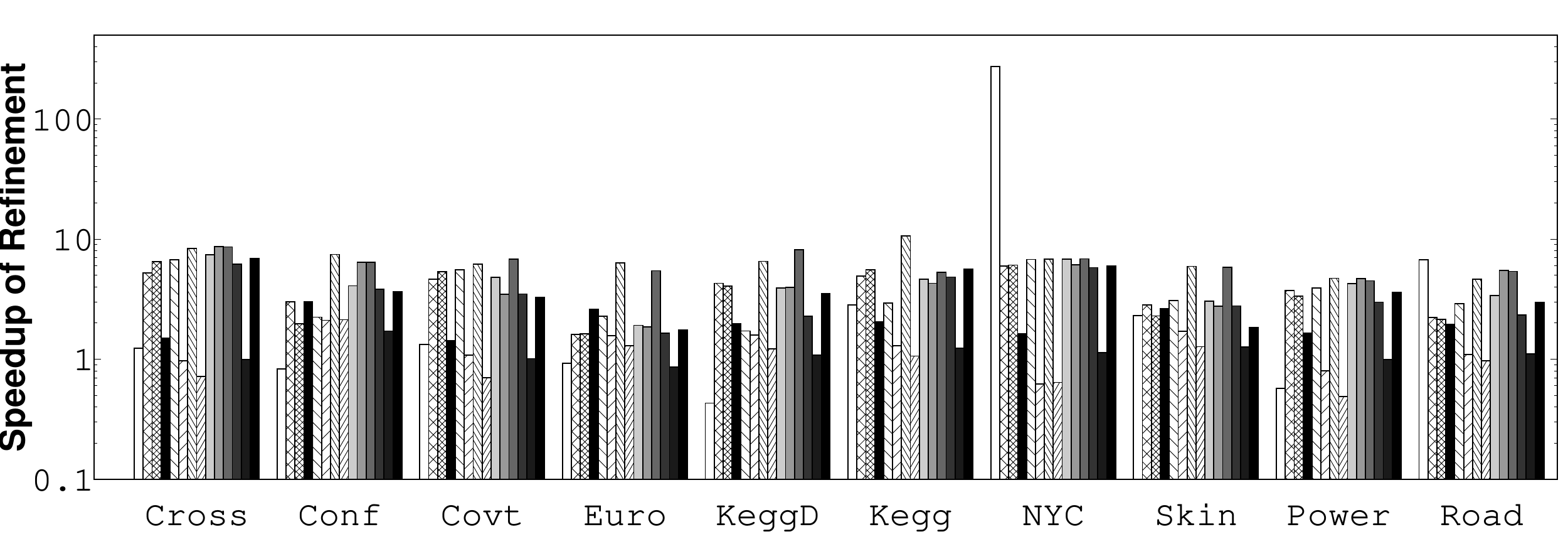}\\\vspace{-0.4em}
			\includegraphics[width=0.9\textwidth,height=0.13\textheight]{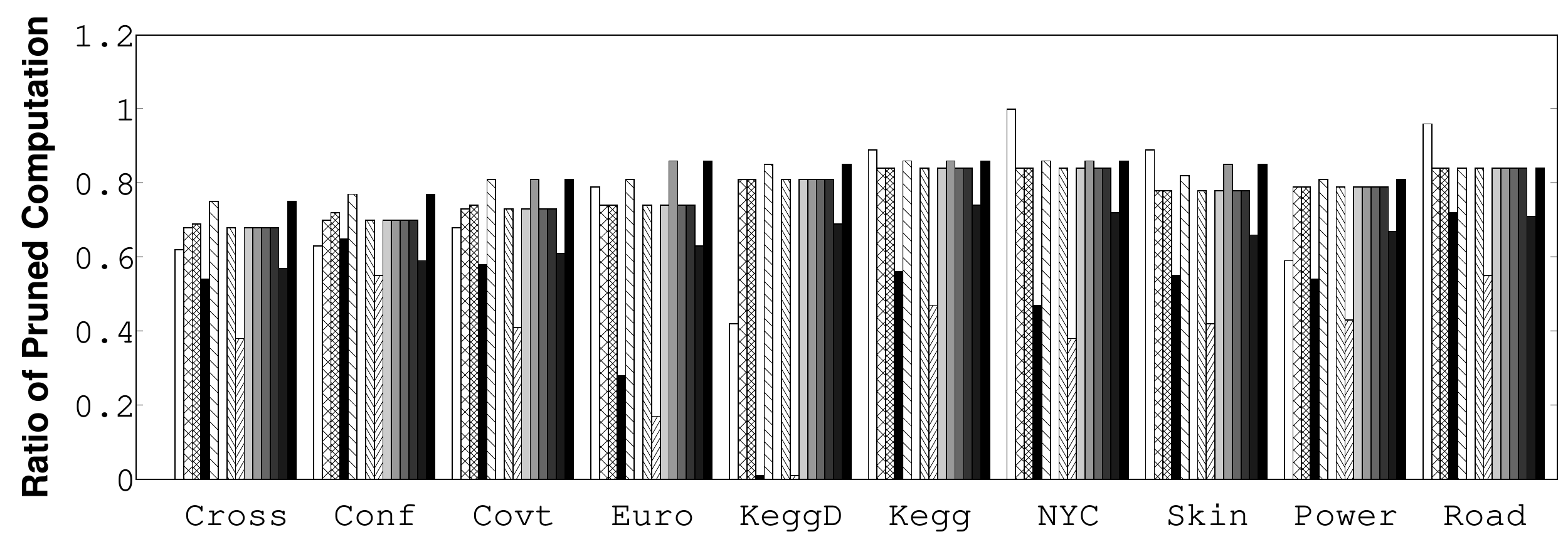}
			\vspace{-1.5em}
			\captionof{figure}{Refinement's speedup and distance calculation pruning ratio ($k=10$).}
			\vspace{-0.5em}
			\label{fig:seq12}
		\end{minipage}
		\begin{minipage}{0.9\textwidth}
			\centering
			\includegraphics[width=0.9\textwidth,height=0.13\textheight]{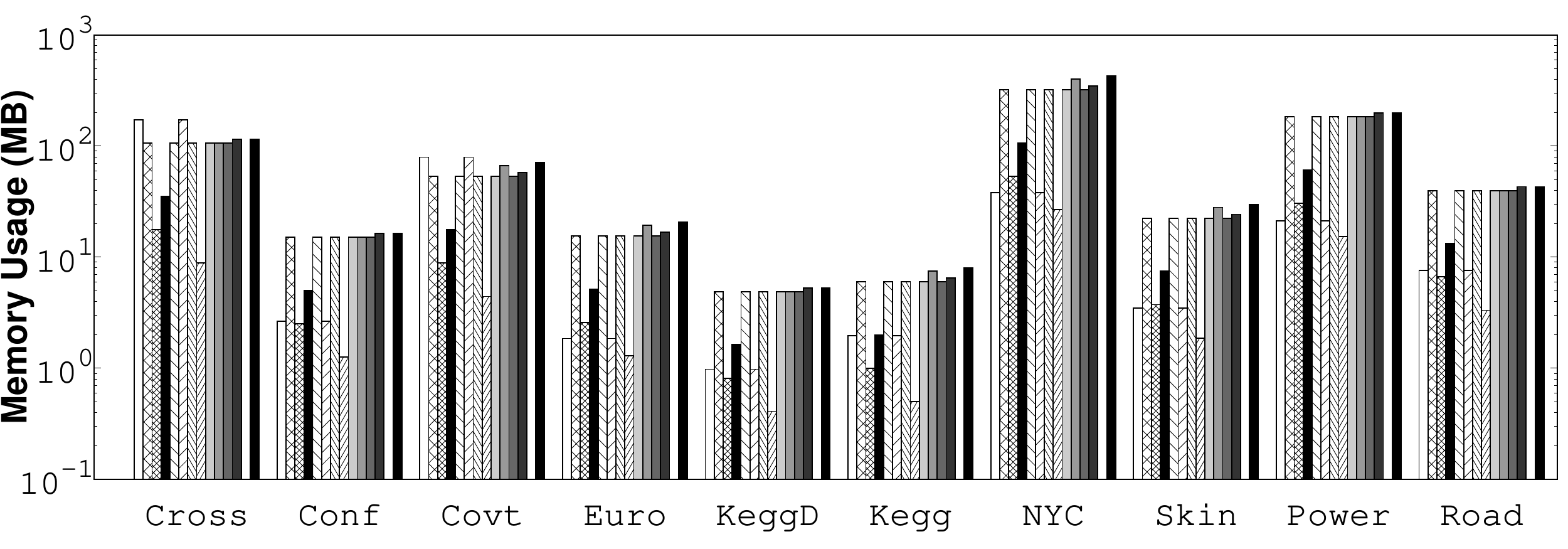}\\\vspace{-0.4em}
			\includegraphics[width=0.9\textwidth,height=0.13\textheight]{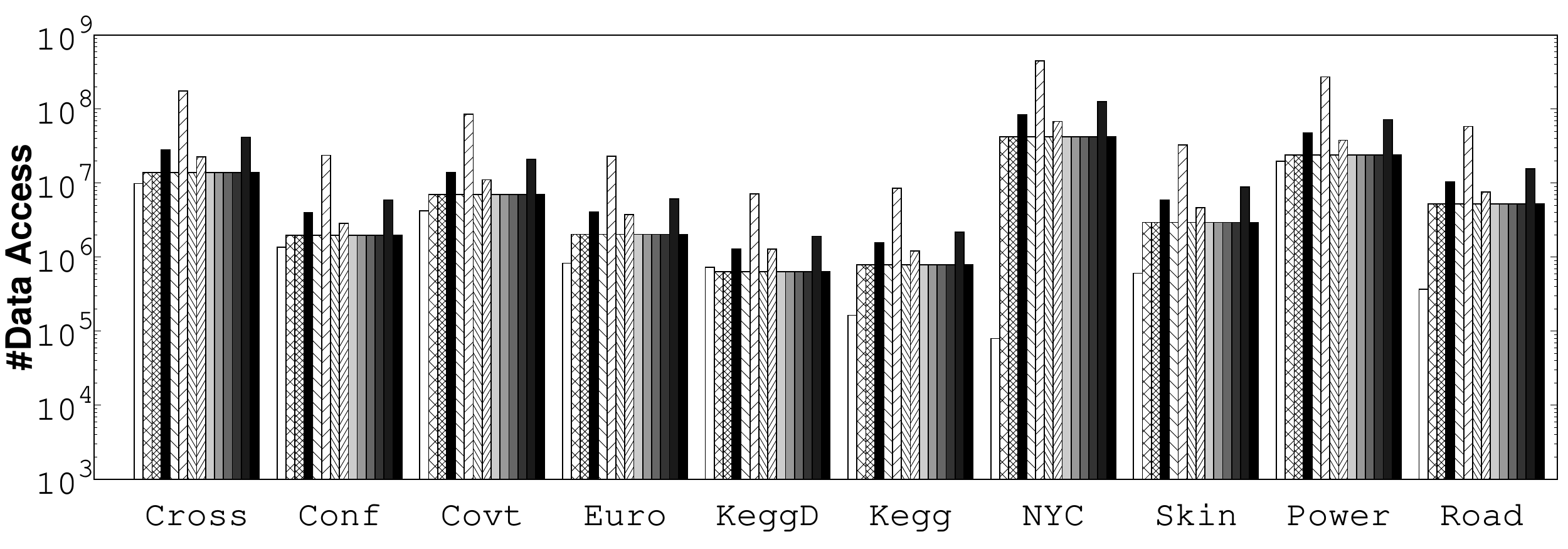}
			\vspace{-1.5em}
			\captionof{figure}{Statistics on the footprint of bound (index) and data accesses ($k=10$).}
			\vspace{-0.5em}
			\label{fig:seq13}
		\end{minipage}
		\begin{minipage}{0.9\textwidth}
			\centering
			\includegraphics[width=0.9\textwidth,height=0.13\textheight]{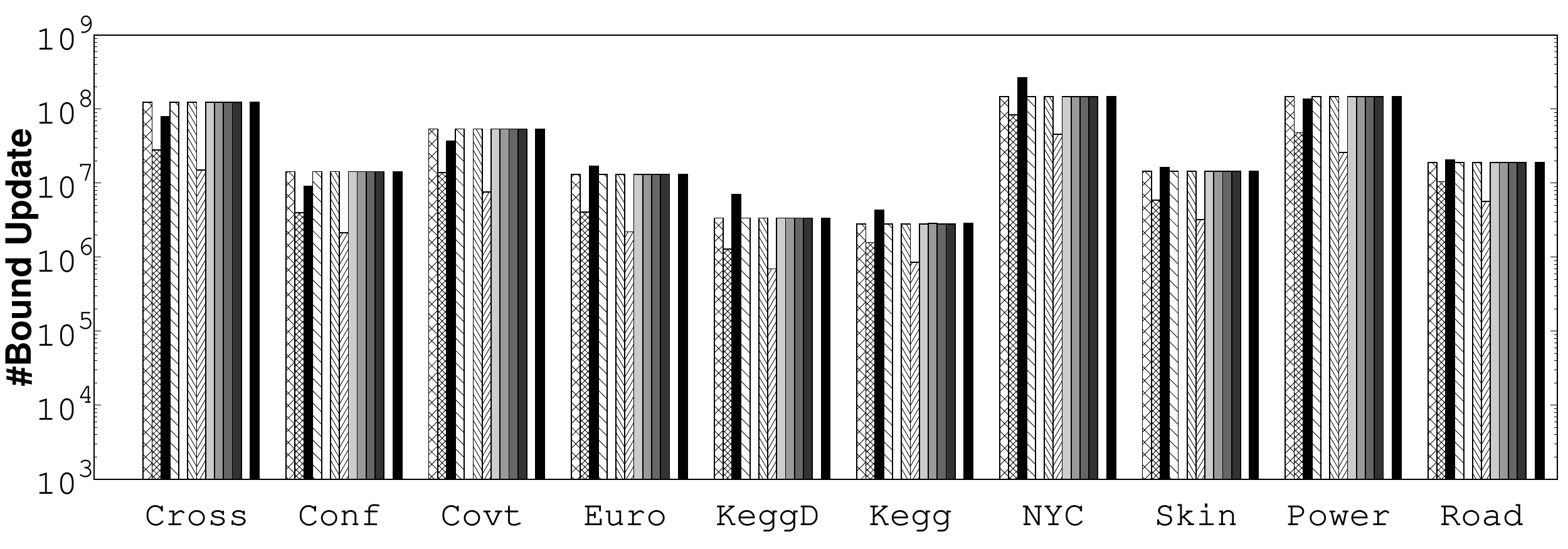}\\\vspace{-0.4em}
			\includegraphics[width=0.9\textwidth,height=0.13\textheight]{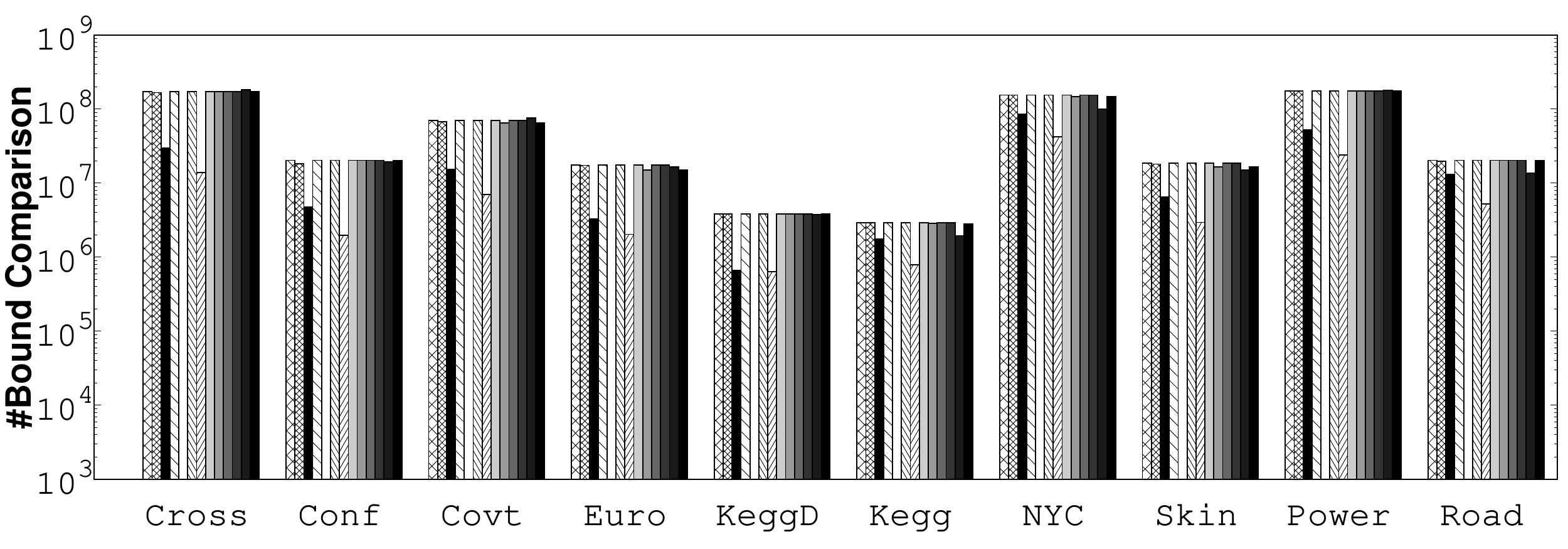}
			\vspace{-1.5em}
			\captionof{figure}{Statistics on the bound accesses and updates ($k=10$).}
			\label{fig:seq14}
		\end{minipage}
\end{table*}

\newpage

\newcommand{\weakcomments}[1]{
	\refstepcounter{cN} 
	\noindent \hangindent=0em 
	\textbf{\textcolor{Maroon}{\underline{Weak Point \thecN}:~}
		\emph{``#1"}	}
}
\newcommand{\comments}[1]{
	\refstepcounter{cN} 
	\noindent \hangindent=0em 
	\textbf{\textcolor{Maroon}{\underline{Comment \thecN}:~}
		\emph{``#1"}	}
}

\newcommand{\response}[1]{
	\\[0.25em] 
	\hangindent=0.3em \textbf{\textcolor{NavyBlue}{\underline{Response}:~}}
	#1
	\vspace{1em} 
}

\newcommand{\newcontent}[1]{
	\\[0.25em] 
	\hangindent=0.3em \textbf{\textcolor{Brown}{\underline{New revision}:~}}
	#1
	\vspace{1em} 
}
\setcounter{section}{9}
\renewcommand\thesection{\Roman{section}}
\renewcommand\thesubsection{\thesection.\Roman{subsection}}
\setcounter{figure}{0}
\renewcommand{\thefigure}{\Alph{figure}}
\pagenumbering{roman}

\end{document}